\documentstyle[12pt,epsfig]{article}
\advance\textheight by \topskip
\begin{document}
\tolerance=100000
\thispagestyle{empty}
\setcounter{page}{0}
\def\lsim{\raisebox{-.1em}{$
\buildrel{\scriptscriptstyle <}\over{\scriptscriptstyle\sim}$}}
\def\gsim{\raisebox{-.1em}{$
\buildrel{\scriptscriptstyle >}\over{\scriptscriptstyle\sim}$}}
\def\preprint{{preprint}}
\begin{flushright}
{TUHEP-TH-07161}\\
{SCUPHY-07002}\\
{SHEP-07-12}\\
{DFTT 40/2009}\\
\end{flushright}

\vspace*{-3.1cm}

\begin{center}
{\Large \bf
Four-lepton LHC events from  \phantom{aaaaaaaa}
\\[0.1cm] 
MSSM Higgs boson decays    \phantom{aaaaaaaa}
\\[0.25cm]
into neutralino and chargino pairs   \phantom{aaaaaaaa} }
\\[0.5 cm]
{\large Mike Bisset$^*$, Jun Li}\\[0.15 cm]
\vskip -0.1cm
{\it Center for High Energy Physics and Department of Physics,}\\
{\it Tsinghua University, Beijing, 100084 P.R. China}
\\[0.5cm]
\vskip -0.17cm
{\large Nick Kersting$^*$}\\[0.15 cm]
\vskip -0.1cm
{\it  Physics Department, Sichuan University, Chengdu, 610065 P.R. China}
\\[0.5cm]
\vskip -0.17cm
{\large Ran Lu$^*$}\\[0.15 cm]
\vskip -0.1cm
{\it  Physics Department, University of Michigan, Ann Arbor, MI 48109, USA}
\\[0.5cm]
\vskip -0.17cm
{\large Filip Moortgat$^*$}\\[0.15 cm]
\vskip -0.1cm
{\it Department of Physics, CERN, CH-1211, Geneva 23, Switzerland}
\\[0.5cm]
\vskip -0.17cm
{\large Stefano Moretti$^{*}$}\\[0.15 cm]
\vskip -0.1cm
{\it School of Physics and Astronomy, University of Southampton,}\\
{\it Highfield, Southampton SO17 1BJ, UK}\\[0.1cm]
{and}\\[0.1cm]
{\it
Dipartimento di Fisica Teorica,
Universit\`a degli Studi di Torino\\
Via Pietro Giuria 1,
10125 Torino,
Italy}
\\[0.25cm]

\end{center}

\begin{abstract}
{\vskip-0.05cm
\noindent\small 
Heavy neutral Higgs boson production and decay into
neutralino and chargino pairs is studied at the Large Hadron Collider
in the context of the minimal supersymmetric standard model. 
Higgs boson decays into the heavier neutralino and chargino 
states, {\it  i.e.},
$H^0,A^0 \rightarrow \widetilde{\chi}_2^0\widetilde{\chi}_3^0,
\widetilde{\chi}_2^0\widetilde{\chi}_4^0,
\widetilde{\chi}_3^0\widetilde{\chi}_3^0,
\widetilde{\chi}_3^0\widetilde{\chi}_4^0,
\widetilde{\chi}_4^0\widetilde{\chi}_4^0$ as well as
$H^0,A^0 \rightarrow \widetilde{\chi}_1^{\pm}\widetilde{\chi}_2^{\mp},
\widetilde{\chi}_2^+ \widetilde{\chi}_2^-$ 
(all leading to four-lepton plus missing transverse energy final states),
is found to improve the possibilities of discovering such Higgs states 
beyond those previously identified by considering  
$H^0,A^0 \rightarrow \widetilde{\chi}_2^0 \widetilde{\chi}_2^0$ decays 
only.  In particular, $H^0,A^0$ bosons with quite heavy masses,
approaching ${\sim}800\, \hbox{GeV}$ in the so-called `decoupling region'
where no clear SM signatures for the heavier MSSM Higgs bosons are 
known to exist, can now be discerned, for suitable but not 
particularly restrictive configurations of the low energy supersymmetric 
parameters. The high $M_A$ discovery reach for the $H^0$ and $A^0$ may 
thus be greatly extended. 
Full event-generator level simulations, including realistic detector 
effects and analyses of all significant backgrounds, are performed to
delineate the potential $H^0,A^0$ discovery regions.
The wedgebox plot technique is also utilized to further analyze the 
$4 \ell$ plus missing transverse energy signal and background 
events.  
This study marks the first thorough and reasonably complete analysis of 
this important class of MSSM Higgs boson signature modes.
In fact, this is the first time discovery regions including all 
possible neutralino and chargino decay modes of the Higgs bosons have
ever been mapped out.
}
\end{abstract}
\newpage

\noindent

\parindent=15pt

\section{Introduction}

Among the most investigated extensions of the standard model
(SM) are those incorporating supersymmetry (SUSY), and among these
the one with the fewest allowable number of new particles and
interactions, the minimal supersymmetric standard model (MSSM), has
certainly received considerable attention.  Yet, when prospective signals
at the Large Hadron Collider (LHC) of the new particle states within the
MSSM are considered, there is still much that needs clarification.  
Nothing underscores this more than the MSSM electroweak symmetry
breaking (EWSB) Higgs sector.  Included therein is a quintet
of Higgs bosons left from the two
$SU(2)_{\hbox{\smash{\lower 0.25ex \hbox{${\scriptstyle L}$}}}}$ 
Higgs doublets after EWSB
 (see \cite{guide,Anatomy2} for more details):  a charged pair, $H^\pm$,
the neutral $CP$-odd $A^0$ and the neutral $CP$-even $h^0$ and
$H^0$ (with $M_h < M_H$).  The entire Higgs sector ({\it i.e.},
masses and couplings to ordinary matter) can be described
at tree-level by only two independent parameters: the mass of one
of the five Higgs states ({\it e.g.}, $M_A$) and
the ratio of the vacuum expectation values of the
two Higgs doublets (denoted by $\tan\beta$). 
These must be augmented to include significant radiative corrections
which most notably raise the upper limit on 
the mass of the light Higgs boson from $M_h \leq M_{Z}$ at
tree-level to ${\lsim}~140\, \hbox{GeV}$ ($150\, \hbox{GeV}$)
with inclusion of corrections up to two loops and assuming a 
stop-sector scale of $M_{SUSY} = 1\, \hbox{TeV}$ ($2\, \hbox{TeV}$)
and $m_t = (178.0 \pm 4.3)\, \hbox{GeV}$ according to \cite{HHW_PR},
or $\lsim 135\, \hbox{GeV}$ with $m_t = (172.6 \pm 1.4)\, \hbox{GeV}$
by \cite{Sven2007} (stop mass range not specified).
This definite upper bound will allow experimentalists to definitively 
rule out such a minimal SUSY scenario at the LHC if such a light Higgs 
state is not observed.  
Thus, the possible production and decay modes of the $h^0$ state 
have understandably 
been investigated in quite some detail \cite{Anatomy2}.  
In contrast, the possibilities 
for the other heavier neutral MSSM Higgs bosons have not been so thoroughly
examined.  Yet it is crucial that the avenues for discovery of these
other MSSM Higgs bosons be well understood since, even if a candidate for
$h^0$ discovery is experimentally identified, it may be indistinguishable 
from a SM Higgs boson 
(this corresponds to the so-called `decoupling
region', with $M_H, M_A \gg 200\, \hbox{GeV}$ and for intermediate to large
values of $\tan\beta$ \cite{Anatomy2,decouple}).
Then the additional identification of heavier Higgs bosons may well 
be required to establish that there is in fact an extended Higgs sector
beyond the single doublet predicted by the SM.

Finding signatures for these heavier MSSM Higgs bosons has proved to be 
challenging.  Unlike the lone Higgs boson of the SM of similar mass,
couplings of these MSSM Higgs bosons to SM gauge bosons are either absent
at tree level (for $A^0$) or strongly suppressed over much of the allowed
parameter space (for $H^0$).  Thus, identification of $A^0$ and $H^0$ via 
their decays into known SM particles relies chiefly on decays of said Higgs 
bosons into the heaviest fermions available, namely, tau leptons and bottom
quarks\footnote{$H^0,A^0$ top quark couplings are suppressed relative to 
a SM Higgs boson of the same mass.}.  
Identification of hadronic decays/jet showers of these third generation 
fermions may be problematic in the QCD-rich environment of the 
LHC\footnote{In addition, jet-free events from Higgs boson decays to 
tau-lepton pairs where both tau-leptons in turn decay leptonically 
also come with considerable background-separation challenges 
\cite{taulepdec}.}, so that it is very questionable that the entire 
parameter space can be covered with just SM-like signatures.  
Fortunately, in the MSSM there is an alternative:  
decays of these Higgs bosons into sparticles, in particular the 
charginos and neutralinos\footnote{In the remainder,
charginos and neutralinos collectively will be abbreviated by `--inos'.} 
formed from the EW gauginos and Higgsinos.  
Higgs boson couplings to certain --ino states may be substantial, and 
these heavy sparticles may themselves decay ---
except for $\widetilde{\chi}_1^0$ which is assumed to be the stable
lightest supersymmetric particle (LSP) --- in readily-identifiable
ways (such as into leptons) to provide a clean experimental signature.

A number of previous articles \cite{PRD1,PRD2,CMS1,CSS_FrIt,HAinv}
as well as at least one Ph.D. thesis \cite{thesis} have focused on the signal 
potential of the decays of the heavier neutral MSSM Higgs bosons into 
neutralinos and charginos:
\begin{eqnarray}
H^0,A^0 \rightarrow \widetilde{\chi}_a^+ \widetilde{\chi}_b^-,
\widetilde{\chi}_i^0 \widetilde{\chi}_j^0 \;\;\;\;\;\;\;
(a,b = 1,2, \;\; i,j = 1,2,3,4).
\label{allproc}
\end{eqnarray}
Therein only subsequent --ino decays into leptons 
(which will be taken to mean electrons and/or muons, 
$\ell=e,\mu$) were considered, as this is preferable 
from the standpoint of LHC detection.  
Since relatively light sleptons can greatly enhance 
\cite{EPJC1,EPJC2,BaerTata}
the branching ratios (BRs) for such decays, 
the properties of the slepton sector of the
MSSM also need to be specified.
All of the previous works concentrated almost\footnote{The decays 
$H^0,A^0 \rightarrow \widetilde{\chi}_1^+ \widetilde{\chi}_1^-,
\widetilde{\chi}_1^0 \widetilde{\chi}_2^0$ were also studied in 
\cite{PRD1} but found to be unproductive due to large backgrounds to 
the resulting di-lepton signals.}
exclusively on the decays
$H^0,A^0 \rightarrow \widetilde{\chi}_2^0 \widetilde{\chi}_2^0$.
In addition, the subsequent neutralino decays
$\widetilde{\chi}_2^0 \rightarrow \widetilde{\chi}_1^0 \ell^+ \ell^-$
were typically
presumed to proceed via three-body decays with an
off-mass-shell intermediate $Z^{0*}$ or slepton, neglecting the 
possibility of the intermediate $Z^0$ or slepton being on-mass-shell
(\cite{HidThr} and \cite{KnK} delve in considerable depth
into the distinctions between these cases).

In this work\footnote{A preliminary account of this analysis is given in
Ref.~\cite{LesHouches2003}.}, {\em all} the decays in (\ref{allproc})
are incorporated. In fact, as the presumed mass of a Higgs boson grows,
more such decay modes will become accessible.  Therefore, if decay channels
to the heavier -inos are significant, they may provide signatures for 
heavier neutral Higgs bosons (with masses well into the aforementioned 
decoupling region).  When heavier --ino states are included, it
also becomes easier to construct model spectra with slepton masses
lying below those of the heavier --inos.  
Thus, in this work, intermediate sleptons are allowed to be both on- and 
off-mass-shell (same for the $Z^{0(*)}$)\footnote{Similar 
studies for charged Higgs boson decays into a neutralino and 
a chargino, where the charged Higgs boson is produced in association with
a $t$ or $\bar{t}$ quark are done in \cite{EPJC1,EPJC2}
(see also Refs.~\cite{LesHouches1999,LesHouches2001}).}. 
More background channels are also emulated than in previous studies.
The Higgs boson production modes considered herein are
$gg \rightarrow H^0, A^0$ (gluon-fusion) 
and $q\bar{q} \rightarrow H^0, A^0$ (quark-fusion).
(The second mode is dominated by the case $q=b$.)

This work is organized as follows. The next section provides an
overview of the MSSM parameter space through calculation
of inclusive rates for the relevant production and decay processes
contributing to the signal. Sect.\ 3 then specializes these
results to the more restrictive minimal supergravity (mSUGRA) scenario
for SUSY breaking.  Sect.\ 4 gives the numerical results for the
signal and background processes based upon Monte Carlo (MC) 
simulations of parton shower (PS) and hadronization as well as detector
effects.  This includes mapping out discovery regions for the LHC.
The recently-introduced `wedgebox' method of \cite{Cascade}, which is
reminiscent of the time-honored Dalitz plot technique, is utilized 
in Sect.\ 5 to extract information about the --ino mass spectra and 
the --ino couplings to the Higgs bosons.  Finally, the last section 
presents conclusions which can be drawn from this study.

\section{MSSM parameter space}

As noted above, $M_A$ and $\tan\beta$ may be chosen as the 
MSSM inputs characterizing the MSSM Higgs bosons' decays into 
SM particles\footnote{Several other MSSM inputs also enter into the 
radiatively-corrected MSSM Higgs boson masses and couplings of the MSSM 
Higgs bosons to SM particles,
namely, inputs from the stop sector --- the soft SUSY-breaking stop
trilinear coupling $A_t$ plus the stop masses --- and the Higgs/Higgsino 
mixing mass $\mu$.  In the present work the stop masses are assumed to be 
heavy ($\approx 1\, \hbox{TeV}$) whereas $A_t$ is fixed to zero.
The $\mu$ parameter is not crucial for the SM decay modes;
however, it will become so when decays to --inos are considered.}.
But when Higgs boson decays to --inos are included, new MSSM inputs
specifying the --ino sector also become crucial.
To identify the latter, the already mentioned Higgs/Higgsino mixing mass, 
$\mu$, and the SUSY-breaking 
$SU(2)_{\hbox{\smash{\lower 0.25ex \hbox{${\scriptstyle L}$}}}}$
gaugino mass, $M_{2}$, in addition to $\tan\beta$, are required.  
The SUSY-breaking 
$U(1)_{\hbox{\smash{\lower 0.25ex \hbox{${\scriptstyle Y}$}}}}$
gaugino mass, $M_{1}$, is assumed to be determined from $M_{2}$ via
gaugino unification
({\it i.e.}, $M_{1} = \frac{5}{3}\tan^2\theta_W M_{2}$).
This will fix the tree-level --ino masses (to which the radiative
corrections are quite modest) along with their couplings to the Higgs
bosons.  

Inputs (assumed to be flavor-diagonal) from the slepton sector are the
left and right soft slepton masses for each of the three generations   
(selectrons, smuons, and staus) and the trilinear `$A$-terms' which come
attached to Yukawa factors and thus only $A_{\tau}$ has a potential
impact.
{\it A priori}, all six left and right mass inputs (and $A_{\tau}$) are
independent.  However, in most models currently advocated, one has
$m_{\widetilde{e}_R} \simeq m_{\widetilde{\mu}_R}$ and
$m_{\widetilde{e}_L} \simeq m_{\widetilde{\mu}_L}$.   
Herein these equalities are assumed to hold.

\subsection{Experimental limits}

To maximize leptonic --ino BR enhancement, sleptons should be made as
light as possible.  But direct searches at LEP \cite{W1LEP2,HiggsLEP2} place
significant limits on slepton masses:
$m_{\widetilde{e}_1} \ge 99.0\, \hbox{GeV}$,
$m_{\widetilde{\mu}_1} \ge 91.0\, \hbox{GeV}$,
$m_{\widetilde{\tau}_1} \ge 85.0\, \hbox{GeV}$
(these assume that the slepton is not nearly-degenerate with the LSP)
and $m_{\widetilde{\nu}} \ge 43.7\, \hbox{GeV}$
(from studies at the $Z^0$ pole).
Furthermore, the sneutrino masses are closely tied to the left soft mass
inputs, and, to avoid extra controversial assumptions, only regions of the 
MSSM parameter space where the LSP is the lightest neutralino rather than a 
sneutrino will be considered\footnote{Further, if a sneutrino were the LSP 
and thus presumably the main constituent of galactic dark matter, its strong 
couplings to SM EW gauge bosons would lead to event rates 
probably inconsistent with those observed by Super-Kamiokande.
In contrast, the coupling of an --ino to SM EW gauge bosons
can be tuned to obtain rates consistent with current experimental
limits.}.
To optimize the --ino leptonic BRs without running afoul of the
LEP limits, it is best\footnote{Unless this leads to
$m_{\widetilde{\nu}} < m_{\widetilde{\chi}_2^0} < 
m_{\widetilde{\ell}^{\pm}}$,
in which case $\widetilde{\chi}_2^0$ decays to charged leptons will be 
suppressed with respect to
 $\widetilde{\chi}_2^0$ decays to neutrinos, to avoid which 
having $m_{\widetilde{\ell}_{\scriptscriptstyle R}} \,  
< \, m_{\widetilde{\ell}_{\scriptscriptstyle L}}$ is preferred.} to set
$m_{\widetilde{\ell}_{\scriptscriptstyle R}}
= m_{\widetilde{\ell}_{\scriptscriptstyle L}}$.
If all three generations have the same soft inputs
(with $A_\tau =A_\ell = 0$), then the
slepton sector is effectively reduced to one optimal input value 
(defined as 
$m_{\widetilde{\ell}_{\scriptscriptstyle soft}} 
\, \equiv m_{\widetilde{\ell}_{\scriptscriptstyle L,R}}$).
However, since --ino decays into tau-leptons are generally not anywhere 
near as beneficial as those into electrons or muons, it would be
even better if the stau inputs were significantly above those of the
first two generations.  This would enhance the --inos' BRs into
electrons and muons.  In the general MSSM, one is of course free to choose
the inputs as such.  Doing so would also weaken restrictions from LEP,   
especially for high values of $\tan\beta$.  Fig.\ 1 in \cite{EPJC2} 
shows values for this optimal slepton mass over the $M_2$--$\mu$ plane
relevant to the --ino sector for $\tan\beta = 10,20$. 
Setting the soft stau mass inputs $100\, \hbox{GeV}$ above those of the 
other soft slepton masses, as will often be done herein, complies 
with current experimental constraints and moderately enhances the signal 
rates \cite{LesHouches2003}.

\subsection{The signal inclusive cross sections}

Figs.\ 1, 2 and 3 show the LHC rates (in fb) for 
$\sigma(pp \rightarrow H^0)$ $\times$ BR$(H^0 \rightarrow 4\ell N)$
$+$
$\sigma(pp \rightarrow A^0)$ $\times$ BR$(A^0 \rightarrow 4\ell N)$,
where $N$ is any number (including zero) of invisible neutral 
particles (in the MSSM these are either neutrinos or 
$\widetilde{\chi}_1^0$ LSPs) 
obtained for $\tan\beta = 5$, $10$, and $20$, respectively\footnote{These 
figures are generated using private codes; however, these have been 
cross-checked against those of the ISASUSY package of ISAJET \cite{ISAJET} 
and the two are generally consistent, exceptions being a few coding errors 
in ISASUSY and the latter's inclusion of some mild radiative corrections 
for the slepton and --ino masses which are not incorporated into the 
codes used here. 
These caveats are noteworthy since results from the output 
of the ISASUSY code will be used as input for the simulation work that 
follows.  These small distinctions may cause a shift in the parameter space 
locations of particularly-abrupt changes in the rates due to encountered 
thresholds, though the gross features found in this section and in the
ISASUSY-based simulation studies are in very good agreement.
Finally, note that higher-order corrections to the Higgs boson --ino --ino
couplings are incorporated into neither ISASUSY nor the private code.
A recent study\cite{radHiggsinoino} indicates that these generally enhance the 
partial decay widths by ${\cal O}$10\%; enhancement to BRs may be even more.
This would make rates reported in this work on the conservative low side.}.
(Hereafter this sum of processes will be abbreviated by 
$\sigma(pp \rightarrow H^0,A^0)$ $\times$ 
BR$(H^0,A^0 \rightarrow 4\ell N)$.)  Each figure gives separate scans of 
the $\mu$ {\it vs.} $M_2$  plane most relevant to the --ino sector for
(from top to bottom) $M_A = 400$, $500$, and $600\, \hbox{GeV}$
--- covering the range of Higgs boson masses of greatest interest 
\cite{LesHouches2003}.
This is in the region of the MSSM parameter space where observation of 
$h^0$ alone may be insufficient to distinguish the MSSM Higgs sector from
the SM case ({\it i.e.}, the decoupling region). The darkened zones seen 
around the lower, inner corner of each plot are the regions excluded by the 
experimental results from LEP.
\begin{figure}[!t]  
\begin{center}
\vspace*{-4.9truecm}
\hspace*{-0.8truecm}
\epsfig{file=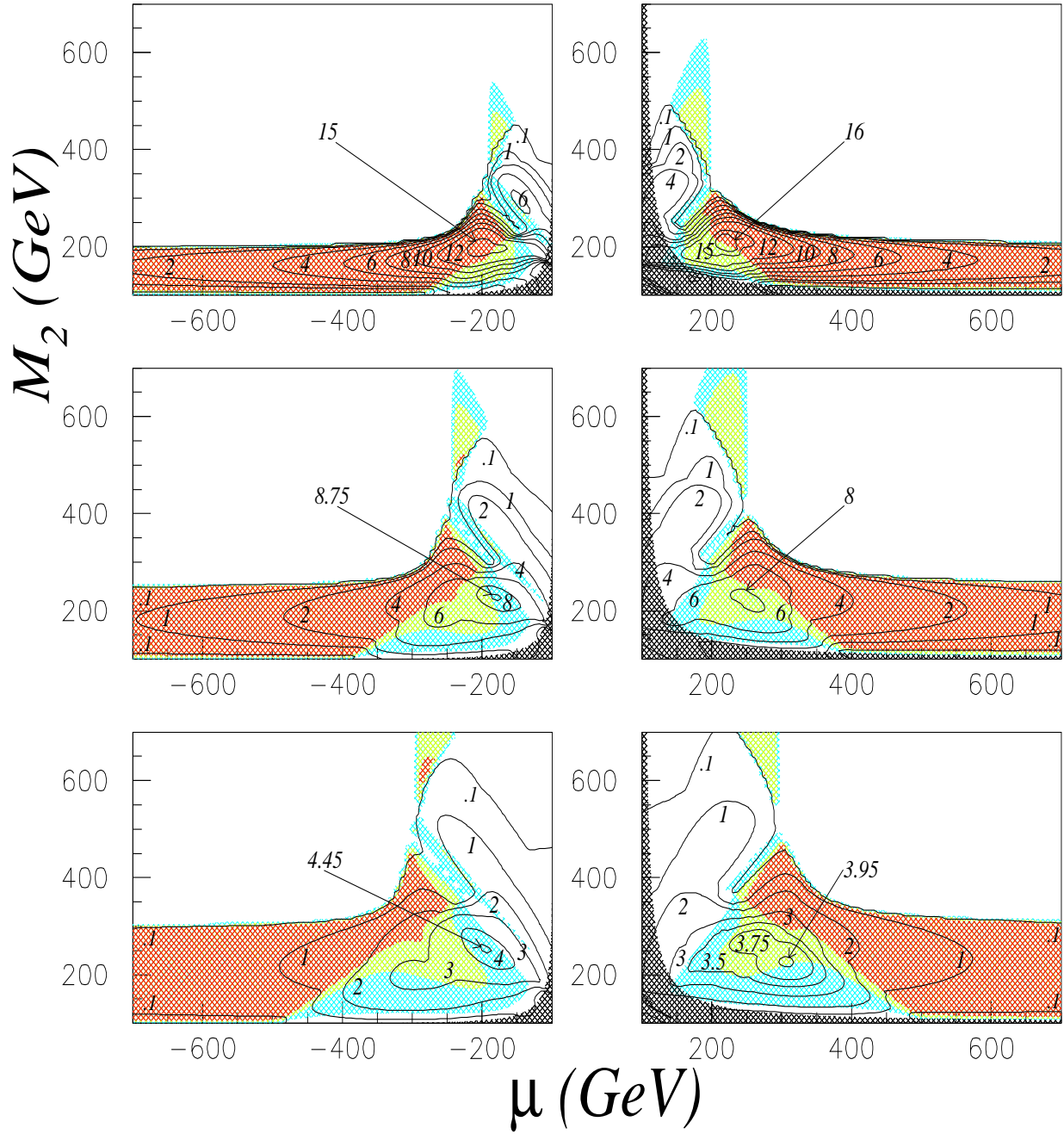,height=225mm,width=185mm}
\end{center}
\vspace*{-1.10truecm}
\caption{
$\sigma(pp \rightarrow H^0,A^0)$ $\times$
BR$(H^0,A^0 \rightarrow 4\ell N)$ (in fb), where $\ell = e^{\pm}$ or
${\mu}^{\pm}$ and $N$ represents invisible final state particles, also
showing where the percentage from $H^0,A^0 \rightarrow 
\widetilde{\chi}^0_2 \widetilde{\chi}^0_2$ is
$>$ 90\% (red), 50\% -- 90\% (yellow), 10\% -- 50\% (light blue),
$<$ 10\% (white), with
$\tan\beta = 5$, $M_A = 400\, \hbox{GeV}$ (top),
$500\, \hbox{GeV}$ (middle), $600\, \hbox{GeV}$ (bottom).
Optimized slepton masses (with stau inputs raised $100\, \hbox{GeV}$)
are used, and with
$m_t = 175\, \hbox{GeV}$, $m_b = 4.25\, \hbox{GeV}$,
$m_{\widetilde q} = 1\, \hbox{TeV}$, 
$m_{\widetilde g} = 800\, \hbox{GeV}$,
$A_\tau=A_{\ell} = 0$. The cross-hatch shaded areas are excluded by LEP.}
\label{tb5color}
\vspace*{1.0truecm}
\end{figure}

\begin{figure}[!t]  
\begin{center}
\vspace*{-4.9truecm}
\hspace*{-0.8truecm}
\epsfig{file=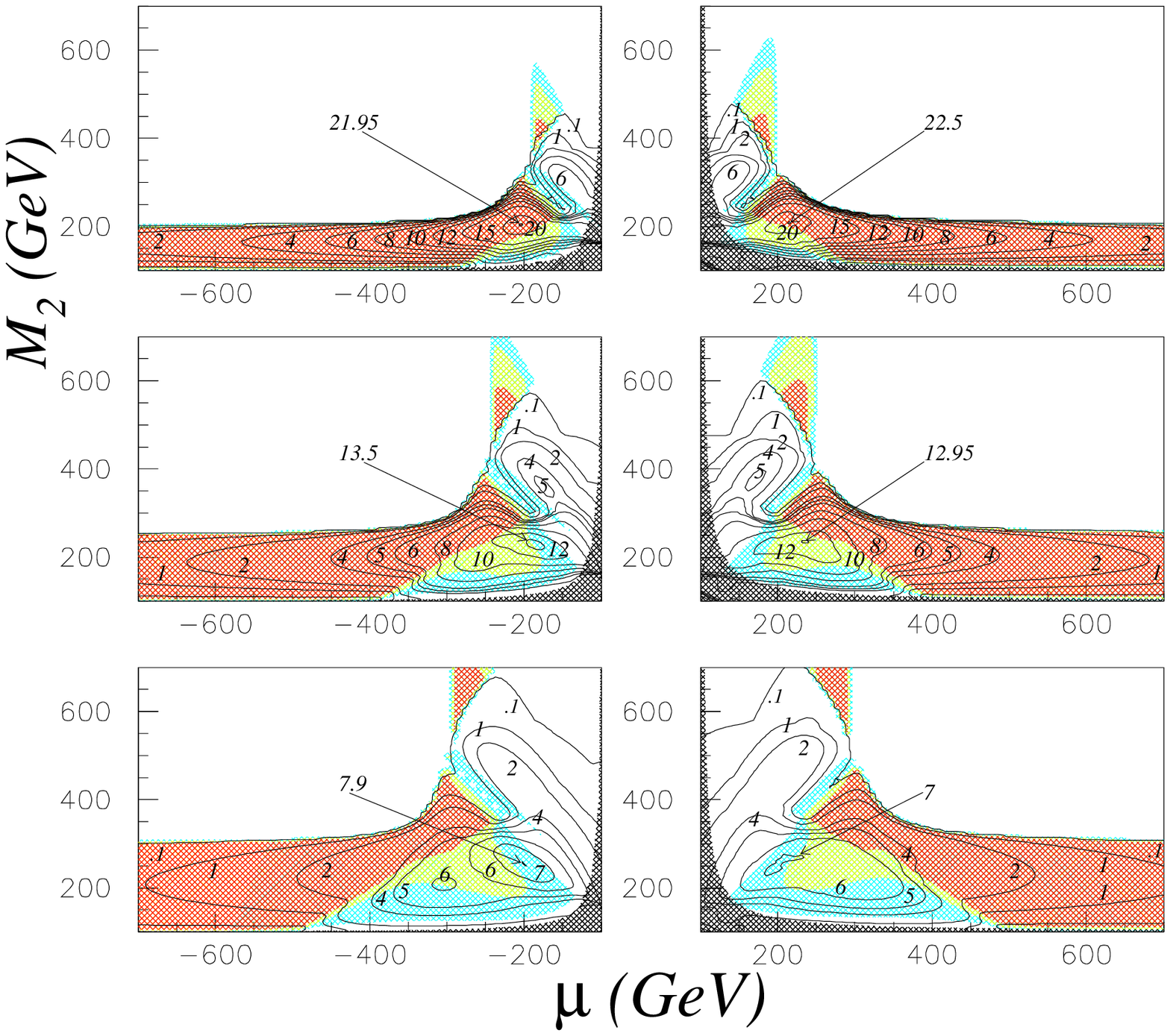,height=225mm,width=185mm}
\end{center}
\vspace*{-1.10truecm}
\caption{
$\sigma(pp \rightarrow H^0,A^0)$ $\times$
BR$(H^0,A^0 \rightarrow 4\ell N)$ (in fb), where $\ell = e^{\pm}$ or
${\mu}^{\pm}$ and $N$ represents invisible final state particles, also
showing where the percentage from $H^0,A^0 \rightarrow
\widetilde{\chi}^0_2 \widetilde{\chi}^0_2$ is
$>$ 90\% (red), 50\% -- 90\% (yellow), 10\% -- 50\% (light blue),
$<$ 10\% (white), with
$\tan\beta = 10$, $M_A = 400\, \hbox{GeV}$ (top),
$500\, \hbox{GeV}$ (middle), $600\, \hbox{GeV}$ (bottom).
Optimized slepton masses (with stau inputs raised $100\, \hbox{GeV}$)
are used, and with
$m_t = 175\, \hbox{GeV}$, $m_b = 4.25\, \hbox{GeV}$,
$m_{\widetilde q} = 1\, \hbox{TeV}$, 
$m_{\widetilde g} = 800\, \hbox{GeV}$,
$A_\tau=A_{\ell} = 0$. The cross-hatch shaded areas are excluded by LEP.}
\label{tb10color}
\vspace*{1.0truecm}
\end{figure}

\begin{figure}[!t]
\begin{center}
\vspace*{-4.9truecm}
\hspace*{-0.8truecm}
\epsfig{file=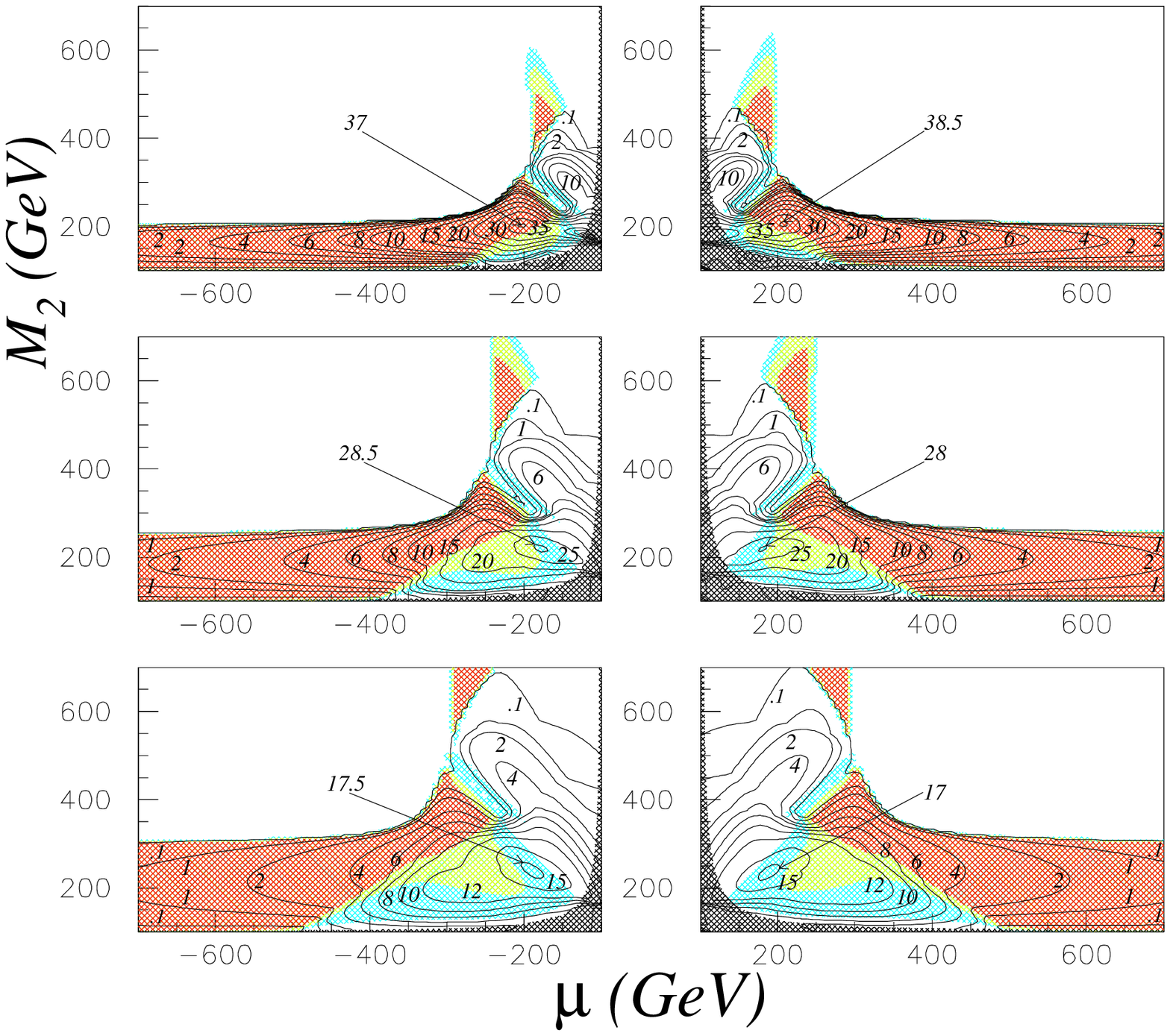,height=225mm,width=185mm}
\end{center}
\vspace*{-1.10truecm}
\caption{
$\sigma(pp \rightarrow H^0,A^0)$ $\times$
BR$(H^0,A^0 \rightarrow 4\ell N)$ (in fb), where $\ell = e^{\pm}$ or
${\mu}^{\pm}$ and $N$ represents invisible final state particles, also
showing where the percentage from $H^0,A^0 \rightarrow
\widetilde{\chi}^0_2 \widetilde{\chi}^0_2$ is
$>$ 90\% (red), 50\% -- 90\% (yellow), 10\% -- 50\% (light blue),  
$<$ 10\% (white), with
$\tan\beta = 20$, $M_A = 400\, \hbox{GeV}$ (top),
$500\, \hbox{GeV}$ (middle), $600\, \hbox{GeV}$ (bottom).
Optimized slepton masses (with stau inputs raised $100\, \hbox{GeV}$)
are used, and with
$m_t = 175\, \hbox{GeV}$, $m_b = 4.25\, \hbox{GeV}$,
$m_{\widetilde q} = 1\, \hbox{TeV}$, 
$m_{\widetilde g} = 800\, \hbox{GeV}$,  
$A_\tau=A_{\ell} = 0$. The cross-hatch shaded areas are excluded by LEP.}
\label{tb20color}
\vspace*{1.0truecm}
\end{figure}

First observe that these `raw' 
or `inclusive' ({\it i.e.}, before applying selection cuts 
to the basic event-type) rates may be sufficiently large.  For an 
integrated luminosity of $100\, \hbox{fb}^{-1}$, the peak raw event number
is around 4000(1700) events for
$M_A = 400$($600$) GeV and $\tan\beta=20$, irrespective of the sign of 
$\mu$.  Also observe that low values of $| \mu |$ and $M_2$
yield the highest signal rates, though significant event numbers are also 
found when one but not the other of these parameters is increased 
(especially $| \mu |$; rates do fall rapidly when $M_2$
increases much beyond $500\, \hbox{GeV}$).  These numbers are substantial 
(especially at high $\tan\beta$) and, if experimental efficiencies are 
good, they may facilitate a much more accurate determination of some masses or 
at least mass differences in the -ino spectrum as well as the Higgs-ino 
mass differences than those achieved in previous studies based solely 
on $H^0,A^0 \rightarrow \widetilde{\chi}_2^0 \widetilde{\chi}_2^0$ decays. 

Note the color coding of the three figures depicting what percentage of 
the signal events are coming from Higgs boson decays to
$\widetilde{\chi}_2^0 \widetilde{\chi}_2^0$:  $> 90$\% in the red zones,
from $90$\% down to $50$\% in the yellow zones, from
$50$\% to $10$\% in the blue zones, and $< 10$\% in uncolored regions.  
If the events are not coming from 
$\widetilde{\chi}_2^0 \widetilde{\chi}_2^0$, 
then they are almost always from Higgs boson decays including heavier 
neutralinos, {\it i.e.},
$H^0,A^0 \rightarrow \widetilde{\chi}_2^0 \widetilde{\chi}_3^0,
\widetilde{\chi}_2^0 \widetilde{\chi}_4^0,
\widetilde{\chi}_3^0 \widetilde{\chi}_3^0,
\widetilde{\chi}_3^0 \widetilde{\chi}_4^0,
\widetilde{\chi}_4^0 \widetilde{\chi}_4^0$
(possibly also with contributions from
$H^0,A^0 \rightarrow \widetilde{\chi}_1^{\pm} \widetilde{\chi}_2^{\mp},
\widetilde{\chi}_2^+ \widetilde{\chi}_2^- $ which are also
taken into account here). Also note that the main source of events 
at the optimal location in the --ino parameter space
shifts from $\widetilde{\chi}_2^0 \widetilde{\chi}_2^0$ to
heavier --ino pairs as $M_A$ grows from $400$ to $600\, \hbox{GeV}$.
Irrespective of the heavier Higgs boson masses, Higgs boson decays
to $\widetilde{\chi}_2^0 \widetilde{\chi}_2^0$ are the dominant
source of signal events in regions with low $M_2$ values and moderate to
high values of $| \mu |$.  But for low to moderate $M_2$ values and
low values of  $| \mu |$, the dominant source of signal events shifts to 
the previously-neglected decays into the heavier --inos.  Thus, inclusion
of these neglected modes opens up an entirely new sector of the MSSM 
parameter space for exploration.  Furthermore, the parameter space
locations with the maximum number of signal events also shifts to these
new sectors as the masses of the Higgs bosons rise.  Therefore, the 
regions in MSSM parameter space wherein 
$\sigma(pp \rightarrow H^0,A^0)$ $\times$ 
BR$(H^0,A^0 \rightarrow 4\ell N)$ 
processes can be utilized in the search for the heavier MSSM Higgs bosons 
will certainly expand substantially with inclusion of these additional 
decay channels.

The rates illustrated in Figs.\ 1--3 incorporate indirect decay
modes.  That is, if the Higgs boson decays into a pair of neutralinos, and
then one or both of these `primary' neutralinos decay into other
neutralinos (or other sparticles or the light Higgs boson or both on-
and off-mass-shell 
SM gauge bosons) which in turn give rise to leptons ({\em with no 
additional colored daughter particles}),
then the contribution from such a decay chain is taken into account.
This remains true no matter how many decays there are in the chain
between the primary --ino and the $4\ell N$ final state, the only 
restrictions being that each decay in the chain must be a tree-level
decay with at most one virtual intermediate state (so $1$ to $3$ decay
processes are included but not $1$ to $4$ decays, {\it etc.}). (As already
intimated, the intermediate state is expected to be an 
on- or off-mass-shell SM gauge boson or slepton, charged or neutral.)  
The decay modes omitted due to these restrictions are never expected to be 
significant.  
Thus, effectively all tree-level decay chains allowable within the MSSM 
have been taken into account.  Potential contributions from literally 
thousands of possible decay chains are evaluated and added to the results.

Inspection of Figs.\ 1--3 supports selection of the following  
representative points in the MSSM parameter space to be employed 
repeatedly in this work.  These are:
\vskip 0.6cm
{\bf Point 1}.
$M_A = 500\, \hbox{GeV}$,
$\tan\beta = 20$,
$M_1 = 90\, \hbox{GeV}$, $M_2 = 180\, \hbox{GeV}$,
$\mu$ = $-500\, \hbox{GeV}$,
\newline
\phantom{aaaaaaaaaaa}
$m_{\widetilde{\ell}_{soft}} = 
m_{\widetilde{\tau}_{soft}} = 250\, \hbox{GeV}$,
$m_{\widetilde{g}} = m_{\widetilde{q}} = 1000\, \hbox{GeV}$.
\vskip 0.45cm
{\bf Point 2}.
$M_{A} = 600\, \hbox{GeV}$
$\tan\beta = 35$,
$M_1 = 100\, \hbox{GeV}$ $M_2 = 200\, \hbox{GeV}$
$\mu = -200\, \hbox{GeV}$,
\newline
\phantom{aaaaaaaaaaa}
$m_{\widetilde{\ell}_{soft}} = 150\, \hbox{GeV}$,
$m_{\widetilde{\tau}_{soft}} = 250\, \hbox{GeV}$,
$m_{\widetilde{g}} = 800\, \hbox{GeV}$,
$m_{\widetilde{q}} = 1000\, \hbox{GeV}$.
\vskip 0.6cm
\parindent=0pt
(Also recall that $m_{\widetilde{\ell}_{soft}} 
\equiv m_{\widetilde{\ell}_{\scriptscriptstyle R}}
= m_{\widetilde{\ell}_{\scriptscriptstyle L}}$ and
$A_\tau=A_\ell=0$.)
Point 1 represents a case where most of the signal events result from 
$H^0,A^0 \rightarrow
\widetilde{\chi}_2^0 \widetilde{\chi}_2^0$ decays\footnote{This choice of
parameters, including the degenerate soft selectron, smuon and stau 
inputs, also corresponds to one of the choices adopted in 
\cite{CMS1}.},
whereas Point 2 is a case where decays including heavier
-inos make the dominant contribution.  Here $\tan\beta$ has been 
set fairly high to enhance rates, as Figs.\ 1--3 suggest.

\parindent=15pt

In Fig.\ 4, the parameter values of Point 1 (left plot) and Point 2
(right plot) are adopted, save that the parameters $M_A$ and $\tan\beta$
are allowed to vary, generating plots in the  $M_A$ {\it vs.} $\tan\beta$ 
plane.  Color shading on the left-side plot clearly shows that the 
$\widetilde{\chi}_2^0 \widetilde{\chi}_2^0$ decay modes totally dominate in 
the production of $4\ell$ signal events for this choice of $M_2$, $\mu$ 
-ino inputs out to $M_A \simeq 700\, \hbox{GeV}$.  Similarly, the 
right-side plot shows that for the --ino inputs of Point 2 
the previously neglected decay modes to heavier --inos dominate, save for 
a relatively small region around $M_A$ $\sim$ $350$-$450\, \hbox{GeV}$ and 
$\tan\beta \sim 2$-$10$. Color coding as in Figs.\ 1--3.

It will be noteworthy to compare the declines in raw rates with increasing 
$M_A$ and decreasing $\tan\beta$ shown here to the corresponding 
$M_A$ {\it vs.} $\tan\beta$ discovery region plots based 
on detailed simulation analyses presented in the analysis section to 
follow.
\begin{figure}[!t]
\begin{center}
\vspace*{-2.1truecm}
\epsfig{file=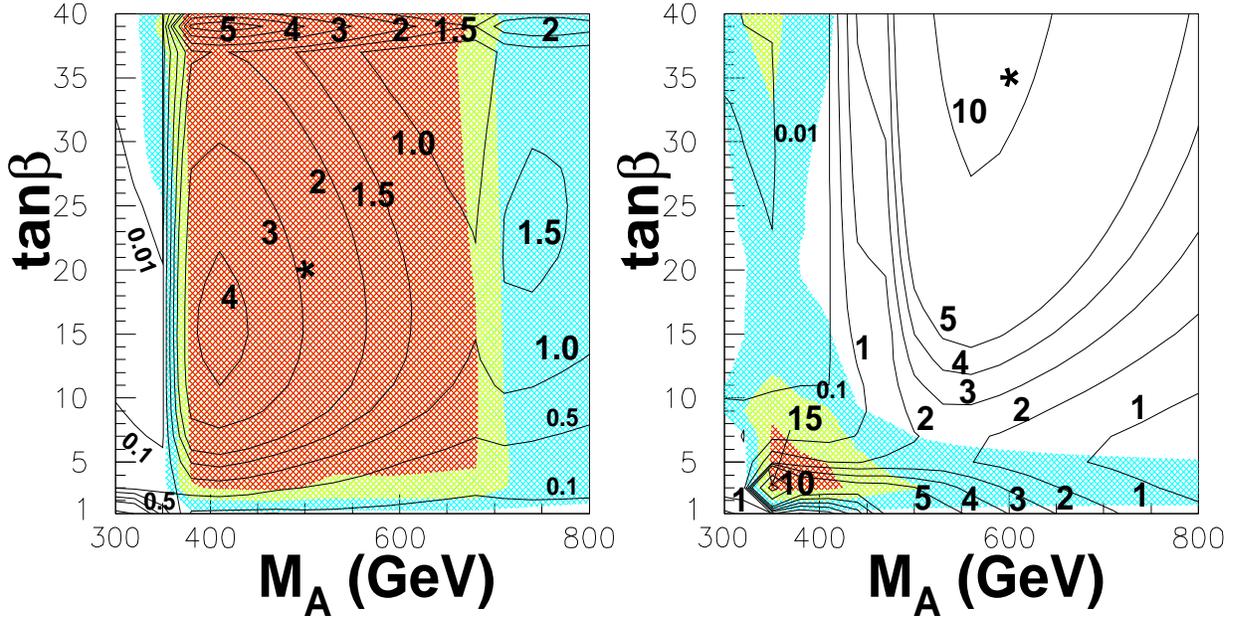,height=190mm,width=180mm}
\end{center}
\vspace*{-9.7truecm}
\caption{
$\sigma(pp \rightarrow H^0,A^0)$ $\times$
BR$(H^0,A^0 \rightarrow 4\ell N)$ (in fb), where $\ell = e^{\pm}$ or
${\mu}^{\pm}$ and $N$ represents invisible final state particles
for Point 1 (left side):
$M_1 = 90\, \hbox{GeV}$, $M_2=180\, \hbox{GeV}$,
$\mu$ = $-500\, \hbox{GeV}$,
$m_{\widetilde{\ell}_{\scriptscriptstyle soft}} = 
m_{\widetilde{\tau}_{\scriptscriptstyle soft}} = 250\, \hbox{GeV}$,
$m_{\widetilde{g}} = m_{\widetilde{q}} = 1000\, \hbox{GeV}$;
and Point 2 (right side):
$M_1 = 100\, \hbox{GeV}$, $M_2= 200\, \hbox{GeV}$,
$\mu = -200\, \hbox{GeV}$,
$m_{\widetilde{\ell}_{\scriptscriptstyle soft}} /
m_{\widetilde{\tau}_{\scriptscriptstyle soft}} = 150 / 250\, \hbox{GeV}$,
$m_{\widetilde{g}} / m_{\widetilde{q}} = 800/1000\, \hbox{GeV}$.
Color coding as in Figs.\ 1--3.
}
\label{fig:cmsNlj-sig}
\end{figure}

Fig.\ 5 illustrates how results depend on the slepton mass(es).
In the upper plot, showing the overall rate,
$\sigma(pp \rightarrow H^0,A^0)$ $\times$ 
BR$(H^0,A^0 \rightarrow 4\ell N)$,
as a function of 
$m_{\widetilde{\ell}_{\scriptscriptstyle soft}} \equiv
m_{\widetilde{\ell}_{\scriptscriptstyle L,R}}$,
one generally sees the naively expected decline in the rate as 
$m_{\widetilde{\ell}_{\scriptscriptstyle soft}}$ increases.  If the 
--inos decay through on- or off-mass-shell sleptons, then the decay products 
always include leptons (and usually charged leptons).  However, as the 
sleptons become heavier (first becoming kinematically inaccessible as 
on-mass-shell intermediates and then growing increasingly disfavored as 
off-mass-shell intermediates), the EW gauge bosons become the dominant 
intermediates, in which case a large fraction of the time the decay products 
will be non-leptons, and so the BR to the $4\ell$ final state drops.  
The plot though also reveals an often far more complex dependence on 
$m_{\widetilde{\ell}_{\scriptscriptstyle soft}}$, with rapid 
oscillations in the rate possible for modest changes in 
$m_{\widetilde{\ell}_{\scriptscriptstyle soft}}$.

\begin{figure}[!t]
\begin{center}
\vspace*{-0.2truecm}
\hspace*{-0.6truecm}
\epsfig{file=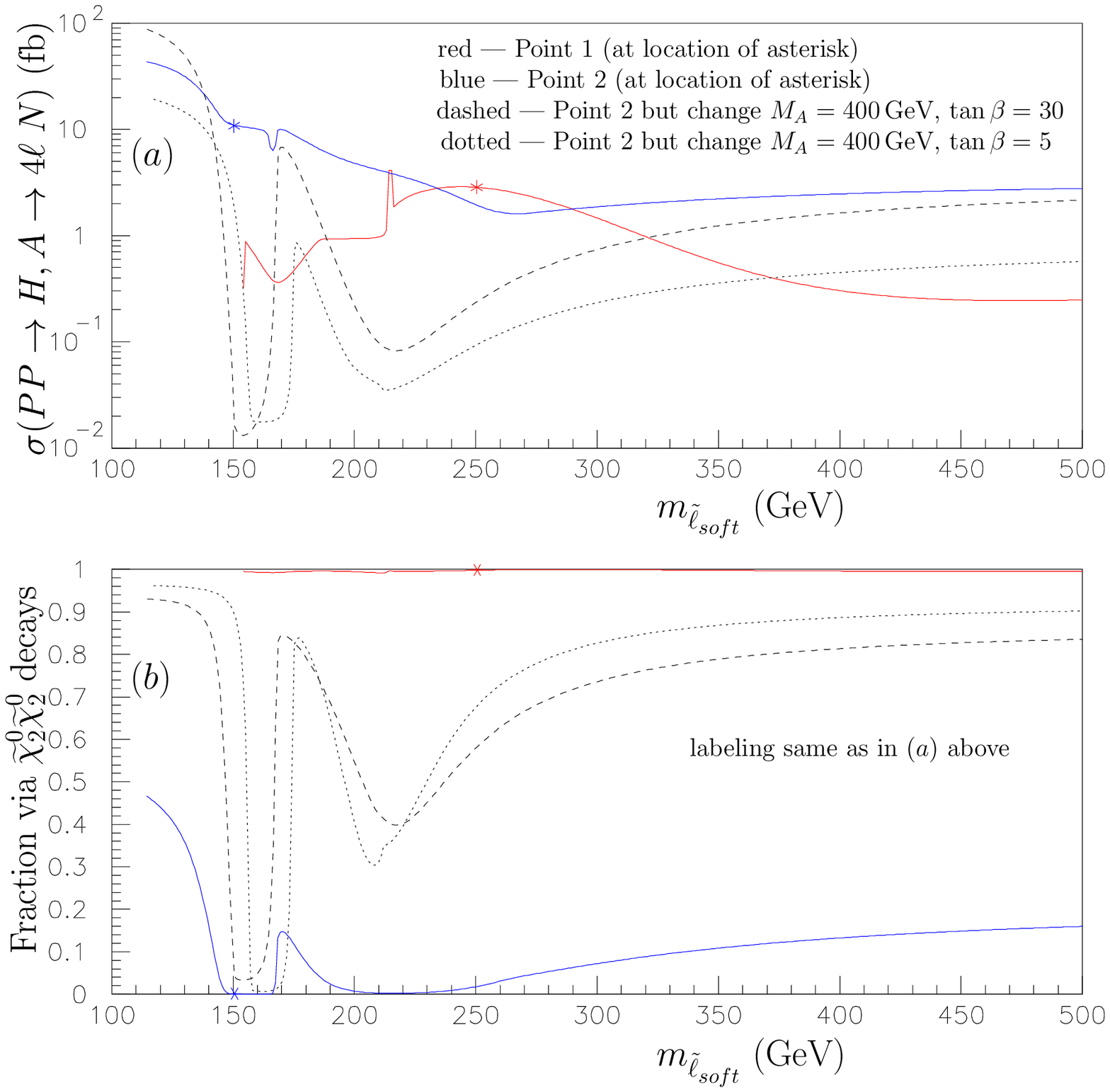,height=185mm,width=160mm}
\end{center}
\vspace*{-0.5truecm}
\caption{  Dependence on slepton mass.
$(a)$ $\sigma(pp \rightarrow H^0,A^0)$ $\times$
BR$(H^0,A^0 \rightarrow 4\ell N)$ (in fb), where $\ell = e^{\pm}$ or
${\mu}^{\pm}$ and $N$ represents invisible final state particles,
{\it vs.}
$m_{\widetilde{\ell}_{\scriptscriptstyle soft}}  \equiv 
m_{\widetilde{\ell}_{\scriptscriptstyle L,R}}$
for MSSM parameter Point 1 (red) and Point 2 (blue) as well as 
some variations based on Point 2 (black).  Asterisks mark the 
$m_{\widetilde{\ell}_{\scriptscriptstyle soft}}$ values to be used for 
Points 1 and 2 later in this work. 
$(b)$ percentage of the inclusive rate from
$\widetilde{\chi}_2^0 \widetilde{\chi}_2^0$ decays
{\it vs.\ } $m_{\widetilde{\ell}_{\scriptscriptstyle soft}}$, with other 
labeling as in $(a)$.
}
\label{fig:mslepton}
\end{figure}

Note again that Point 1, drawn in red in Fig.\ 5, represents a case 
where most of the signal events result from $H^0,A^0 \rightarrow 
\widetilde{\chi}_2^0 \widetilde{\chi}_2^0$ decays, 
whereas Point 2, drawn in blue, is a case where decays including heavier 
--inos make the dominant contribution.  This is made clear by the lower 
plot where the percentage of the inclusive rate from 
$\widetilde{\chi}_2^0 \widetilde{\chi}_2^0$ decays is plotted 
{\it vs.\ } $m_{\widetilde{\ell}_{\scriptscriptstyle soft}}$.  
In Fig.\ 5, the slepton mass is varied.  But later in this work the
value of $m_{\widetilde{\ell}_{\scriptscriptstyle soft}}$ will be fixed 
at the values given earlier for Points 1.\ and 2.\
(these locations are marked by asterisks in both plots in Fig.\ 5).
These choices are fairly optimal, especially for Point 1.

Points 1.\ and 2.\ show some interesting dependence on 
$m_{\widetilde{\ell}_{\scriptscriptstyle soft}}$.
This dependence can be made more acute though by adjusting the input
parameters.  For instance, the black dotted and dashed curves in 
Fig.\ 5 result from lowering the $M_A$ value of Point 2 to 
$400\, \hbox{GeV}$ and changing $\tan\beta$ from $35$ to
$5$ and $30$, respectively.  Then not only does the inclusive rate 
undergo rapid variation with 
$m_{\widetilde{\ell}_{\scriptscriptstyle soft}}$, but the percentage 
of the inclusive rate from $\widetilde{\chi}_2^0 \widetilde{\chi}_2^0$ 
decays fluctuates rapidly as well.  Points 1.\ and 2.\ were selected for 
further analysis later in this work in part because the results are not
strongly affected by a small shift in the value of 
$m_{\widetilde{\ell}_{\scriptscriptstyle soft}}$.
However, apparently this is not true for all points in MSSM parameter 
space. 

Finally, notice that the 
overall normalization of both processes $gg \rightarrow H^0,A^0$ and 
$b\bar{b} \rightarrow H^0,A^0$ is of $2 \rightarrow 1$ 
lowest-order\footnote{There is an alternative $2 \rightarrow 3$ approach 
based on MC implementation of 
$gg/q\bar q \rightarrow b\bar{b} H^0 , b\bar{b} A^0$ diagrams.  The
results of these two approaches have been compared and contrasted in
Ref.~\cite{2to1vs2to3}.  A full MC implementation for the $2 \rightarrow 3$
approach based on $gg \rightarrow ggH^0,ggA^0$ and related modes
(eventually yielding two jets in the final state alongside
$H^0$ or $A^0$) \cite{gluglucontrib}) is as-of-yet unavailable though in
public event generators.  It is therefore more consistent to solely
employ {\it complete} $2\to1$ emulations and not {\it incomplete}
 $2\to3$ ones.}.
Each of these gluon- and quark-fusion partonic contributions is separately 
convoluted with an empirical set of PDFs (CTEQ 6M \cite{CTEQ6} in this 
case) to obtain predictions at the proton-proton level, for which the total 
center-of-mass energy is $\sqrt{s}=14\, \hbox{TeV}$.  
The cross-section thus defined is computed using the MSSM implementation 
\cite{HWMSSM} of the HERWIG program \cite{HERWIG}
(as available in
Version 6.5 \cite{HERWIG65},
with the exception of the choices $m_t = 175\, \hbox{GeV}$ and 
$m_b = 4.25\, \hbox{GeV}$ for the top and bottom quark masses)
and the MSSM input information produced by ISASUSY
(through the ISAWIG \cite{ISAWIG} and HDECAY \cite{HDECAY} interfaces).
Sometimes a Higgs boson will be produced in association with jets, and
thus, as discussed in Ref.~\cite{2to1vs2to3}, 
what percentage of the time a Higgs boson is produced with hadronic activity 
passing jet selection criteria (as will be applied in the analysis section) 
is (possibly) sensitive to the type of emulation ($2 \rightarrow 1$ or 
$2 \rightarrow 3$) being employed.  
Note though that in Figs.\ 1--4 colored fermions are not allowed in the --ino 
decay chains. 
This is in fact inconsistent and leads to an over-(under-)estimate of 
the hadronically-quiet (inclusive, allowing jets) $4 \ell$ rates
(the under-estimation of the inclusive rates is expected to be modest
due to the price of extra BRs in the decay chains of the neglected 
channels).  To attempt to correct for this by factoring in results from 
the simulation runs might obscure what is meant by 'raw' rates, so
this minor inconsistency is simply tolerated in these estimates.

\subsection{Signal-to-background rates}

The signal, taken here to be events resulting from heavy MSSM
Higgs bosons decaying into --ino pairs, is not the only relevant
quantity in this analysis that depends on the position in the MSSM 
parameter space --- backgrounds from other MSSM processes will also
vary from point to point.  Fig.\ 1 of \cite{Cascade} shows the
competing processes for --ino pair-production via Higgs boson 
decays\footnote{One could also consider signals from Higgs boson
decays to other sparticles, especially sleptons.  This was discussed
in \cite{Higgsslep}, which demonstrated that the heavier MSSM Higgs 
boson decays to sleptons only have sufficient BRs for low values of 
$\tan\beta$ ($\lsim \, 3$).}: 'direct' --ino production ({\it i.e.}, 
{\it via} a s-channel gauge boson) and --inos produced in 'cascade' 
decays of squarks and gluinos.  The latter is considered in some detail 
in \cite{Cascade}, 
but will be removed from consideration here by making the assumption 
throughout this work that gluinos and squarks are heavy ({\it circa} 
$1\, \hbox{TeV}$). 
However, since the signal depends on them, (all) the --inos cannot be 
made heavy\footnote{The sleptons also cannot be made arbitrarily heavy. 
Direct slepton pair-production, as studied in \cite{directslep}, will 
{\em generally} lead to dilepton final states rather than the $4\ell$ final 
state desired here.  The smaller contributions from these processes are 
included in the analyses to follow.}, 
and the masses of the EW gauge bosons are known, so the direct channel 
background cannot be easily removed by restricting the analysis to some 
subset of the parameter space by means of such a straight-forward 
assumption.

In fact, the location in the parameter space where the raw signal rate is 
largest sometimes differs from that where the ratio of the signal to the 
leading background from direct --ino production is largest.  
For instance, the plot in Fig.\ 2 ($\tan\beta = 10$) for 
$M_A = 600\, \hbox{GeV}$ shows a maximum in the inclusive rate at
approximately $(\mu, M_2) = (-200\, \hbox{GeV},~250\, \hbox{GeV})$.
On the other hand, the  signal-to-background ratio ($S/B$)
is largest at $\approx(-250\, \hbox{GeV},~500\,\hbox{GeV})$. 
The production cross-section for the Higgs bosons is the same at
both points.  Thus, to understand why the two locations differ so
much the BR$(H^0,A^0 \rightarrow 4\ell N)$ and the direct --ino production 
$\times$ BR$($ -inos $\rightarrow 4\ell N)$ need to be studied.  The 
former drops from ${\sim}6$\% to ${\sim}2$\% in moving from the inclusive
rate maximum to the $S/B$ maximum (thus cutting the overall signal rate 
by a factor of 3).  The background at the inclusive rate maximum is 
mostly  $\widetilde{\chi}^0_2 \widetilde{\chi}^0_3$, 
$\widetilde{\chi}^0_3 \widetilde{\chi}^0_4$ 
and $\widetilde{\chi}^\pm_2 \widetilde{\chi}^\mp_2$
with respective production cross-sections (and BRs into $4\ell N$ final 
states) of
$4\times 10^{-2}\, \hbox{pb}$ ($18$\%),
$1\times 10^{-2}\, \hbox{pb}$ ($8$\%) and
$1\times 10^{-2}\, \hbox{pb}$ ($2$\%).
At the point where the $S/B$ is a maximum, these (still dominant) 
backgrounds rates shift to
$1\times 10^{-2}\, \hbox{pb}$ ($16$\%),
$1\times 10^{-4}\, \hbox{pb}$ ($27$\%) and
$1\times 10^{-2}\, \hbox{pb}$ ($2$\%),
respectively.  So the ${\widetilde\chi}^0_2{\widetilde\chi}^0_3$ 
production rate drops by a factor of $4$ while 
${\widetilde\chi}^0_3{\widetilde\chi}^0_4$
production almost vanishes (which is the main factor), mostly because of 
increased phase space suppression due to larger --ino masses:
$m_{{\widetilde\chi}^0_2}(m_{{\widetilde\chi}^0_3})
[m_{{\widetilde\chi}^0_4}]\{m_{{\widetilde\chi}^\pm_2}\}$ changes from
$118(180)[212]\{289\}\, \hbox{GeV}$ at the rate maximum to 
$219(257)[273]\{515\}\, \hbox{GeV}$ at the $S/B$ maximum.  
The result is that the overall background rate drops by a factor of 5.  
In short, the $S/B$ improves because the direct --ino pair-production 
cross-section falls more rapidly than the signal BR into $4\ell N$ 
final states.
Analogous plots to those in Figs.\ 1--3 studying the $S/B$ variation
across the parameter space are not presented.  Instead, discovery regions
for selected --ino input parameter sets will be given in Sect.\ 4.  
While favorable MSSM points have been chosen for the simulation analyses,
they were not selected to maximize the $S/B$.  Therefore, this channel 
may work even better at points other than those analysed in detail herein.
  
\section{mSUGRA parameter space}

\begin{figure}[!t]
\begin{center}
\vspace*{-4.40truecm}
\epsfig{file=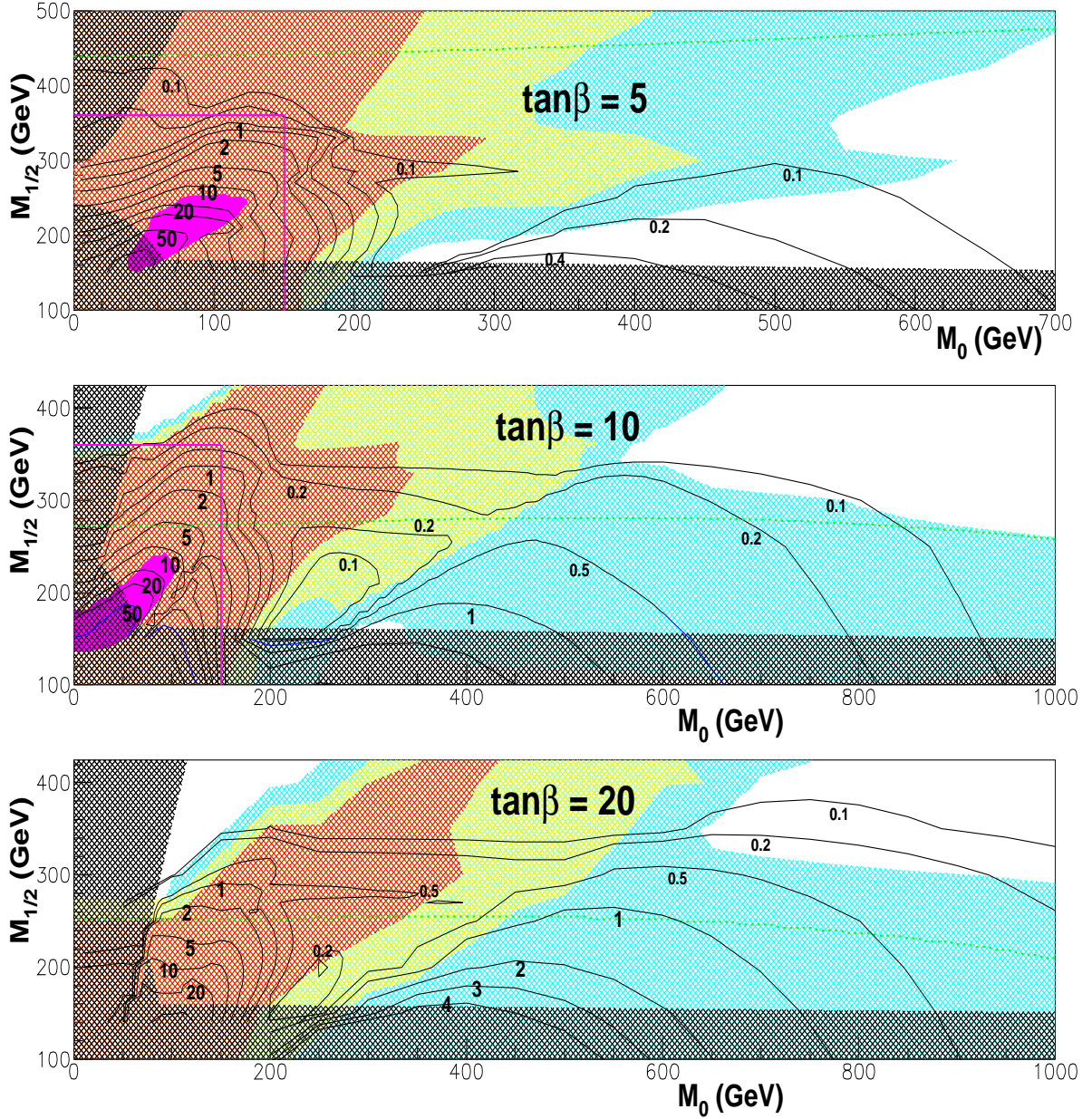,height=220mm,width=180mm}
\end{center}
\vspace*{-1.8truecm}
\caption{
$\sigma(pp \rightarrow H^0,A^0)$ $\times$
BR$(H^0,A^0 \rightarrow 4\ell N)$ (in fb), where $\ell = e^{\pm}$ or
${\mu}^{\pm}$ and $N$ represents invisible final state particles
for $\tan\beta = 5,10,20$ in the mSUGRA 
$M_{{0}}$ {\it vs.}\  $M_{{{1}/{2}}}$ plane, with ${\rm{sgn}} ( \mu ) = +1$ 
and $A_{{0}} = 0$.  Colors depict the 
percentage of events stemming from $H^0,A^0 \rightarrow
\widetilde{\chi}^0_2 \widetilde{\chi}^0_2$
$>$ 90\% (red), 50\% -- 90\% (yellow), 10\% -- 50\% (light blue),
$<$ 10\% (white).  The dark shaded regions are
excluded by theoretical considerations or   
LEP measurements (save constraints from LEP 
Higgs-strahlung which roughly reach up to the dashed green curves
with considerable uncertainty --- see text).
Also shown in purple are the CMS TDR (Fig.\ 11.32) 
$5\sigma$ discovery regions 
(assuming $L_{int} = 30\, \hbox{fb}^{-1}$) 
for $H^0,A^0 \rightarrow \widetilde{\chi}^0_2 \widetilde{\chi}^0_2$.
The solid purple lines show the extent of the plots in Fig.\ 11.32. 
}
\label{fig:mSUGRAHino}
\end{figure}

Augmenting the general MSSM with additional assumptions about the 
unification of SUSY inputs at a very high mass scale
yields the more restrictive 'mSUGRA' models.  Here the number of free
input parameters is much reduced (hence the popularity of such scenarios
for phenomenological analyses), with said free parameters generally set as
$\tan\beta$,
a universal gaugino mass defined at the Grand Unification Theory (GUT)
scale ($M_{{{1}/{2}}}$),
a universal GUT-level scalar mass
($M_{{0}}$),
a universal GUT-level trilinear scalar mass term
($A_{{0}}$),
and the sign of $\mu$ (henceforth, sgn$(\mu)$).
As already noted, the signal has a strong preference for low values of
$| \mu |$.  Yet in mSUGRA scenarios, $| \mu |$ is not a free parameter,
as it is closely tied to the masses of the scalar Higgs bosons
{\it via} the $M_{{0}}$ input.  An earlier study of charged Higgs boson 
decays into a neutralino and a chargino \cite{EPJC2} demonstrated that 
this was sufficient to preclude detection of a $3\ell$ $+$ top-quark 
signal from such processes over the entire reach of the unexcluded mSUGRA 
parameter space.  
Here, with the heavier neutral MSSM Higgs bosons, the situation is
not so discouraging.
Fig.\ \ref{fig:mSUGRAHino} shows the values for
$\sigma(pp \rightarrow H^0,A^0)$ $\times$  
BR$(H^0,A^0 \rightarrow 4\ell N)$
obtained for $\tan\beta = 5,10,20$ and  $\mu > 0$.
Two disconnected regions of unexcluded parameter space appear
where the expected number of events (for $100\, \hbox{fb}^{-1}$ of
integrated luminosity) is in the tens to hundreds (or even thousands).  
Interestingly, one of these 
(which includes discovery regions depicted in the CSM TDR
\cite{CMSTDRSUSY}\footnote{Note: virtually all mSUGRA parameter space plots 
in the TDR showing excluded regions are for $\tan\beta = 10$; the exceptions
being the $\tan\beta = 5$ plot in Fig.\ 11.32 and the
$\tan\beta = 35$ plots in Figs.\ 13.12 \& 13.13; and the
$\tan\beta = 35$ plots seem to inaccurately have the $\tan\beta = 10$
exclusion zones.  These latter plots and others in Chapter 13 do show
a chargino lower mass limit (green dotdashed curve) and other supercollider
experimental bounds which are more consistent with the excluded regions
shown in the ATLAS TDR (and in the present work).}) is where 
$\widetilde{\chi}^0_2 \widetilde{\chi}^0_2$ is the dominant source of 
$4\ell$ events while the other is where decays of the heavier --inos 
dominate. For $\tan\beta=5$, rates in the 
$\widetilde{\chi}^0_2 \widetilde{\chi}^0_2$ 
region are much larger than in the heavier --inos region.  However, for
$\tan\beta=20$, rates in the two regions become more comparable. 

Also shown as solid purple zones on the $\tan\beta=5$ and $\tan\beta=10$ 
plots are $5\sigma$ discovery regions from the 
CMS TDR (Fig.\ 11.32) \cite{CMSTDRSUSY}.
These CMS TDR discovery regions assume an integrated luminosity of just 
$30\, \hbox{fb}^{-1}$, 
and thus would have certainly been considerably larger if a base luminosity of 
$100\, \hbox{fb}^{-1}$ was used instead.
This CMS TDR analysis was at a technical level comparable to that in this work, 
{\em but only considered MSSM Higgs boson decays into 
$\widetilde{\chi}^0_2 \widetilde{\chi}^0_2$ pairs}.
Thus, the CMS TDR analysis would not pick up the region where heavier --ino 
decays dominate (in fact the plots in Fig.\ 11.32 in the CMS TDR only showed 
the regions delineated by the solid purple lines in 
Fig.\ \ref{fig:mSUGRAHino}). 
Given that the somewhat lower rates of the higher $M_0$, 
heavier --ino decays-dominated region may be compensated by assuming a 
larger integrated luminosity,
as well as perhaps finding a higher selection efficiency due to harder 
daughter leptons,
it is difficult to infer from the CMS TDR $5\sigma$ $30\, \hbox{fb}^{-1}$
discovery regions whether or not disjoint discovery regions may develop
in this novel region of the parameter space.
This is currently under investigation \cite{msugrawork}.

The excluded regions shown in Fig.\ \ref{fig:mSUGRAHino} merit some explanation.
Note that in each plot the discovery region from the CMS TDR cuts into the 
excluded region, 
whereas in Fig.\ 11.32 of the CMS TDR they do not touch the (more limited)
excluded regions shown.  This is mainly because the excluded regions in 
Fig.\ 11.32 of the CMS TDR only mark off regions where the 
$\widetilde{\chi}^0_1$ is not the LSP
(because the mass of the lighter stau is lower --- this removes the 
upper left corner of the plots) and where EWSB is not obtained (along the 
horizontal axis), while ignoring other experimental constaints --- such as 
the lower limit on the lighter chargino's mass from the LEP experiments.  
Such additional experimental constraints are included, for instance, in the 
excluded regions shown in Fig.\ 20-1 of the ATLAS TDR \cite{ATLASTDRSUSY}
\footnote{Note: virtually all mSUGRA parameter space plots in the TDR
showing excluded regions are for $\tan\beta = 10$ and for
(the now ruled-out) $\tan\beta = 2$.}.  
These experimental constraints have been updated to represent
the final limits from the LEP experiments, accounting for the gross 
differences between the excluded regions depicted in the ATLAS TDR and those 
in the present work\footnote{Raising the lower bound on the chargino mass 
from the {\it circa} 1998 \cite{PPDB1998} 
LEP-1.5-era ${\sim}65\, \hbox{GeV}$ to ${\sim}100\, \hbox{GeV}$ raises the 
approximately horizontal boundary for higher $M_0$ values, while the rise of 
the bounds for the slepton masses from ${\sim}45\, \hbox{GeV}$ to 
$m_{\widetilde{{e}}_1}, m_{\widetilde{{\mu}}_1}, m_{\widetilde{{\tau}}_1} 
\simeq 99\, \hbox{GeV}, 91\, \hbox{GeV}, 85\, \hbox{GeV}$ 
adds the quarter-circle-like bite seen in the lower-left corner of the 
$\tan\beta = 10$ plot in Fig.\ \ref{fig:mSUGRAHino} (which is absent in the 
ATLAS TDR plots).}.
Somewhat crude\footnote{For reasons detailed in \cite{DDK}, foremost among 
which is the uncertainty in the calculation of $M_h$.  Herein the Higgs boson 
mass formul\ae\ of ISAJET \cite{ISAJET} and \cite{thesis} are employed. 
Results here are roughly consistent with Figs.\ 1 \& 2 of \cite{DDK} 
(2006 paper). Note that in the case of mSUGRA, unlike in the general MSSM 
examples in the current work, 
the stop and other squark parameters --- which make the main 
contributions to the quite significant radiative corrections 
to $M_h$ --- are determined from the few mSUGRA inputs without the need 
to set values by hand for assorted soft SUSY-breaking masses.
Certainly, in mSUGRA, the LEP bounds on light Higgs boson production are 
strongly-tied to rates for heavy Higgs boson to sparticle decay channels, 
though this correlation will not be intensively examined in this work.
} 
estimates for the regions excluded by the LEP searches for MSSM Higgs bosons 
are indicated separately by the dashed green lines based on the empirical 
formula developed by Djouadi, Drees and Kneur \cite{DDK}.  
Finally, it must be emphasized that constraints from 
lower-energy experiments (in particular from $b \rightarrow s \gamma$) and 
from cosmological considerations (such as LSP dark matter annihilation rates) 
are {\em not} herein considered.  
In the far more restricted parameter domain of mSUGRA models it is more 
difficult to circumvent such constraints, and they can exclude considerable 
portions of the allowed parameter space shown in the figures (for further 
details, see \cite{extra-constrs}).   

As was done with the general MSSM parameter space, Fig.\ 5 enables 
selection of a couple of representative mSUGRA points for
simulation studies.  These are:
\vskip 0.6cm  
{\bf Point A}.
$M_0 = 125\, \hbox{GeV}$,
$M_{1/2} = 165\, \hbox{GeV}$, 
$\tan\beta = 20$,
${\rm sgn(\mu)} = +1$, $A_0 = 0$.
\vskip 0.45cm 
{\bf Point B}.
$M_0 = 400\, \hbox{GeV}$,
$M_{1/2} = 165\, \hbox{GeV}$,
$\tan\beta = 20$,
${\rm sgn(\mu)} = +1$, $A_0 = 0$.
\vskip 0.6cm  
\parindent=0pt
Point A is dominated by 
$H^0,A^0 \rightarrow \widetilde{\chi}_2^0 \widetilde{\chi}_2^0 
\rightarrow 4\ell$ decays 
(which account for more than 99\% of the inclusive signal event rate 
before cuts) while in Point B the corresponding rates
are below 30\% (the largest signal event channel is now 
$H^0,A^0 \rightarrow \widetilde{\chi}_1^{\pm} \widetilde{\chi}_2^{\mp}
\rightarrow 4\ell$, yielding over 50\% of the events, with significant 
contributions from 
$H^0,A^0 \rightarrow \widetilde{\chi}_2^0 \widetilde{\chi}_2^0,
\widetilde{\chi}_2^0 \widetilde{\chi}_3^0,
\widetilde{\chi}_2^0 \widetilde{\chi}_4^0 \rightarrow 4\ell$).  
Full MC and detector simulations for Points A and B
will be presented in the next section.  
These will show that $4\ell N$ signals remain visible in the mSUGRA 
parameter space, at least at these points.

\parindent=15pt

\section{Simulation analyses}

The HERWIG 6.5 \cite{HERWIG65} MC package (which obtains its
MSSM input information from ISASUSY \cite{ISAJET} through the ISAWIG 
\cite{ISAWIG} and HDECAY \cite{HDECAY} interfaces) is employed coupled
with private programs simulating a typical LHC detector environment
(these codes have been checked against results in the literature).
The CTEQ 6M \cite{CTEQ6} set of PDFs is used and top and bottom quark 
masses are set to $m_t=175\, \hbox{GeV}$ and $m_b=4.25\, \hbox{GeV}$, 
respectively.  

Four-lepton events are first selected according to these criteria: 
\begin{itemize}
\item
Events have exactly four leptons, $\ell=e$ or $\mu$, irrespective of their
individual charges, meeting the following criteria:
\newline
Each lepton must have $|\eta^\ell|<2.4$ and 
$E_T^\ell >7,4$~GeV for $e,\mu$ (see ATLAS TDR \cite{ATLASTDRSUSY}).
\newline
Each lepton must be isolated.
The isolation criterion demands there be no tracks
(of charged particles) with $p_T > 1.5\, \hbox{GeV}$ in a cone of
$r = 0.3\, \hbox{radians}$ around a specific lepton, and also that
the energy deposited in the electromagnetic calorimeter be less than
$3\, \hbox{GeV}$
for $0.05\, \hbox{radians} < r < 0.3\, \hbox{radians}$.
\newline
Aside from the isolation demands, no restrictions are placed at this 
stage on the amount of hadronic activity or the number of reconstructed 
jets in an event.
\end{itemize}
Further,
\begin{itemize}
\item
Events must consist of two opposite-sign, same-flavor lepton pairs.
\end{itemize}

Events thus identified as candidate signal events are then subjected to the 
following cuts:
\begin{itemize}
\item $Z^0$-veto: no opposite-charge same-flavor lepton pairs may 
reconstruct $M_{Z} \pm 10$~GeV.
\item  restrict $E_T^{\ell}$: all leptons must finally have 
$20\, \hbox{GeV} < E_T^\ell < 80\, \hbox{GeV}$.
\item restrict missing transverse energy, $E_T^{\rm{miss}}$: 
events must have
$20\, \hbox{GeV} < E_T^{\rm{miss}} < 130\, \hbox{GeV}$.
\item  cap $E_{T}^{\rm{jet}}$: 
all jets must have $E_T^{\rm{jet}} < 50\, \hbox{GeV}$.
\newline
Jets are reconstructed using a UA1-like iterative
({\it i.e.}, with splitting and merging, see Ref.~\cite{jets} for a
description of the procedure) cone algorithm with fixed size $0.5$,
wherein charged tracks are collected at
$E_T > 1\, \hbox{GeV}$ and $|\eta| < 2.4$ and each reconstructed 
jet is required to have $E_T^{\rm{jet}} > 20\, \hbox{GeV}$.
\end{itemize}
Lastly, application of an additional cut on the four-lepton invariant mass
is investigated:
\begin{itemize}
\item  four-lepton invariant mass (inv.\ m.) cut:
the $4 \ell$ inv.\ m.\ must be
$\leq 240\, \hbox{GeV}\, .$
\end{itemize}
For the signal events, the upper limit for the four-lepton inv.\ m.\
will be  $M_{H,A} - 2M_{{\widetilde\chi}^0_1}$, and thus its value is 
dependent upon the chosen point in MSSM parameter space.  In the actual
experiment, the value of $M_{H,A} - 2M_{{\widetilde\chi}^0_1}$ would be 
{\it a priori} unknown.  So one could ask how a numerical value can be 
chosen for this cut?  If too low a value is selected, many signal events 
will be lost.  On the other hand, if too large a value is chosen, more 
events from background processes will be accepted, diluting the signal.  
One could envision trying an assortment of numerical values for the 
four-lepton inv.\ m.\ upper limit (one of which could for instance be 
the nominal value of $240\, \hbox{GeV}$ noted above) to see which value 
optimized the signal relative to the backgrounds.  However, here sparticle 
production processes are very significant backgrounds (after application of 
the other three cuts, only such processes and residual 
$Z^{0(*)} Z^{0(*)}$ events remain), which, like the signal, may well have 
unknown rates.   Thus, strengthening this cut would lower the total number of 
events without indicating whether the signal to background ratio is going up 
or down --- unless additional information is available from other studies at 
least somewhat restricting the location in MSSM parameter space Nature has 
chosen.  If such information were available, this cut could indeed lead to a 
purer set of signal events.  One could instead consider all events from MSSM
processes to be the signal while the SM processes comprise the background.
However, the aim of this work is to identify the heavier Higgs bosons, 
not merely to identify an excess attributable to SUSY.  

Detailed results are tabulated for the aforementioned two general 
MSSM and two mSUGRA parameter space points.  MSSM Point 1 and mSUGRA Point 
A have the vast majority of their $4\ell$ events from 
$H^0,A^0 \rightarrow \widetilde{\chi}_2^0 \widetilde{\chi}_2^0$, while
MSSM Point 2 and mSUGRA Point B obtain most of their $4\ell$ events from 
Higgs boson decays to heavier --ino pairs 
($\widetilde{\chi}_2^0 \widetilde{\chi}_3^0$,
$\widetilde{\chi}_2^0 \widetilde{\chi}_4^0$
$\widetilde{\chi}_3^0 \widetilde{\chi}_3^0$,
$\widetilde{\chi}_3^0 \widetilde{\chi}_4^0$
and/or
$\widetilde{\chi}_4^0 \widetilde{\chi}_4^0$).
The sparticle spectra\footnote{The older ISASUSY version which inputs 
sparticle masses into HERWIG 6.3 lacks D-terms in the slepton masses, 
meaning the smuon masses in the simulation runs equate to the selectron 
masses given in Table~\ref{tab:masses}.  This has a minor effect upon 
the edges in the Dalitz-like 'wedgebox' plots to be shown later.
See discussion in \cite{Cascade}.} 
for these points are presented in Table~\ref{tab:masses}.

\subsection{MSSM benchmark points}

\begin{table}[!t]
     \caption{Relevant sparticle masses (in GeV) for specific MSSM 
and mSUGRA parameter points studied in the analyses.}   
     \begin{center}
     \begin{tabular}{|l||l|l|l|l|} \hline
    & Point 1  & Point 2 & Point A & Point B \\ \hline
  $M_{A}$ & $500.0$ &  $600.0$ & $257.6$ & $434.9$ \\ \hline
  $M_{H}$ & $500.7$ &  $600.8$ & $257.8$ & $435.3$\\ \hline
  ${\widetilde\chi}^0_1$  
           & \phantom{$1\!\!$} $89.7$ &  $93.9$ & $60.4$ & $60.8$ \\ 
\hline
  ${\widetilde\chi}^0_2$
          & $176.3$ &  $155.6$ & $107.8$ & $108.0$ \\ \hline
  ${\widetilde\chi}^0_3$
           & $506.9$ & $211.8$ & $237.6$ & $232.8$ \\ \hline
  ${\widetilde\chi}^0_4$
          & $510.9$ &  $262.2$ &  $260.0$ & $256.3$ \\ \hline
  ${\widetilde\chi}^\pm_1$
          & $176.5$ &  $153.5$ & $106.8$ & $106.8$  \\ \hline
  ${\widetilde\chi}^\pm_2$
          & $513.9$ &  $263.2$ & $260.0$ & $258.2$  \\ \hline\hline
  $m_{\widetilde{\nu}}$ 
          & $241.6$ & $135.5$ & $154.8$ & $407.9$  \\ \hline
  $m_{\widetilde{e}_1}$ 
          & $253.8$ & $156.3$ & $145.7$ & $406.1$  \\ \hline
  $m_{\widetilde{\mu}_1}$
          & $252.0$ & $154.3$ & $145.6$ & $406.1$  \\ \hline
  $m_{\widetilde{e}_2}$   
          & $254.4$ & $157.2$ & $174.1$ & $415.7$  \\ \hline
  $m_{\widetilde{\mu}_2}$ 
          & $256.2$ & $159.2$ & $174.2$ & $415.7$  \\ \hline
  $m_{\widetilde{e}_2} - m_{\widetilde{e}_1}$
         & \phantom{$1$} $0.59$ & \phantom{$1$} $0.96$
         & \phantom{$1$} $28.46$ & $9.56$ \\ \hline
  $m_{\widetilde{\mu}_2} - m_{\widetilde{\mu}_1}$
         & \phantom{$1$} $4.20$ & \phantom{$1$} $4.81$
         & \phantom{$1$} $28.62$ & $9.63$   \\ \hline
       \end{tabular}      
    \end{center}
 \label{tab:masses}
\end{table}

Table 2 shows results for MSSM Point 1, a
$H^0,A^0 \rightarrow \widetilde{\chi}_2^0 \widetilde{\chi}_2^0$-dominated
point.  Note that, after cuts, signal events do make up the majority of 
events in the sample.  The only remaining backgrounds are from direct 
neutralino/chargino pair-production\footnote{Herein final states involving 
a sparton and a chargino/neutralino are included together with the results
for $\widetilde{\chi} \widetilde{\chi}$, as designed in HERWIG, though for 
the points studied here the latter overwhelmingly dominate the 
former.} (denoted by $\widetilde{\chi}\widetilde{\chi}$), from slepton 
pair-production (denoted by $\widetilde{\ell}$, $\widetilde{\nu}$) and 
from $Z^{0(*)} Z^{0(*)}$ production.
\begin{table}[!t]\label{tab:MSSMPt1}
    \caption{Event rates after the successive cuts
defined in the text for MSSM Point 1 
(assuming an integrated luminosity of $100\, \hbox{fb}^{-1}$).}
\begin{center}
\begin{tabular}{|l||c|c|c|c|c|c|c|} \hline  
   Process & $4\ell$ events 
& ${\ell}^+{\ell}^-{\ell}^{{\scriptscriptstyle ( \prime )}+}
                   {\ell}^{{\scriptscriptstyle ( \prime )}-}$
& $Z^0$-veto & $E_T^{\ell}$ & $E_T^{\rm{miss}}$& $E_{T}^{\rm{jet}}$ 
& $4\ell$ inv.\ m. \\ \hline
    $\widetilde{q}$, $\widetilde{g}$                                   
&  118  &   64   &   49  &   19  &   1   &   0   &    0  \\ \hline
    $\widetilde{\ell}$,$\widetilde{\nu}$                        
& 100   &   65   &   46  &   30  &  23   &  13   &    7  \\ \hline
 $\widetilde{\chi} \widetilde{\chi},
\widetilde{q}/\widetilde{g} \widetilde{\chi}$     
&  34   &   17   &  13   &   10  &   5   &   2   &    1  \\ \hline
    $tH^-$ + c.c.                                               
&  0    &    0   &   0   &    0  &   0   &   0   &    0  \\ \hline
    $Z^{0(*)} Z^{0(*)}$                                          
& 1733  &  1683  &  43   &   39  &   5   &   4   &    4  \\ \hline
    $t\bar{t} Z^{0(*)}$                                                
&  47   &   23   &   2   &    1  &   1   &   0   &    0  \\ \hline
    $t\bar{t} h^0$                                               
&   4   &    2   &   2   &    1  &   1   &   0   &    0  \\ \hline
    $H^0,A^0$ signal                                            
& 20,32 &  18,31 & 14,26 & 13,25 & 11,22 & 8,17  & 6,13  \\ \hline
    \end{tabular}
    \end{center}
\end{table}
The number of events obtained from $A^0$ decays after cuts is about twice 
the number obtained from $H^0$ decays.  This is despite the fact that the 
$H^0$ and $A^0$ production cross sections are the same within $1$\%.  
The ratio of $A^0$ to $H^0$ events at this point can be compared to 
that for inclusive rates (with no cuts) which may be calculated using 
the BRs obtained from ISASUSY\footnote{These were normalized using HERWIG 
production cross-sections, though here this is of scant importance since 
the $H^0$ and $A^0$ production cross-sections are almost the same.
Also, for consistency with the HERWIG simulation analysis, ISASUSY
Version 7.56 was used to generate the BRs.}.  
Including all possible decay chains, ISASUSY numbers predict 
$A^0:H^0 = 1.83:1.00$ ($64.7$\% $A^0$ events).  
This is in reasonable agreement with $A^0:H^0 = 1.6:1.0$ 
($61.5$\% $A^0$ events) obtained from the $4\ell$ before cuts entries 
in the first column of Table 2.
The different $H^0$ and $A^0$ event rates may then be traced back to 
differences in the 
$H^0/A^0$-$\widetilde{\chi}_2^0$-$\widetilde{\chi}_2^0$ couplings
(as opposed to the enhancing or opening up of other $H^0$ decay modes, 
such as for instance $H^0 \rightarrow h^0 h^0$).    
Study of the inclusive rates based on the ISASUSY BRs also confirmed that
over $99$\% of the four-lepton signal events resulted from
$H^0/A^0 \rightarrow \widetilde{\chi}_2^0 \widetilde{\chi}_2^0$ decays.  
The percentage of $A^0 \to 4\ell$ events surviving the subsequent cuts is 
about $10$\% larger than the percentage of $H^0 \to 4\ell$ events 
surviving.

\begin{figure}[!t]
\centerline{}
\begin{center}
\epsfig{file=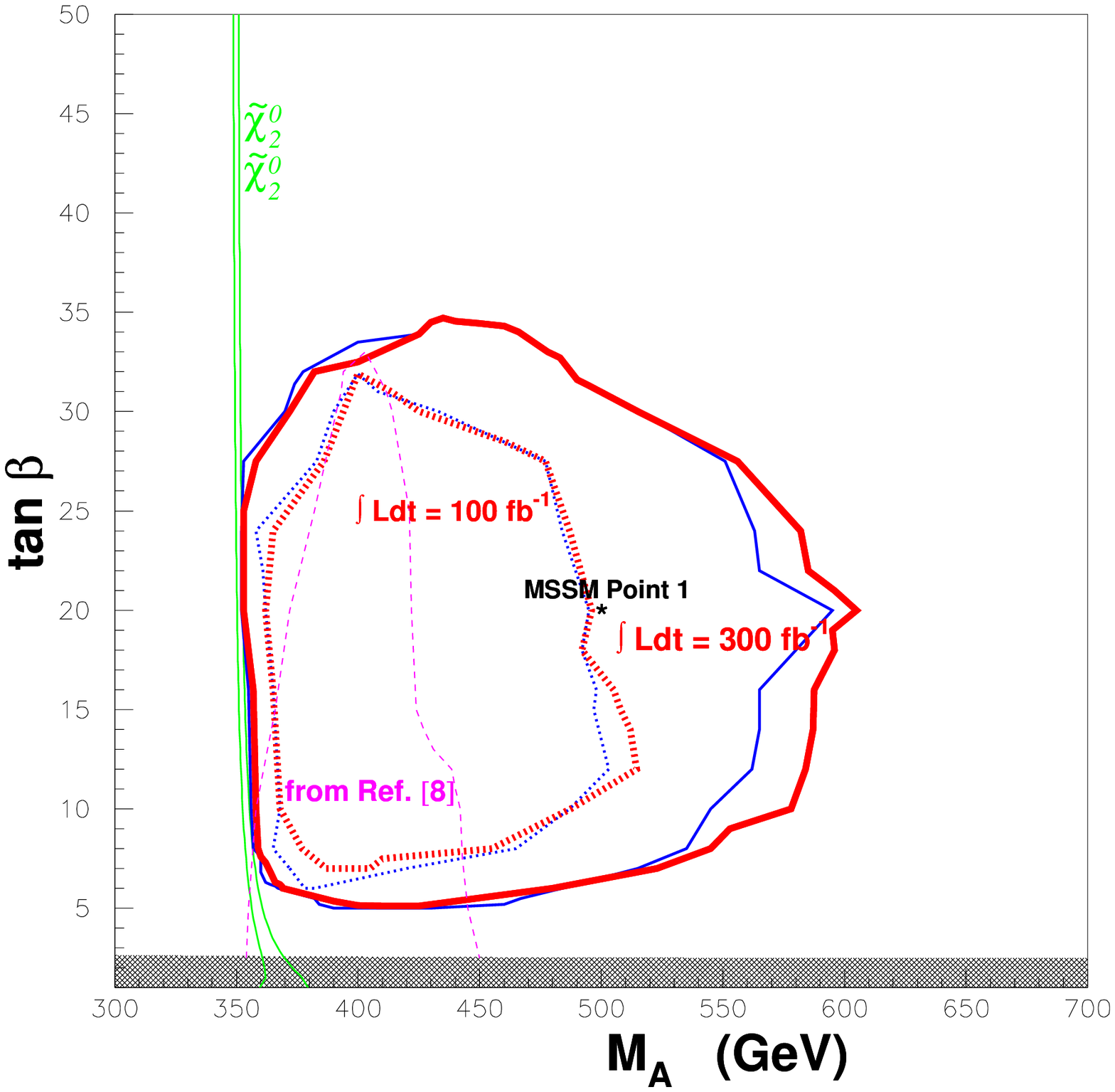,height=170mm,width=170mm}
\end{center}
\vskip -0.9cm  
\caption{
Discovery region in red in $(M_A, \tan\beta)$ plane for --ino/slepton  
parameters 
$\mu = -500\, \hbox{GeV}$, 
$M_2 = 180\, \hbox{GeV}$, $M_1 = 90\, \hbox{GeV}$, 
$m_{\widetilde{\ell}_{soft}} =
m_{\widetilde{\tau}_{soft}} = 250\, \hbox{GeV}$
as in MSSM Point 1 (whose location is marked by a black asterisk).  
Here Higgs boson decays to $\widetilde{\chi}_2^0 \widetilde{\chi}_2^0$ 
totally dominate.
Solid (dashed) red border delineates the discovery region for 
$L_{int} = 300\, \hbox{fb}^{-1}$ ($100\, \hbox{fb}^{-1}$).
The two green curves are
$M_A,M_H - 2m_{\widetilde{\chi}_2^0}$.
Also shown in light purple are analogous results from a previous study 
\cite{CMS1}
for $100\, \hbox{fb}^{-1}$.  The blue contours add the extra cut on the 
four-lepton inv.\ m.\ for the nominal cut-off value of 
$240\, \hbox{GeV}$.
}
\label{fig7:discovery}
\vskip -0.5cm
\end{figure}

Fixing the --ino input parameters $M_2$ \& $\mu$ and the slepton \& 
squark inputs to be those of MSSM Point 1, $\tan\beta$ and $M_A$ were 
then varied to map out a Higgs boson discovery region in the 
traditional ($M_A$, $\tan\beta$) plane.  
This is shown in red in Fig.\ \ref{fig7:discovery}, where the solid 
(dashed) red border delineates the discovery region assuming an integrated 
luminosity of $300\, \hbox{fb}^{-1}$ ($100\, \hbox{fb}^{-1}$).
The exact criteria used for demarcating the discovery region is that
there be at least $10$ signal events and that the
$99$\%-confidence-level upper limit on the background is smaller than the
$99$\%-confidence-level lower limit on the signal plus background.
Mathematically, the latter condition translates into the
formula \cite{MSSMhgamgam}:
\begin{equation}
N_{\hbox{signal}} > (2.32)^2 \left[ 1 + 
\frac{2 \sqrt{N_{\hbox{bckgrd}}}}{2.32} 
\right] \;\; ,
\label{disccrit}
\end{equation}
where $N_{\hbox{signal}}$ and $N_{\hbox{bckgrd}}$ are the expected 
number of signal and background events, respectively.  As with MSSM 
Point 1, direct neutralino/chargino pair-production, slepton pair production 
and SM $Z^{0(*)} Z^{0(*)}$ are the only background processes remaining 
after cuts 
(the actual number of surviving background events varies modestly with 
$\tan\beta$) at all points tested, with slepton pair production continuing 
as the dominant background.  Taking into account these backgrounds, 
$24$-$28$ ($38$-$45$) signal events are required to meet the criteria for 
$100\, \hbox{fb}^{-1}$ ($300\, \hbox{fb}^{-1}$) of integrated luminosity, 
depending on the value of $\tan\beta$, if the four-lepton inv.\ m.\ cut 
is not employed.  
Adding in this last optional cut changes the required numbers to $19$-$22$ 
($28$-$34$) signal events and shifts the discovery region boundaries to 
those shown as blue (dashed blue) curves in Fig.\ \ref{fig7:discovery}.  
This places MSSM Point 1 just outside the upper $M_A$ edge of
the $100\, \hbox{fb}^{-1}$ discovery region (whether or not the four-lepton 
inv.\ m.\ cut is used).  Lowering $M_A$ to $400\, \hbox{GeV}$ raises
the number of signal events from $25$ to $36$.
Note that Fig.\ 4 (left-side plot) predicts that 
$H^0,A^0 \rightarrow \widetilde{\chi}_2^0 \widetilde{\chi}_2^0$ decays 
will generate the bulk of the signal throughout the discovery region.  
The lower $M_A$ edge of the discovery region closely follows where
the (dominant) $\widetilde{\chi}^0_2 \widetilde{\chi}^0_2$ decay becomes 
kinematically accessible, {\it i.e.}, 
$M_A \, \ge \, 2m_{\widetilde{\chi}_2^0}$.
The $A^0$ contribution outweighing the $H^0$ contribution was found
to be a general result valid for almost\footnote{Inside of the discovery 
region (for $300\, \hbox{fb}^{-1}$), a couple points along the high
$M_A$ -- lower $\tan\beta$ edge were found where the rate from $H^0$ very 
slightly exceeded that from $A^0$.} 
all points in the $(M_A, \tan\beta)$-plane tested:  events from $A^0$ 
equaled or outnumbered those from $H^0$.  
Note from Table 2 that MSSM Point 1 at $M_A = 500\, \hbox{GeV}$ and 
$\tan\beta = 20$ yielded $A^0:H^0 = 2.1:1.0$ ($68$\% $A^0$ events) {\em 
after} all cuts save the four-lepton inv.\ m.\ cut (as comparison to 
the numbers in the preceding paragraph indicate, $A^0$ events tend to do 
slightly better at surviving the cuts, though little reason could be found 
for this small effect).   Lowering $M_A$ to $400\, \hbox{GeV}$ shifts this 
ratio to $A^0:H^0 = 3.9:1.0$ ($81$\% $A^0$ events).  

The preponderance of $A^0$ events is generally greatest for 
lower values of $M_A$.  For $M_A \, \lsim \, 375\, \hbox{GeV}$,
$90$-$100$\% of the signal events are from $A^0$.  Since 
$M_A < M_H$ and $M_A \, \simeq \, 2m_{\widetilde{\chi}_2^0}$ this is 
mainly a threshold effect.  The $A^0$ event percentage drops to 
around $70$\% when $M_A \, \simeq \, 415\, \hbox{GeV}$.  For higher 
$M_A$ values inside the $100\, \hbox{fb}^{-1}$ discovery region
(outside the $100\, \hbox{fb}^{-1}$ discovery region but inside the 
$300\, \hbox{fb}^{-1}$ discovery region), this percentage ranges from 
${\sim}70$\% down to ${\sim}55$\%
(${\sim}60$\% down to ${\sim}50$\%), save for the upper tip 
where $\tan\beta \, \gsim \, 30$ wherein the $A^0$ percentage remains 
above $70$\% or even $80$\%.

Inclusion of the four-lepton inv.\ m.\ cut with the nominal cut-off
value of $240\, \hbox{GeV}$ shifts the discovery region boundaries in 
Fig.\ \ref{fig7:discovery} from the red curves to the blue ones.
There are slight gains for low $M_A$ values at high and low 
values for $\tan\beta$; however, the high $M_A$ edges also recede
somewhat.  Note also that the highest and lowest $\tan\beta$ values
which fall inside the discovery region are virtually unaltered.
Though the cut's effect on the expanse of the discovery region
is quite modest, inclusion of this cut at included points with lower $M_A$ 
values can certainly raise the $\hbox{signal}$ : $\hbox{background}$. For 
instance, at ($M_A$, $\tan\beta$) = ($400\, \hbox{GeV}$, $20$), 
this ratio goes from $37:19$ without the $4\ell$ inv.\ m.\ cut 
to $37:12$ with it.
However, shifting $M_A$ to $500\, \hbox{GeV}$ as in MSSM Point 1 
is enough to remove any advantage, as can be seen in Table 2.

Input parameters for MSSM Point 1 were also chosen to match a point 
studied in a previous analysis \cite{CMS1} --- which only looked at 
$\widetilde{\chi}_2^0 \widetilde{\chi}_2^0$ Higgs boson 
decays\footnote{A different simulation
of the quark-fusion channel involving $b$ (anti)quarks (in the CMS note 
the simulation was performed using 
$gg \rightarrow b \bar{b} H^0, b \bar{b} A^0$) is adopted here.  
In addition, the MC analysis in \cite{CMS1} was done with PYTHIA version 
5.7 \cite{PYTHIA5.7},
which only implemented an approximated treatment of the SUSY sector, while
herein ISASUSY is used in conjunction with HERWIG (though intrinsic 
differences between the two generators in the implementation of the
PS and hadronization stages should be minimal in our context).
Also, the background processes $tH^-$ + c.c., $t\bar t Z$ and
$t\bar t h$, which were not emulated in \cite{CMS1}, in this study were 
checked to yield no background events throughout Fig.\ 7.}.
The light purple contour shown in the plot is the result from this older study
(see the blue contour in Fig.\ 19 therein). 
Results in the present case for the most part agree with those of that 
previous study, though in the current analysis the discovery region 
extends to somewhat higher values of $M_A$ and dies for $\tan\beta$ 
values below ${\sim}5$.  The latter is primarily due to low $\tan\beta$
strong enhancement of the $H^0$($A^0$)-$t$-$\bar{t}$ coupling,
which is proportional to $\csc\beta$ ($\cot\beta$), increasing the 
$H^0,A^0 \rightarrow t \bar{t}$
BRs at the expense of the --ino BRs\footnote{The partial widths for 
$H^0$ and $A^0$ decays to --inos also drop by roughly a factor of 2 in 
going from $\tan\beta=6$ to $\tan\beta=2$ (at $M_A = 450\, \hbox{GeV}$), and 
the $H^0 \rightarrow h^0 h^0$ and $A^0 \rightarrow h^0 Z^{0(*)}$ widths 
increase by about a factor of $2$.  These also lower the signal rate.  
On the other hand, decay widths to $b$-quarks and tau-leptons also drop by a 
bit over a factor of $2$, helping the signal.  These effects are 
overwhelmed by an almost order-of-magnitude enhancement in the 
$H^0$ and $A^0$ to $t \bar{t}$ decay widths.}.  
BR$(H^0 \rightarrow t\bar{t})$ (BR$(A^0 \rightarrow t\bar{t})$)
rises from around $0.30$ to $0.68$ to $0.93$ 
($0.51$ to $0.79$ to $0.96$)
as $\tan\beta$ runs from $6$ to $4$ to $2$.
  
For MSSM Point 2, Higgs boson decays to the heavier neutralinos and 
charginos neglected in previous studies produce most of the signal events.
Table~\ref{tab:percents2} gives the percentage contributions to the 
signal events among the $H^0,A^0$ decay modes based on an inclusive rate
study using BR results from ISAJET (ISASUSY) 7.58 normalized with HERWIG 
cross-sections.  This parton-level analysis merely demands exactly four
leptons in the (parton-level) final state.
According to this inclusive rates study, Higgs boson decays to
$\widetilde{\chi}_2^0  \widetilde{\chi}_2^0$ now contribute less
than one hundredth of one percent of the signal events, in stark
contrast to MSSM Point 1 where such decays accounted for virtually all
of the signal events.   Applying all the cuts at the full event-generator 
level does not alter this.  Said numerical results with the application of 
the successive cuts for MSSM Point 2 are given in Table~\ref{tab:MSSMPt2}.

\begin{table}[!t]  
     \vskip -0.1cm
     \caption{Percentage of $H^0,A^0 \rightarrow 4\ell N$ events 
(excluding cuts) coming from various --ino channels
   for MSSM Point 2. (Other channels are negligible.)}
\begin{center}
\begin{tabular}{lc}
$H^0 \rightarrow \widetilde{\chi}_3^0  \widetilde{\chi}_4^0$ &
$31.5$\%  \\
$A^0 \rightarrow \widetilde{\chi}_4^0  \widetilde{\chi}_4^0$ &
$31.1$\%  \\
$A^0 \rightarrow \widetilde{\chi}_3^0  \widetilde{\chi}_4^0$ &
$13.4$\%  \\
$H^0 \rightarrow \widetilde{\chi}_4^0  \widetilde{\chi}_4^0$ &
$8.4$\%  \\
$H^0 \rightarrow \widetilde{\chi}_1^{\pm}  \widetilde{\chi}_2^{\mp}$ &
$6.9$\%  \\
$A^0 \rightarrow \widetilde{\chi}_1^{\pm}  \widetilde{\chi}_2^{\mp}$ &
$4.3$\%  \\ 
$A^0 \rightarrow \widetilde{\chi}_3^0  \widetilde{\chi}_3^0$ &
$1.9$\%  \\
$H^0 \rightarrow \widetilde{\chi}_3^0  \widetilde{\chi}_3^0$ &
$0.8$\%  \\
$H^0 \rightarrow \widetilde{\chi}_2^+  \widetilde{\chi}_2^-$ &
$0.75$\%  \\
$H^0 \rightarrow \widetilde{\chi}_2^+  \widetilde{\chi}_2^-$ &
$0.6$\%  \\
all other contributions &  
$< 0.5$\%  \\
\end{tabular}
\end{center}
 \label{tab:percents2}
\end{table}
Note that the four-lepton inv.\ m.\ cut,
with the nominal numerical value of $240\, \hbox{GeV}$, removes about 
$74$\% of the signal events while only slightly reducing the number of 
background events.  This clearly shows that this cut, while helpful for
points with lower $M_A$ values in Fig.\ \ref{fig7:discovery}, is  
quite deleterious at MSSM Point~2.  Without the $4\ell$ inv.\ m.\ cut,
an integrated luminosity of $25\, \hbox{fb}^{-1}$ is sufficient to meet
the discovery criteria; while with the $4\ell$ inv.\ m.\ cut, 
an integrated luminosity of ${\sim}130\, \hbox{fb}^{-1}$ is required.
Choosing a higher numerical cut-off would lead to a viable cut for this 
point; however, it may prove impossible to {\it a priori} decide on an 
appropriate value for the actual experimental analysis (see earlier 
discussion).

\begin{table}[!t]\label{tab:MSSMPt2}
    \caption{Event rates after successive cuts as
defined in the text for MSSM Point 2 
\newline
(assuming 100 fb$^{-1}$).}
    \begin{center}
\begin{tabular}{|l||c|c|c|c|c|c|c|} \hline
   Process & $4\ell$ events
& ${\ell}^+{\ell}^-{\ell}^{{\scriptscriptstyle ( \prime )}+}
                   {\ell}^{{\scriptscriptstyle ( \prime )}-}$
& $Z^0$-veto & $E_T^{\ell}$ & $E_T^{\rm{miss}}$& $E_{T}^{\rm{jet}}$
& $4\ell$ inv.\ m. \\ \hline
    $\widetilde{q}$, $\widetilde{g}$                                    
&    817  & 332     & 197    &  96   & 21   &   0   &  0  \\ \hline
    $\widetilde{\ell}$,$\widetilde{\nu}$                                
&     12  &  5      & 4     &   4   &   2   &   2   &  2  \\ \hline 
    $\widetilde{\chi} \widetilde{\chi},
\widetilde{q} / \widetilde{g} \widetilde{\chi}$ 
&    123  & 74      &   32   &  17  &  13   &  10   &  4  \\ \hline
    $tH^-$ + c.c.                                               
&    76   & 38      &   22   &  15  &   9   &   3   &  1    \\ \hline 
    $Z^{0(*)} Z^{0(*)}$                               
& 1733   &  1683    &   43   &  39  &   5   &   4   &  4  \\ \hline   
    $t\bar{t} Z^{0*}$                                                  
&  47   &   23      &    2   &   1  &   1   &   0   &  0  \\ \hline
    $t\bar{t} h^0$                                                  
&      4  &   1     &    1   &   1  &   1   &   0   &  0  \\ \hline
    $H^0,A^0$ signal                                                
& 189,179 & 156,149 & 64,80 & 55,64 & 43,50 & 32,37 & 9,9  \\ \hline
    \end{tabular}
    \end{center}
\vspace{0.25cm}
\end{table}

Table \ref{tab:MSSMPt2} gives a ratio of $A^0 \to 4\ell$ events to 
$H^0 \to 4\ell$ events (before additional cuts) of $A^0:H^0 = 1:1.05$  
($48.6$\% $A^0$ events).  ISASUSY BR studies of the inclusive four-lepton 
event rates at this point also predict that $H^0$ will produce more signal 
events than $A^0$ this time, with $A^0:H^0 = 1:1.36$ ($42.4$\% $A^0$ 
events).  Exact agreement between the two methods is certainly not expected, 
and it is at least reassuring that both predict more 
$H^0 \to 4\ell$ events (unlike at MSSM Point 1).  
The percentage of $A^0 \to 4\ell$ events surviving the subsequent cuts
is again slightly larger than that for $H^0 \to 4\ell$ events 
($21$\% {\it vs.} $17$\%, excluding the four-lepton inv.\ m.\ cut).
Note that the $Z^0$-veto takes a larger portion out of the signal event 
number for MSSM Point 2 than it did for MSSM Point 1, with only about $50$\% 
surviving for the former while about $80$\% survive for the latter.  
This is understandable since, for MSSM Point 1, virtually all events were 
from $\widetilde{\chi}_2^0\widetilde{\chi}_2^0$ pairs, and
$\widetilde{\chi}_2^0$ is not heavy enough to decay to 
$\widetilde{\chi}_1^0$ via an on-mass-shell $Z^0$.  For MSSM Point 2, on 
the other hand, a variety of heavier --inos are involved, and the mass 
differences between $\widetilde{\chi}_3^0$ or
$\widetilde{\chi}_4^0$ and $\widetilde{\chi}_1^0$ do exceed $M_Z$.

\begin{figure}[!t]
\centerline{}
\begin{center}
\epsfig{file=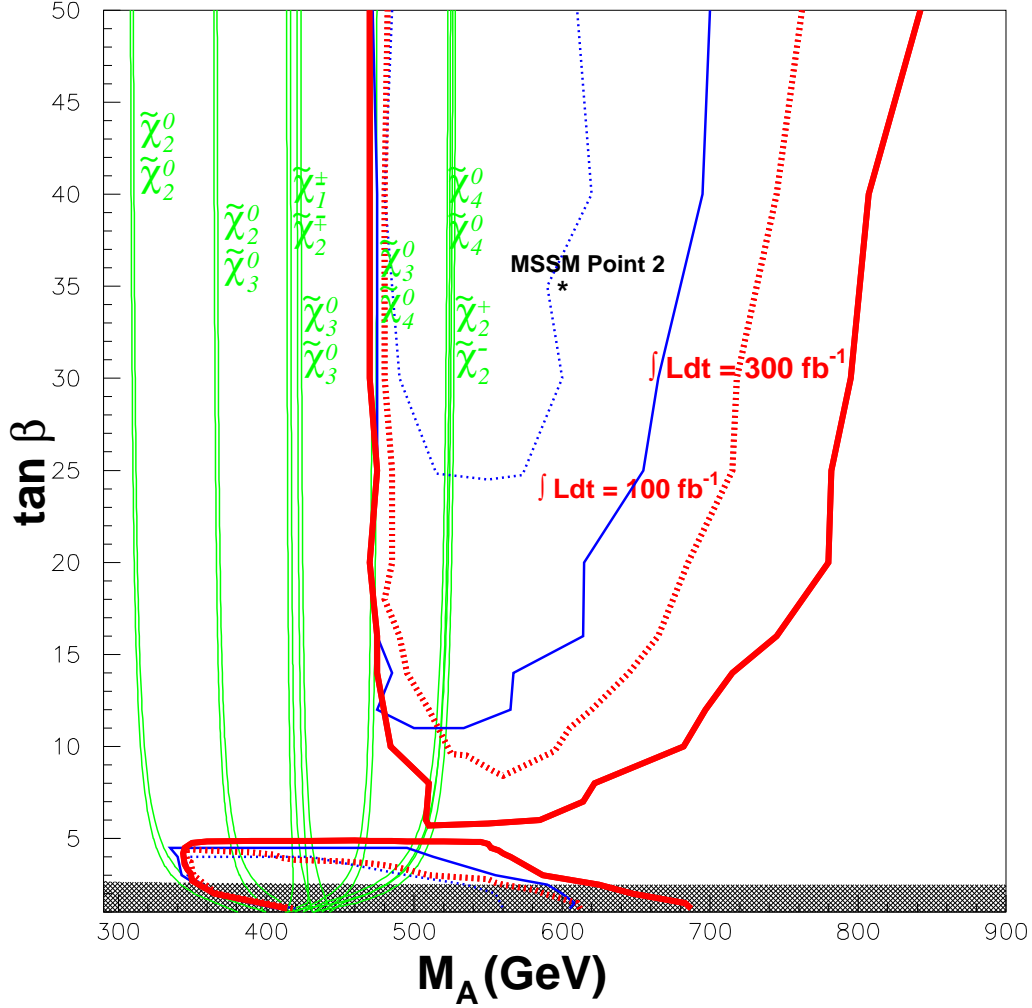,height=150mm,width=150mm}
\end{center}
\vskip -0.9cm
\caption{   
Discovery region in red in $(M_A, \tan\beta)$ plane for --ino/slepton
parameters   
$\mu = -200\, \hbox{GeV}$,
$M_2 = 200\, \hbox{GeV}$, $M_1 = 100\, \hbox{GeV}$,
$m_{\widetilde{\ell}_{soft}} = 150\, \hbox{GeV}$,
$m_{\widetilde{\tau}_{soft}} = 250\, \hbox{GeV}$
as in MSSM Point 2 (whose location is marked by an black asterisk).
Here Higgs boson decays to a variety of higher mass --inos
(see text) constitute the majority of the signal events.
Solid (dashed) red border delineates the discovery region for
$L_{int} = 300\, \hbox{fb}^{-1}$ ($100\, \hbox{fb}^{-1}$).
The green curves are 
$M_A,M_H - m_{\widetilde{\chi}_i^0}m_{\widetilde{\chi}_j^0}$
and 
$M_A,M_H - m_{\widetilde{\chi}_k^{\pm}}m_{\widetilde{\chi}_2^{\mp}}$
($i,j=2,3,4$; $k=1,2$).
The blue contours add the extra cut on the four-lepton inv.\ m.\
for the nominal cut-off value of $240\, \hbox{GeV}$.
}
\label{fig8:discovery}
\end{figure}

Again the --ino input parameters $M_2$ \& $\mu$ and the slepton \&
squark inputs are fixed, this time to be those of MSSM Point 2, and
$\tan\beta$ and $M_A$ allowed to vary to map out the Higgs boson discovery 
region in the ($M_A$, $\tan\beta$) plane (using the same criteria as in 
Fig.\ \ref{fig7:discovery}) shown in red in Fig.\ \ref{fig8:discovery}. 
As before, the solid (dashed) red border delineates the discovery region 
assuming an integrated luminosity of $300\, \hbox{fb}^{-1}$ ($100\, 
\hbox{fb}^{-1}$).  Assuming that the four-lepton inv.\ m.\ cut
is omitted, MSSM Point 2 lies firmly inside the $100\, \hbox{fb}^{-1}$   
discovery region (with the $15$ sparticle/charged Higgs boson 
+ $4$ $Z^{0(*)}Z^{0(*)}$ event background, 
Relation (\ref{disccrit}) requires $26$ signal events to be 
included in the $100\, \hbox{fb}^{-1}$ discovery region, while $69$ signal 
events are expected).  Note that Fig.\ 4 (right-side plot) predicts that
$H^0,A^0 \rightarrow \widetilde{\chi}_2^0 \widetilde{\chi}_2^0$ decays 
will only generate a substantial number of signal events when $\tan\beta$ 
and $M_A$ are small (the red and yellow zones in the plot), with decays 
to heavier --inos dominating elsewhere.  This leads to a disjoint 
discovery region in Fig.\ \ref{fig8:discovery}, consisting of a smaller 
mainly $\widetilde{\chi}_2^0 \widetilde{\chi}_2^0$-dominated portion for 
lower values of $\tan\beta$ and $M_A$ and a novel larger portion at 
considerably higher $M_A$ values that stretches up to $\tan\beta$ values 
well above $50$.  Note the distance 
between the lower $M_A$ edge of this larger portion of the discovery region 
and the curves for $M_A,M_H - 2m_{\widetilde{\chi}_2^0}$.  In concurrence 
with the percentage contributions for MSSM Point 2 given above, the lower 
$M_A$ edge of the discovery region abuts the 
$M_A,M_H - m_{\widetilde{\chi}_3^0} - m_{\widetilde{\chi}_4^0}$ curves
(shown in green in Fig.\ \ref{fig8:discovery}),
{\em for} $\tan\beta \, \gsim \, 10$.  
The situation for $\tan\beta\, \lsim \, 10$
and $450\, \hbox{GeV} \, \lsim \, M_A \, \lsim 700\, \hbox{GeV}$ 
(in both the upper and lower disjoint portions of the discovery region)
is more complicated, with $\widetilde{\chi}_2^0 \widetilde{\chi}_2^0$ and
several other decays making significant contributions. 

The discovery region shown in Fig.\ \ref{fig8:discovery} represents a
significant extension of LHC MSSM Higgs boson detection capabilities
to quite high Higgs boson masses.  With $300\, \hbox{fb}^{-1}$ of 
integrated luminosity, there is some stretch of $M_A$ values covered
for almost all values of $\tan\beta$ ($1 < \tan\beta < 50$), the
exception being $4 \, \lsim \, \tan\beta \, \lsim \, 6$.  
If the integrated luminosity is dropped to $100\, \hbox{fb}^{-1}$, 
the higher $M_A$ portion of the discovery region recedes up to  
$\tan\beta \, \gsim \, 8$-$10$, still lower than the $300\, 
\hbox{fb}^{-1}$ 
discovery regions from MSSM Higgs boson decays to 
third generation SM fermions found in the ATLAS \cite{ATLASsource} and 
other \cite{htautau} simulations.
The new discovery region has considerable overlap with the so-called
decoupling zone, where the light MSSM Higgs boson is difficult to
distinguish from the Higgs boson of the SM, and, 
{\em up to now, no signals of the other MSSM Higgs bosons were known}.

Though the number of signal events swells to over $50$ ($30$) per
$100\, \hbox{fb}^{-1}$ for $\tan\beta \, \lsim \, 2$ ($4$), the 
background from --ino pair-production via EW gauge bosons is also 
becoming quite large, and thus more integrated luminosity is required for 
the excess from Higgs boson decays to meet the (\ref{disccrit}) criterion.
Note how an `excess' attributed to the Higgs boson signal could
alternatively be accounted for by the MSSM background if the value
of $\tan\beta$ is lowered.  (Note also though that restrictions from LEP
experiments exclude the most sensitive region of extremely low 
$\tan\beta$ values.) 
As in  Fig.\ \ref{fig7:discovery}, the low $M_A$ edge of the lower portion
of the discovery region in Fig.\ \ref{fig8:discovery} abuts the 
$M_A,M_H - 2m_{\widetilde{\chi}_2^0}$ curves.

Yet for $M_A$ in the vicinity of $350\, \hbox{GeV}$ to $450\, \hbox{GeV}$, 
the discovery regions in Fig.\ \ref{fig7:discovery}
and Fig.\ \ref{fig8:discovery} resemble mirror images of each other:
the former lies exclusively above $\tan\beta \simeq 5$ while the
latter lies exclusively below $\tan\beta \simeq 5$.  The reasons
behind this stark contrast, though a bit complicated, critically depend 
on the different inputs to the slepton sector.
In Fig.\ \ref{fig8:discovery},
for $M_A \, \lsim \, 470\, \hbox{GeV}$, 
Higgs boson decays to other heavier
-inos are kinematically inaccessible, and,
for higher $\tan\beta$ values,
$\widetilde{\chi}_2^0$ decays almost exclusively via sneutrinos into
neutrinos and the LSP, yielding no charged leptons.
This is not the case in this region of Fig.\ \ref{fig7:discovery}
--- here $\widetilde{\chi}_2^0$ undergoes three-body decays via 
off-mass-shell sleptons and $Z^{0*}$ with substantial BRs into
charged leptons.  The situation for Fig.\ \ref{fig8:discovery}
changes as $\tan\beta$ declines below ${\sim}10$ since
$\widetilde{\chi}_2^0$ BRs to charged sleptons, while still much 
smaller than those to sneutrinos, grow beyond the percent level
--- sufficient to generate a low $\tan\beta$ discovery region in 
Fig.\ \ref{fig8:discovery}.
One might expect analogous behavior in  Fig.\ \ref{fig7:discovery};
however, in the low $\tan\beta$ region of Fig.\ \ref{fig7:discovery}
the partial widths
$\Gamma (H^0,A^0 \rightarrow \widetilde{\chi}_2^0 \widetilde{\chi}_2^0)$
are much smaller, especially for $A^0$, than they are in this region of
Fig.\ \ref{fig8:discovery} and decline with falling $\tan\beta$,
whereas in Fig.\ \ref{fig8:discovery}
$\Gamma (H^0 \rightarrow \widetilde{\chi}_2^0 \widetilde{\chi}_2^0)$
actually increases (though only moderately) as $\tan\beta$ falls.
The $\widetilde{\chi}_2^0 \widetilde{\chi}_2^0$ partial widths coupled 
with the subsequent $\widetilde{\chi}_2^0$ decays to charged leptons
are large enough in the case of Fig.\ \ref{fig8:discovery} so that the 
signal is not overwhelmed by the rising 
$\Gamma(H^0,A^0 \rightarrow t \bar{t})$ partial widths
as it is in the case of Fig.\ \ref{fig7:discovery}.  
Also, in Fig.\ \ref{fig8:discovery} but not in Fig.\ \ref{fig7:discovery},
as $M_A$ increases beyond ${\sim}450\, \hbox{GeV}$, contributions from
other --ino pairs besides $\widetilde{\chi}_2^0 \widetilde{\chi}_2^0$
become significant and further enhance the low $\tan\beta$ $4\ell$ signal 
rate.

Differences in the discovery regions at very high $\tan\beta$ values are 
also attributable to the slepton input parameters.  
In Fig.\ \ref{fig8:discovery}, the discovery region reaches up well beyond 
$\tan\beta = 50$, while in Fig.\ \ref{fig7:discovery} the discovery region 
is curtailed, ending before reaching $\tan\beta = 35$.  Since the soft slepton 
mass inputs for all three generations are degenerate for MSSM Point 1, for 
high $\tan\beta$ values in Fig.\ \ref{fig7:discovery} splitting effects 
with the staus drive one of the physical stau masses well below the selectron 
and smuon masses.  
This leads to lots of --ino decays including tau leptons, virtually 
shutting down the decays to electrons and muons.  Since the soft stau mass 
inputs are elevated well above the other slepton inputs for MSSM Point 2, 
this high $\tan\beta$ cap is removed in Fig.\ \ref{fig8:discovery}.

Comments made above for MSSM Point 2 about the increased severity of the 
$Z^0$-line cut and the inappropriateness of the four-lepton inv.\ m.\ 
cut (with the numerical cut-off set to $240\, \hbox{GeV}$) are also 
applicable to points throughout the larger portion of the discovery region 
of Fig.\ \ref{fig8:discovery}.  
As can be seen from the blue curves in Fig.\ \ref{fig8:discovery},
inclusion of the $240\, \hbox{GeV}$ $4\ell$ inv.\ m. cut eliminates about 
half of the $300\, \hbox{fb}^{-1}$ discovery region and far more than half
of the $100\, \hbox{fb}^{-1}$ region, including all points between
$\tan\beta \simeq 8$ and $\tan\beta \simeq 25$ for the latter.  

Also, in contrast to the discovery region of Fig.\ \ref{fig7:discovery},
in large segments of the Fig.\ \ref{fig8:discovery} discovery region the
number of signal events from $H^0$ decays exceed those from $A^0$ decays.
First consider the smaller, low $\tan\beta$, portion of the disjoint 
discovery region.  Herein, to the right of the
$M_A,M_H - m_{\widetilde{\chi}_3^0} - m_{\widetilde{\chi}_4^0}$ curves
(shown in green in Fig.\ \ref{fig8:discovery}),  
the percentage of $A^0$ events ranges from 
${\sim}30$-${\sim}40$\% (${\sim}25$-${\sim}30$\%) 
for $\tan\beta \, \gsim \, 2$ ($\lsim \, 2$).
To the left of these curves, the $A^0$ event percentage grows to 
${\sim}45$-${\sim}60$\% for $\tan\beta \, \gsim \, 2$; increasing further 
to ${\sim}70$-${\sim}80$\% near the region's upper left tip ($M_A$ in the 
vicinity of $350\, \hbox{GeV}$ and $\tan\beta$ around $3$ to $4.5$)
where the signal is dominated by Higgs-mediated 
$\widetilde{\chi}_2^0 \widetilde{\chi}_2^0$ production.

In the novel and larger high $\tan\beta$ portion of the 
discovery region in Fig.\ \ref{fig8:discovery}, where the 
$\widetilde{\chi}_2^0 \widetilde{\chi}_2^0$
contribution is minor to insignificant, the $H^0$ and $A^0$
contributions to the signal events stay within
$20$\% of each other
(with the $A^0$ event percentage ranging from 
${\sim}40$-${\sim}60$\%) to the right of the 
$M_A,M_H - 2m_{\widetilde{\chi}_4^0}$ curves. 
In the finger-like projection between the nearly-vertical
$M_A,M_H - m_{\widetilde{\chi}_3^0} - m_{\widetilde{\chi}_4^0}$
and $M_A,M_H - 2m_{\widetilde{\chi}_4^0}$ curves 
the $A^0$ percentage drops to $< \, 25$\% (after cuts, 
excluding the $4\ell$ inv.\ m.\ cut)\footnote{Here are some results from  
specific points in this region:  for $M_A = 510\, \hbox{GeV}$ and
$\tan\beta = 10,16,25,40$, the percentage of $A^0$ signal
events (again, after cuts, excluding the $4\ell$ inv.\ m.\ cut), is  
$23$\%, $17$\%, $11.5$\%, $21$\%.},
meaning that the number of events from $H^0$ to those from $A^0$ 
exceeds $3$ to $1$.
The $H^0$ dominance in this zone stems from the
$H^0$-$\widetilde{\chi}_3^0$-$\widetilde{\chi}_4^0$ coupling
($H^0$-$\widetilde{\chi}_3^0$-$\widetilde{\chi}_3^0$ coupling)
being two to three times larger (smaller) than the
$A^0$-$\widetilde{\chi}_3^0$-$\widetilde{\chi}_4^0$ coupling
($A^0$-$\widetilde{\chi}_3^0$-$\widetilde{\chi}_3^0$ coupling),  
combined with the fact that the
$\widetilde{\chi}_3^0 \widetilde{\chi}_4^0$ decays are about twice as
likely to produce $4\ell$ events as
those of $\widetilde{\chi}_3^0 \widetilde{\chi}_3^0$.
This of course means that $\widetilde{\chi}_4^0$ has a higher leptonic BR
than $\widetilde{\chi}_3^0$.  This in turn is due to
$\widetilde{\chi}_3^0$ decaying into $\widetilde{\chi}_1^0 Z^0$ about
half the time ($Z^0$ gives lepton pairs ${\sim}7$\% of the time), while
$\widetilde{\chi}_4^0$ almost never decays this way, instead having larger
BRs to charged sleptons [and $\widetilde{\chi}_1^{\pm} W^{\mp}$] which
always [${\sim}21$\% of the time] yield charged lepton pairs.
The situation changes quickly once the
$H^0,A^0 \rightarrow \widetilde{\chi}_4^0 \widetilde{\chi}_4^0,
\widetilde{\chi}_2^+ \widetilde{\chi}_2^-$ thresholds are (almost
simultaneously, see Fig.\ \ref{fig8:discovery}) crossed, thereafter for
higher $M_A$ values the $A^0$ and $H^0$ contributions remain reasonably 
close to each other as already stated.

As with points in  Fig.\ \ref{fig7:discovery}, direct chargino/neutralino
pair-production and slepton pair-production together with 
$Z^{0(*)} Z^{0(*)}$ production make up most of the 
background surviving the cuts.  Now, however, these are joined by a minor 
segment due to $t H^- \; + \; \hbox{c.c.}$ production, which depends on 
$M_A$ in addition to $\tan\beta$. 

Results showed $g b \rightarrow t H^-  \; + \; \hbox{c.c.}$ could yield 
several events at points in the discovery region.  Since the presence of a 
charged Higgs boson would also signal that there is an extended Higgs 
sector, these events could easily have been grouped with the signal rather 
than with the backgrounds.  Clearly though the set of cuts used in this 
work is not designed to pick out such events.  The jet cut typically 
removes roughly two-thirds to three-quarters of these events.  Here though 
it is interesting to note that, despite the presence of a top quark, the jet
cut does not remove all such events (unlike results found for squark and
gluino events and four-lepton $t\bar{t}X$ events).  A more effective set 
of cuts for $t H^-$, $\bar{t} H^+$ events is developed in \cite{EPJC2}, 
wherein substantially larger numbers of charged Higgs boson events survive 
the cuts therein at favorable points in the MSSM parameter space.  It is 
also worth noting though that the reach of the discovery region (at a 
favorable point in the MSSM parameter space) for the 
$H^0,A^0 \to 4\ell$ signal as described in this work surpasses that of the 
charged Higgs boson discovery regions found in \cite{EPJC2}.
(or in any other previous work on Higgs boson decays to sparticles).

An aspect to be mentioned in this connection, already highlighted in 
Ref.~\cite{LesHouches2003}, is the somewhat poor efficiency for the 
signals following the $Z^0$-veto, especially when combined with the fact 
that the $Z^{0(*)} Z^{0(*)}$ background survives the same constraint.
On the one hand, a non-negligible number of events in the signal decay 
chains leading to $4\ell N$ final states actually proceed via (nearly) 
on-mass-shell $Z^0$ bosons, particularly for MSSM Point 2, in which the 
mass differences $m_{\widetilde{\chi}_i^0} - m_{\widetilde{\chi}_1^0}$ 
($i=3,4$) can be very large, unlike the case 
$m_{\widetilde{\chi}_2^0} - m_{\widetilde{\chi}_1^0}$ 
for MSSM Point 1 (and in previous studies limited to only 
$\widetilde{\chi}_2^0 \widetilde{\chi}_2^0$ decay modes).  
On the other hand, the rather large intrinsic $Z^0$ width 
(when compared to the experimental resolution expected for di-lepton 
invariant masses) combined with a substantial production cross-section 
implies that $Z^{0(*)} Z^{0(*)}$ events will not be totally rejected 
by the $Z^0$-veto.
Altogether, though, the suppression is much more dramatic for the 
$Z^0Z^0$ background than for the signal, and so this cut is retained
(though the $Z^0$-veto will be dropped in some instances in the context 
of the forthcoming wedgebox analysis).
Also, varying the size of the $10\, \hbox{GeV}$ window around $M_Z$  
did not improve the effectiveness of this cut.

\subsection{mSUGRA benchmark points}

Turning attention briefly to the results within the more 
restrictive mSUGRA framework for SUSY-breaking, results for
mSUGRA Point A and mSUGRA Point B (as defined in Sect.\ 3)
are presented in Tables~5--6.
Mass spectra for these parameter sets are given in Table~\ref{tab:masses}.
For mSUGRA Point A ample signal events are produced and survive the cuts 
to claim observation of the Higgs boson at $100\,\hbox{fb}^{-1}$.  
The largest background is from direct slepton production, with 
direct neutralino/chargino production also contributing significantly,
whereas SM backgrounds are virtually nil.
Note how the $E_{T}^{\rm{jet}}$ cap suffices to eliminate the background from 
colored sparticle (squarks and gluinos) production. 

Recall that for mSUGRA Point A the signal is dominated by
$H^0,A^0 \rightarrow \widetilde{\chi}_2^0 \widetilde{\chi}_2^0$ decays,
whereas for mSUGRA Point B heavier --inos make major contributions.
Thus, a wedgebox plot analysis of the former is expected to show a
simple box topology, while in the case of the latter there
unfortunately may be too few events (even with $300\,\hbox{fb}^{-1}$
of integrated luminosity) to clearly discern a pattern.
For mSUGRA Point B, $9$($10$) signal events survive after all cuts
(save the $4\ell$ inv.\ m.\ cut), while $6$ background events
survive, assuming $100\,\hbox{fb}^{-1}$ of integrated luminosity.
This is insufficient to claim a discovery by the criterion of
Relation (\ref{disccrit}).  However, when the integrated luminosity
is increased to $300\,\hbox{fb}^{-1}$, then the raw number of signal 
events suffices to cross the discovery threshold.   Unfortunately 
though, for mSUGRA Point B the background from colored sparticle
production is not removed by the upper limit imposed on $E_{T}^{\rm{jet}}$.  
One can however
stiffen the $E_{T}^{\rm{jet}}$ cut, capping the allowable jet transverse energy
at $30\, \hbox{GeV}$ rather than $50\, \hbox{GeV}$ and thus eliminate much 
of this background without diminishing the signal rate significantly.
Then, with $300\,\hbox{fb}^{-1}$ of integrated luminosity the discovery 
criteria can be met.

An earlier ATLAS study \cite{SlavaZmushko01,ATLASTDRSUSY} also sought to map 
out the discovery reach of the Higgs boson to neutralino four-lepton signature 
within the mSUGRA framework transposed onto the ($M_A$, $\tan\beta$) plane.  
Though some statements to the contrary are included in this ATLAS study, it 
does seem to have been focused on the $\widetilde{\chi}_2^0\widetilde{\chi}_2^0$ 
contributions (analogous to previously-discussed general MSSM studies of this 
signature), thus apparently omitting parameter sets such as mSUGRA Point B
considered herein.  Thus, the viability of mSUGRA Point B indicates an
enlargement of the signal discovery region to higher values of $M_A$
(and the mSUGRA parameter $M_0$) at intermediate values of $\tan\beta$ 
({\it i.e.},in the `decoupling' region) from that reported in this ATLAS study
(akin to the enlargements shown in the general MSSM case, though the
extent of this enlargement in the case of mSUGRA models will not be quantified
herein).

\begin{table}[!h]\label{tab:mSUGRAPtA}
    \caption{Event rates after the successive cuts
defined in the text for mSUGRA Point A
\newline
(assuming an integrated luminosity of $100\, \hbox{fb}^{-1}$).}
\begin{center}
\begin{tabular}{|l||c|c|c|c|c|c|c|} \hline  
   Process & $4\ell$ events
& ${\ell}^+{\ell}^-{\ell}^{{\scriptscriptstyle ( \prime )}+}
                   {\ell}^{{\scriptscriptstyle ( \prime )}-}$
& $Z^0$-veto & $E_T^{\ell}$ & $E_T^{\rm{miss}}$& $E_{T}^{\rm{jet}}$
& $4\ell$ inv.\ m. \\ \hline
    $\widetilde{q}$, $\widetilde{g}$
&  927   &  504    &   312  &    280  &   174   &   0  &    0  \\ \hline
    $\widetilde{\ell}$,$\widetilde{\nu}$
& 326     &  178    &  145    &  117   &  100   &  71    & 58   \\ \hline
 $\widetilde{\chi} \widetilde{\chi},
\widetilde{q} / \widetilde{g} \widetilde{\chi}$
&  567   &  294   &  203  &   179  &  121   &  29   &   21  \\ \hline
    $tH^-$ + c.c.   
&  1    &    0   &   0   &    0  &   0   &   0   &    0  \\ \hline
    $Z^{0(*)} Z^{0(*)}$
&1733   &  1683  &  43   &   39  &   5   &   4   &    4  \\ \hline
    $t\bar{t} Z^{0(*)}$
&  47   &   23   &   2   &    1  &   1   &   0   &    0  \\ \hline
    $t\bar{t} h^0$
&   4   &    2   &   2   &    1  &   1   &   0   &    0  \\ \hline
    $H^0,A^0$ signal
& 46,140 &  40,123 & 38,122 & 38,120 & 30,83 & 24,66  & 24,66  \\ \hline
    \end{tabular}
    \end{center}
\end{table}

\begin{table}[!h]\label{tab:mSUGRAPtB}
    \caption{Event rates after the successive cuts
defined in the text for mSUGRA Point B 
\newline
(assuming an integrated luminosity of $100\, \hbox{fb}^{-1}$).}
\begin{center}
\begin{tabular}{|l||c|c|c|c|c|c|c|} \hline  
   Process & $4\ell$ events 
& ${\ell}^+{\ell}^-{\ell}^{{\scriptscriptstyle ( \prime )}+}
                   {\ell}^{{\scriptscriptstyle ( \prime )}-}$
& $Z^0$-veto & $E_T^{\ell}$ & $E_T^{\rm{miss}}$& $E_{T}^{\rm{jet}}$ 
& $4\ell$ inv.\ m. \\ \hline
    $\widetilde{q}$, $\widetilde{g}$                                   
&  4504   &  2598    &   1911  &    1672  &   917   &   12   &    12  \\ \hline
    $\widetilde{\ell}$,$\widetilde{\nu}$                        
& 309     &   169    &  134    &   110   &  94    &  67    &   57  \\ \hline
 $\widetilde{\chi} \widetilde{\chi},
\widetilde{q} / \widetilde{g} \widetilde{\chi}$
&  579   &   302   &  206   &   174  &   115   &   32   &   27  \\ \hline
    $tH^-$ + c.c.                                
&  1    &    1   &   0   &   0  &  0   &   0   &    0  \\ \hline
    $Z^{0(*)} Z^{0(*)}$                                          
&1733   &  1683  &  43   &  39   &  5    &  4    &    4  \\ \hline
    $t\bar{t} Z^{0(*)}$
&  47   &   23   &   2   &     1 &   1   &   0   &    0  \\ \hline
    $t\bar{t} h^0$                                   
&   5   &    2   &   1   &   1   &   1   &   0   &    0  \\ \hline
    $H^0,A^0$ signal                                   
& 43,130 &  38,118 & 37,116 & 37,116 & 29,93 & 23,75  & 23,75  \\ \hline
    \end{tabular}
    \end{center}
\end{table}

\section{Wedgebox analysis of Higgs boson decays to --ino pairs}

The wedgebox plot technique was introduced in a previous work \cite{Cascade}
which focused on neutralino pairs produced via colored sparticle production
and subsequent `cascade' decays.  Another work \cite{EWpaper} has just recently
focused on neutralino pairs produced via EW processes, including via a
$Z^{0(*)}$ boson or via $H^0$,$A^0$ production; the former is termed 
`direct' production while the latter is `Higgs-mediated' production.  
A jet cut was found to be fairly efficient in separating these two 
neutralino pair-production modes from cascade production assuming the 
colored gluinos and squarks are fairly heavy.    

To utilize the wedgebox technique, the criteria for the final four-lepton
state are further sharpened by demanding that the final state consist of one 
$e^+ e^-$ pair and one $\mu^+ \mu^-$ pair \footnote{In fact, this extra 
restriction is not strictly necessary, since recent preliminary 
work shows same-flavor four-lepton final states can be 
correctly paired with a reasonably high efficiency for 
at least some processes and some points in the MSSM 
parameter space \cite{Extension}.}.  
The wedgebox plot then consists of the $M(\mu^+ \mu^-)$ invariant mass
plotted versus the $M(e^+ e^-)$ invariant mass for all candidate events.
If a given neutralino, $\widetilde\chi^0_i$,
decays to the LSP, $\widetilde\chi^0_1$, and a charged lepton pair via a
three-body decay mediated by a virtual $Z^{0*}$ or virtual/off-mass-shell
charged slepton, then $M(\ell^+\ell^-)$ is bounded from above by
$m_{\widetilde{\chi}^0_i} - m_{\widetilde{\chi}^0_1}$
(and from below by $0$ if lepton masses are neglected).
Given a sufficient number of events, the wedgebox plot of the signal events
will be composed of a superposition of `boxes' and `wedges' \cite{Cascade}, 
in the $M(e^+ e^-)$-$M(\mu^+ \mu^-)$ plane resulting from 
decay chains of the form:
\begin{equation}
H^0,A^0 \rightarrow \widetilde\chi^0_i \widetilde\chi^0_j
\rightarrow e^+ e^- \mu^+ \mu^- \widetilde\chi^0_1\widetilde\chi^0_1 \; .
\end{equation}
If $\widetilde\chi^0_i$ ($\widetilde\chi^0_j$) decays into an $e^+e^-$
($\mu^+\mu^-$) pair, then $M(e^+ e^-)$ ($M(\mu^+ \mu^-)$) is bounded above
by $m_{\widetilde{\chi}^0_i} - m_{\widetilde{\chi}^0_1}$
($m_{\widetilde{\chi}^0_j} - m_{\widetilde{\chi}^0_1}$).
On the other hand, if $\widetilde\chi^0_i$ ($\widetilde\chi^0_j$) 
decays into a $\mu^+\mu^-$ ($e^+e^-$) pair, then these
$M(e^+ e^-)$ and $M(\mu^+ \mu^-)$ upper bounds are swapped.
Superposition of these two possibilities yields a `box' when 
$i=j$ (which will be called an `$i$-$i$ box') and a 
`wedge' (or `L-shape') when $i \ne j$ (this will be called an  
`$i$-$j$-wedge').

A heavy neutralino, $\widetilde\chi^0_i$, could instead decay to the
$\widetilde\chi^0_1$ $+$ leptons final state via a pair of two-body decays
featuring an on-mass-shell charged slepton of mass\footnote{Note that this 
is the physical slepton mass, not the soft mass input.} 
$m_{\widetilde{\ell}}$.  
Events containing such decays will lead to the same wedgebox pattern
topologies as noted above; however, the upper bound on $M(\ell^+\ell^-)$ 
is modified to 
\cite{Paige}
\begin{equation}
\label{two-body}
M(\ell^+\ell^-) < m_{\widetilde{\chi}_i^0}
\sqrt{ 1 - \left(\frac{m_{\widetilde{\ell}}}
                      {m_{\widetilde{\chi}_i^0}}\right)^{\!\!\! 2}}  
\sqrt{ 1 -
\left(\frac{m_{\widetilde{\chi}_1^0}}
           {m_{\widetilde{\ell}}}\right)^{\!\!\! 2}}  \;\; .
\end{equation}
The $M(\ell^+\ell^-)$ spectrum is basically triangular in this case
and sharply peaked toward the upper bound, while the former three-body decays
yield a similar but less sharply peaked spectrum. 
The two-body decay series alternatively could be via an on-mass-shell 
$Z^0$, resulting in an $M(\ell^+\ell^-) = M_Z$ spike.  

   Additional complications can arise if the heavy neutralino 
$\widetilde\chi^0_i$ can decay into another neutralino $\widetilde\chi^0_j$
($j \ne 1$) or a chargino which subsequently decays to yield the 
$\widetilde\chi^0_1$ final state.  These may introduce new features
to the wedgebox plot:  $\widetilde\chi^0_i$ to $\widetilde\chi^0_j$ ($j \ne 1$) 
decay chains involving
$\widetilde{\chi}_3^0 \rightarrow \ell^+\ell^-\widetilde{\chi}_2^0$,
$\widetilde{\chi}_4^0 \rightarrow \ell^+\ell^-\widetilde{\chi}_2^0$, 
and/or
$\widetilde{\chi}_4^0 \rightarrow   \ell^+\ell^-\widetilde{\chi}_3^0$
will generate additional abrupt event population changes or edges,
termed `stripes,' on the wedgebox plot.  One can imagine quite elaborate decay 
chains, with
$\widetilde{\chi}_4^0 \rightarrow \widetilde{\chi}_3^0 \rightarrow 
\widetilde{\chi}_2^0 \rightarrow \widetilde{\chi}_1^0$ for instance.  
However, such elaborate chains are very unlikely to emerge from any 
reasonable or even allowed choice of MSSM input parameters.
Further, each step in such elaborate decay chains
either produces extra visible particles in the final state
or one must pay the price of the BR to neutrino-containing states.  The
latter tends to make the contribution from such channels insignificant,
while the former, in addition to also being suppressed by the additional
BRs, may also be cut if extra restrictions are placed on the final state
composition in addition to demanding an $e^+e^-$ pair and a $\mu^+\mu^-$
pair.  The aforementioned extra visible particles could be two more 
leptons, meaning that all four leptons come from only one of the initial 
-inos,
$\widetilde{\chi}_i^0 \rightarrow \ell^+\ell^- \widetilde{\chi}_k^0
\rightarrow \ell^+\ell^- {\ell^{\prime}}^+{\ell^{\prime}}^-
\widetilde{\chi}_1^0$,
while the other --ino, which must yield no leptons (or other visible 
final state SM particles forbidden by additional cuts), decays via
$\widetilde{\chi}_j^0 \rightarrow \nu\bar{\nu} \widetilde{\chi}_1^0$
or
$\widetilde{\chi}_j^0 \rightarrow q\bar{q} \widetilde{\chi}_1^0$.
Again though such channels will be suppressed by the additional required BRs.
A further caveat is that decays with extra missing energy (carried off by
neutrinos, for example) or missed particles can further smear the endpoint.
The presence of charginos may also further complicate the wedgebox picture.
Heavier --inos can decay to the LSP $+$ lepton pair final state via a 
chargino, $\widetilde{\chi}_i^0 \rightarrow \ell^+\nu  \widetilde{\chi}_1^-
\rightarrow \ell^+\nu {\ell^{\prime}}^-{\bar{\nu}}^{\prime} 
\widetilde{\chi}_1^0$, or a Higgs boson itself may decay into a chargino pair,
with one chargino subsequently yielding three leptons 
while the other chargino yields the remaining one 
(such events are called `3+1 events' \cite{EWpaper}).  The chargino 
yielding three leptons will typically decay via a $\widetilde{\chi}_2^0$, 
resulting in a re-enforcement of the solely --ino-generated wedgebox 
plot topology.  A single chargino-generated lepton paired with another 
lepton from a different source produces a wedge-like structure but with 
no definite upper bound.  For a more in-depth discussion of these nuances, 
see \cite{EWpaper}.

\begin{figure}[!t]
\centerline{}
\begin{center}
\hskip -0.5cm
\epsfig{file=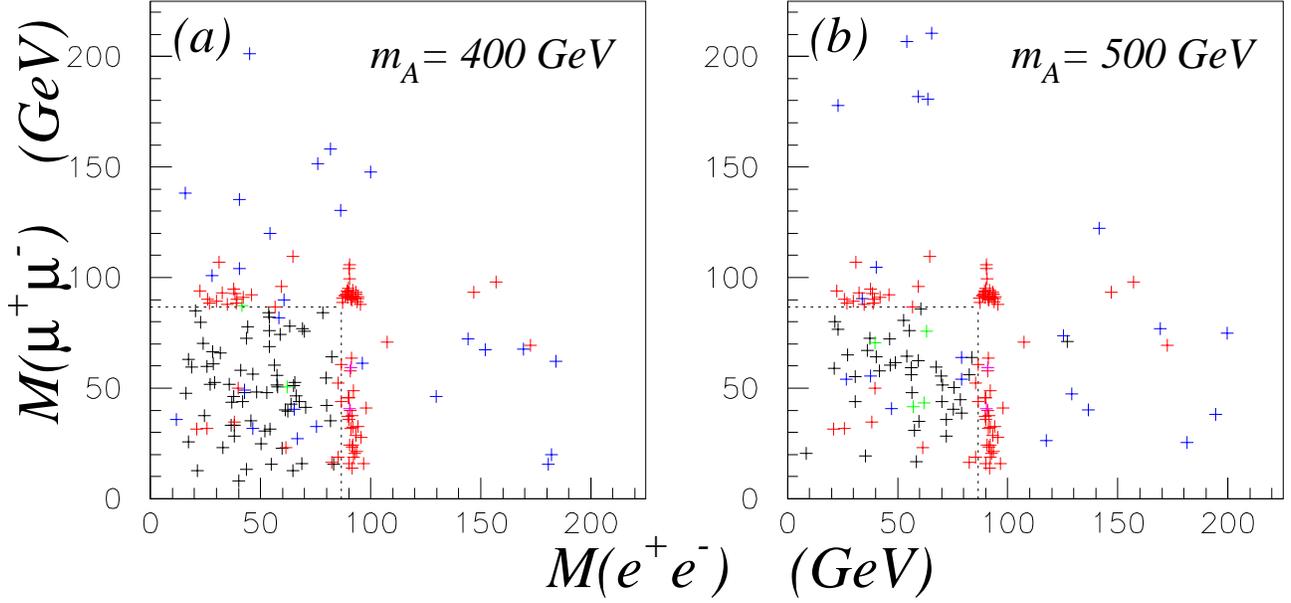,height=80mm,width=170mm}
\end{center}
\vskip -0.2cm
\caption{
Wedgebox plot for MSSM Point 1 inputs with $M_A = 500\, \hbox{GeV}$
$(b)$ and shifting to $M_A = 400\, \hbox{GeV}$ $(a)$, assuming an
integrated luminosity of $300\, \hbox{fb}^{-1}$.  Neither the
$Z^0$-veto cut nor the 4-lepton invariant mass cut are enabled.
Black-colored markers are for the $H^0$ and $A^0$ signal events.
SM background events from $Z^{0(*)} Z^{0(*)}$ (where either one or both 
of the $Z^0$'s are permitted to be off-mass-shell are red), while the two 
surviving $t \bar{t} Z^{0(*)}$ events are purple.  MSSM background 
events from slepton production or direct neutralino/chargino production are 
in blue and green, respectively.
The horizontal and vertical dashed lines forming a box are at the
location $m_{\widetilde{\chi}_2^0} - m_{\widetilde{\chi}_1^0}$.  
MSSM Point 1 --ino and slepton inputs are
$\mu = -500\, \hbox{GeV}$,
$M_2 = 180\, \hbox{GeV}$, $M_1 = 90\, \hbox{GeV}$,
$m_{\widetilde{\ell}_{soft}} =
m_{\widetilde{\tau}_{soft}} = 250\, \hbox{GeV}$.
}
\label{Point1squareWBplot}
\end{figure}

The right-hand plot in Fig.\ \ref{Point1squareWBplot} shows the wedgebox 
plot obtained in the case of MSSM Point 1, assuming an integrated LHC 
luminosity of $300\, \hbox{fb}^{-1}$.  Criteria for event selection are
as given in the previous section, save that the more restrictive demand of 
an $e^+e^-{\mu}^+{\mu}^-$ final state is applied while the $Z^0$-veto and 
four-lepton invariant mass cuts are not applied.  Both signal and 
background events are included; the former are colored black.  The 
latter consist of both SM backgrounds (on- or off-mass-shell 
$Z^0$-boson pair-production --- $Z^{0(*)}Z^{0(*)}$, $83$ events, and 
$t\bar{t}Z^{0(*)}$, largely removed by the missing energy and jet cuts, 
$2$ remaining events; these events are colored red and purple, 
respectively, in Fig.\ \ref{Point1squareWBplot} ) and MSSM sparticle 
production processes
(`direct' neutralino or chargino production, $4$ events, and slepton 
pair-production, $22$ events; such events are colored green and blue, 
respectively, in Fig.\ \ref{Point1squareWBplot}).  No events from colored 
sparticle production survive the cuts, particularly the jet cut --- this 
is a crucial result.
Signal events consist of $14$ $H^0$ events and $25$ $A^0$ 
events, yielding a signal to background of $39:111 = 1:2.85$.  With
$S / \sqrt{{B} } = 3.7$, this is not good enough 
to claim a discovery based on Relation (\ref{disccrit}).  If the input 
CP-odd Higgs boson mass is lowered to $M_A = 400\, \hbox{GeV}$, whose 
wedgebox plot is the left-hand plot of Fig.\ \ref{Point1squareWBplot}, 
then the number of signal events rises to $14$ + $52$ = $66$ $H^0$ and 
$A^0$ events (runs for MSSM backgrounds gave $2$ `direct' 
neutralino-chargino events and $26$ slepton-pair production events), 
yielding $S / \sqrt{{B} } = 6.2$ and satisfying 
Relation (\ref{disccrit}).  Note how the increase is solely due to more 
$A^0$-generated events.  Comparing the $M_A = 500\, \hbox{GeV}$ 
(MSSM Point 1) plot $(b)$ and the $M_A = 400\, \hbox{GeV}$ plot $(a)$ in 
Fig.\ \ref{Point1squareWBplot} shows how the increased number of signal 
events in $(a)$ more fully fills in the $2$-$2$ box whose outer edges
(dashed lines in the figure) are given by 
$m_{\widetilde{\chi}_2^0} - m_{\widetilde{\chi}_1^0}=86.6\, \hbox{GeV}$  
since for these input parameters slepton masses are too high to permit 
$\widetilde{\chi}_2^0$ decays into on-mass-shell sleptons.  

A key observation is that the distributions of the signal and the 
background events differ markedly\footnote{On the other hand, the 
distributions of $A^0$ and $H^0$ events show no substantial systematic 
differences in their distributions' wedgebox plot topologies.}.  
All but one of the signal events lie within the $2$-$2$ box\footnote{Note 
that a similar result is found in Fig.\ 16 of \cite{CMS1}.  There, however, 
only signal events were shown, and, since {\it a priori} only 
$H^0,A^0 \rightarrow \widetilde{\chi}_2^0 \widetilde{\chi}_2^0$
events were considered, the vast array of other potential wedgebox
topologies was not brought to light.}.
The majority of the slepton pair-production
events ($19$ out of $26$ events for $(a)$ and $17$ out of $22$ events for
$(b)$), the dominant MSSM background, lie outside the $2$-$2$ box.
The topology of these `3+1' events is a $2$-$2$ box plus a wedge 
lacking a clear outer edge extending from said box (see \cite{EWpaper}).
The few `direct' neutralino and chargino production events happen to all 
lie within the $2$-$2$ box; however, these events 
are actually due to\footnote{If direct neutralino
pair-production produces a significant number of events, then the dominant
source of said events is always  $\widetilde{\chi}_2^0 
\widetilde{\chi}_3^0$
production; $\widetilde{\chi}_2^0 \widetilde{\chi}_2^0$ production is
heavily suppressed.
See discussion in \cite{EWpaper}.  This leads to the general conclusion
that, with a jet cut in place to remove cascade events from colored
sparticle decays, the appearance of a disproportionately strong
(densely populated) box on a wedgebox plot is highly
indicative of the presence of Higgs-boson-generated events.  The caveat
to this being that chargino production can generate a box-shape in some
rather limited regions of the MSSM input parameter space.
Again, see \cite{EWpaper} for further discussion.}
$\widetilde{\chi}_2 \widetilde{\chi}_3$ 
pair-production and thus, for a larger sample, such events would populate 
a $2$-$3$ wedge with many of the events falling outside of the 
$2$-$2$ box.

SM background events are concentrated on and around lines where either 
$M(e^+ e^-)$ and/or $M(\mu^+ \mu^-)$ equals $M_Z$, which unfortunately 
is close to the outer edges of the $2$-$2$ box.  Using the unfair
advantage of color-coded events, one can correctly choose to place the 
edges of the box so as to exclude most of the SM background events.
Experimentalists may have a more difficult time deciding on wedgebox
edges that lie too close to $M_Z$.  Though, at the price of perhaps 
losing some of the signal events\footnote{Correct edge values from 
which to reconstruct information on the --ino mass spectrum would also 
be lost.}, one could make a selection 
rule of an effective $2$-$2$ box with edges sufficiently within $M_Z$
in such cases.  Correct identification of the outer edge value for the 
$2$-$2$ box removes all but $11$ of the $85$ SM background events.
The signal:background is then $39:20$ for $(b)$ and $66:19$ for $(a)$,
an immense improvement in the purity of the samples --- both points 
now certainly satisfy the Relation (\ref{disccrit}) criterion.  
Accepting only points lying within a box with outer edges at 
$80\, \hbox{GeV}$, more safely eliminating SM $Z^{0(*)} Z^{0(*)}$ 
events, leads to a signal:background of $33:12$ for $(b)$ and $59:14$ 
for $(a)$.
Note that one can also select points lying well outside the $2$-$2$ box to 
get a fairly pure sample (at this point in the parameter space) of slepton 
pair-production events.  Even if one does not know where Nature has
chosen to reside in the MSSM input parameter space, the selection of 
only events occupying one distinct topological feature of the experimental 
wedgebox plot may yield a sample pure enough (though one may not know 
exactly what purified sample one has obtained!) to be amenable to other 
means of analysis (perhaps entailing some addition reasonable hypotheses
as to what sparticles might be involved) \cite{MassRelMeth}.

\begin{figure}[!t]
\centerline{}
\vspace{0.01cm}
\begin{center}
\hskip -0.5cm
\epsfig{file=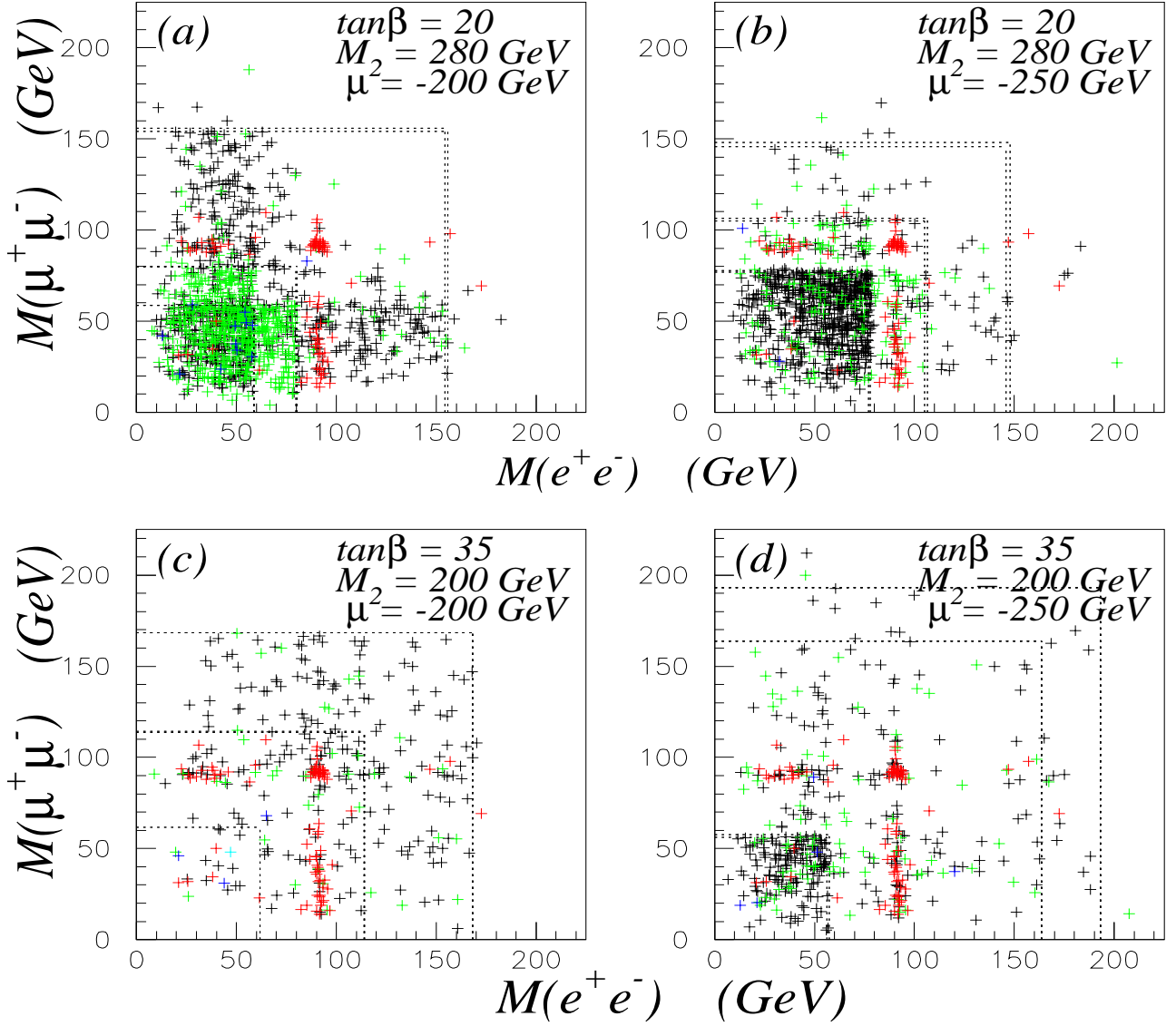,height=150mm,width=170mm}
\end{center}   
\vskip -0.6cm
\caption{
Wedgebox plot for MSSM Point 2 inputs 
$(c)$ and shifting to $M_A = 400\, \hbox{GeV}$ (left), assuming an
integrated luminosity of $300\, \hbox{fb}^{-1}$.  Neither the
$Z^0$-veto cut nor the 4-lepton invariant mass cut are enabled.
Black-colored markers are for the $H^0$ and $A^0$ signal events.  SM
background events from $Z^{0(*)} Z^{0(*)}$ (where either one or both of 
the $Z^0$'s are permitted to be off-mass-shell) are red, while the two 
surviving $t\bar{t}Z^{0(*)}$ events are purple.  MSSM background events 
from slepton production or direct neutralino/chargino production are in 
blue and green, respectively.
The horizontal and vertical dashed lines forming a box are at the
location $M_{\widetilde{\chi}_2^0} - M_{\widetilde{\chi}_1^0}$.
MSSM Point 1 --ino and slepton inputs are
$\mu = -500\, \hbox{GeV}$,
$M_2 = 180\, \hbox{GeV}$, $M_1 = 90\, \hbox{GeV}$,
$m_{\widetilde{\ell}_{soft}} =
m_{\widetilde{\tau}_{soft}} = 250\, \hbox{GeV}$.
Also indicated by dashed lines on the plot are the $2$-$2$, $3$-$3$ and 
$4$-$4$ box edges found from relation (\ref{two-body}) --- save for the 
$2$-$2$ box edges for $(c)$ which are from 
$m_{\widetilde{\chi}_2^0} - m_{\widetilde{\chi}_1^0}$. 
}
\label{Point2wedgeWBplot}
\end{figure}

Fig.\ \ref{Point2wedgeWBplot} in turn examines several related 
choices for input parameter sets, including MSSM Point 2 --- which is plot 
$(c)$ therein, in which $H^0$ and $A^0$ have large BRs into heavier 
--ino pairs such that the majority of the $4\ell$ signal events do not 
arise from $\widetilde{\chi}_2^0 \widetilde{\chi}_2^0$ decays for all points 
save that of plot $(b)$.  
Plot $(d)$ differs from MSSM Point 2, plot $(c)$, only in that 
the Higgsino mixing parameter $\mu$ is shifted from 
$\mu = -200\, \hbox{GeV}$ to $\mu = -250\, \hbox{GeV}$.
Yet even this modest change drastically alters
the topology of the resulting wedgebox plot.  This is
illustrative of how the wedgebox plot may be useful in extracting
fairly detailed information about the --ino spectrum and corresponding 
MSSM input parameters.
In plots $(a)$ and $(b)$ of  Fig.\ \ref{Point2wedgeWBplot}
the EW gaugino input parameters are raised from
$M_2 = 200\, \hbox{GeV}$ in plots $(c)$ and $(d)$ to
$M_2 = 280\, \hbox{GeV}$  (recall the assumption used herein
that the value of $M_1$ is tied to that of $M_2$).  Also
$\tan\beta$ is lowered from $35$ to $20$, while $\mu$ values of  
plots $(c)$ and $(d)$ are retained.  Again, these shifts in
input parameters radically alter the resulting wedgebox topology.
Plots $(a)$ and $(b)$ clearly show wedge-like topologies.
Note again the markedly different event distributions for the
signal and background events in all four plots, but particularly
striking in plot $(a)$.
Note how the four MSSM parameter set points
yielding the wedgebox plots depicted in Fig.\ \ref{Point2wedgeWBplot}
all might crudely be categorized as high $\tan\beta$, low $| \mu |$,
low to moderate $M_2$, and light slepton points.
Yet the associated wedgebox plots come out decidedly different.

Taking advantage of knowing which points in MSSM parameter space are 
being simulated (something the experimentalist cannot know in the 
actual experiment) allows comparison between the assorted calculated 
production rates at the four points and the observed features on the 
wedgebox plots.  Table~\ref{tab:pairperc} gives such theoretical estimates 
based on analysis of ISAJET (ISASUSY) 7.58 results for the four 
points\footnote{Table~\ref{tab:percents2} given previously corresponds 
to column $(c)$ in Table~\ref{tab:pairperc} with the $H^0$ and $A^0$ 
contributions listed separately.}.  It must be borne in mind though that 
effects from cuts may alter the percentage contributions found on the 
wedgebox plots from those given in Table~\ref{tab:pairperc}.

\begin{table}[!t]
  \vskip -0.5cm
     \caption{{\small
   Percentage contributions to $H^0,A^0 \rightarrow 4\ell$ events
   from the various neutralino and chargino pair-production modes
   for the four MSSM Parameter set points given in 
   Fig.\ \ref{Point2wedgeWBplot}.
   Based upon ISAJET(ISASUSY) 7.58 \cite{ISAJET} with no consideration 
   given to any cuts.  Decays that are kinematically not allowed are 
   marked by a $0$; contributions below $0.001$\% are marked as negligible 
   ($neg$).
   $H^0,A^0 \rightarrow Z^{0(*)} Z^{0(*)}$, $H^0 \rightarrow h^0 h^0$ 
   and $A^0 \rightarrow h^0 Z^{0(*)}$ make negligible contributions in 
   all cases.
    Also given are the number of $H^0$,$A^0$ signal events and the 
    number of background events, assuming
    $300\, \hbox{fb}^{-1}$ of integrated luminosity as in the figure.
    }}
     \begin{center}  
     \begin{tabular}{|c||r@{}l|r@{}l|r@{}l|r@{}l|} \hline
   Decay Pair 
   & \multicolumn{2}{|c}{$(a)$}
   & \multicolumn{2}{|c}{$(b)$}
   & \multicolumn{2}{|c}{$(c)$}
   & \multicolumn{2}{|c|}{$(d)$}
     \\ \hline \hline   
    ${\widetilde\chi}^0_2$ ${\widetilde\chi}^0_2$
            & $18.$&$6$\%
            & $70.$&$6$\%
            & $0.$&$0015$\%
            & $35.$&$0$\%
    \\  \hline
   ${\widetilde\chi}^0_2$ ${\widetilde\chi}^0_3$
            & $0.$&$1$\%
            & $4.$&$5$\%
            & $0.$&$05$\%
            & $13.$&$1$\%
    \\ \hline
   ${\widetilde\chi}^0_2$ ${\widetilde\chi}^0_4$
            & $45.$&$1$\%  
            & $13.$&$0$\%
            & $0.$&$05$\%
            & $1.$&$6$\%
     \\  \hline
   ${\widetilde\chi}^0_3$ ${\widetilde\chi}^0_3$
               & $1.$&$5$\%
               & $0.$&$4$\%
               & $2.$&$7$\%
               & $0.$&$9$\%
     \\  \hline
   ${\widetilde\chi}^0_3$ ${\widetilde\chi}^0_4$
            & $18.$&$1$\%
            & $5.$&$0$\%
            & $45.$&$0$\%
            & $9.$&$5$\%
     \\  \hline
   ${\widetilde\chi}^0_4$ ${\widetilde\chi}^0_4$
           & $0$&
           & $0$&
           & $39.$&$6$\%
           & $7.$&$8$\%
     \\  \hline\hline 
    ${\widetilde\chi}^{\pm}_1$ ${\widetilde\chi}^{\mp}_2$
           & $16.$&$6$\% 
           & $6.$&$5$\% 
           & $11.$&$3$\%
           & $31.$&$8$\%
         \\ \hline
     ${\widetilde\chi}^+_2$ ${\widetilde\chi}^-_2$
           & $0$&
           & $0$&
           & $1.$&$4$\%  
           & $0.$&$3$\%
          \\ \hline \hline
     ${\widetilde\chi}^0_1$ ${\widetilde\chi}^0_3$
           & $0.$&$001$\%
           & $0.$&$005$\%
           & \multicolumn{2}{|c|}{$neg$}
           & $0.$&$05$\%
    \\ \hline   
    ${\widetilde\chi}^0_1$ ${\widetilde\chi}^0_4$
           & $0.$&$02$\% 
           & \multicolumn{2}{|c|}{$neg$}  
           & \multicolumn{2}{|c|}{$neg$}
           & $0.$&$01$\%
    \\ \hline \hline
    $H^0,A^0$ evts.
           &  \multicolumn{2}{|c|}{$305$,$423$}
           &  \multicolumn{2}{|c|}{$276$,$473$}
           &  \multicolumn{2}{|c|}{$122$,$105$}
           &  \multicolumn{2}{|c|}{$182$,$140$}
    \\ \hline
   bckgrd. evts.
           &  \multicolumn{2}{|c|}{$683$}
           &  \multicolumn{2}{|c|}{$257$}
           &  \multicolumn{2}{|c|}{$132$}
           &  \multicolumn{2}{|c|}{$186$}
    \\ \hline
       \end{tabular}
    \end{center}
 \label{tab:pairperc}
\end{table}

The first thing to notice from this table is the virtual absence of 
events stemming from $\widetilde{\chi}_2^0$ to $\widetilde{\chi}_1^0$
decays for MSSM Point 2 = plot $(c)$ relative to the other three
points.  This is due to the fact that, for this input parameter set,
the sparticle spectrum satisfies the condition that 
$m_{\widetilde{\nu}} < m_{\widetilde{\chi}_2^0} < 
m_{\widetilde{\ell}^{\pm}}$,
meaning that $\widetilde{\chi}_2^0$ mainly decays via an 
on-mass-shell sneutrino `spoiler' mode, 
$\widetilde{\chi}_2^0 \rightarrow
\widetilde{\nu} \bar{\nu} \rightarrow \widetilde{\chi}_1^0 \nu \bar{\nu}$,
and its BR into a pair of charged leptons is highly suppressed.
For the other three points, 
$m_{\widetilde{\chi}_2^0} > 
m_{\widetilde{\ell}^{\pm}},m_{\widetilde{\nu}}$.
Actually, of the four wedgebox plots shown in
Fig.\ \ref{Point2wedgeWBplot}, the one for MSSM
Point 2 most closely resembles a simple box.  
However, Table \ref{tab:pairperc} indicates that (before cuts)
$45.0$\% of the events are from
$\widetilde{\chi}_3^0 \widetilde{\chi}_4^0$,
 $39.6$\% of the events are from
$\widetilde{\chi}_4^0 \widetilde{\chi}_4^0$,
and $12.7$\% of the events are from
$\widetilde{\chi}_1^{\pm} \widetilde{\chi}_2^{\mp}, \,
\widetilde{\chi}_2^+ \widetilde{\chi}_2^-$.

In Fig.\ \ref{Point2wedgeWBplot}, charged sleptons are now light enough so 
that the neutralino to slepton decay chains, which make significant 
contributions to the four-lepton signal events, may proceed via 
on-mass-shell charged sleptons.  So while the outer edges of the 
$2$-$2$ box in Fig.\ \ref{Point1squareWBplot} was determined by the 
$\widetilde{\chi}_2^0$-$\widetilde{\chi}_1^0$ mass difference, here
Relation \ref{two-body} 
brings the slepton masses into play\footnote{Unfortunately,
the physical slepton masses input into HERWIG 6.5 are generated by
ISASUSY 7.58 \cite{ISAJET}, which neglects a left-right 
mixing term $\propto$ $m^2_\ell \mu^2\tan^2\beta$ (see \cite{Cascade}).
While this term is negligible for selectrons, it does shift the physical 
smuon masses by as much as a few GeV.  Neglecting this term results in 
degenerate soft slepton inputs leading to degenerate physical selectron 
and smuons masses (so the smuon masses for MSSM Point 2 given in 
Table \ref{tab:masses} are changed into the mass values given there for 
the selectrons), which in turn may noticeably under-estimate the 
mass splitting between smuons and thus the thickness of the edges
shown on the plots.  Later versions of ISAJET correct this oversight,
as do private codes employed in Sect.\ 2.}.
In plot $(a)$, virtually all $\widetilde{\chi}_i^0$ to 
$\widetilde{\chi}_1^0$ decays proceed via on-mass-shell sleptons, but
only the $\widetilde{\chi}_4^0$ to $\widetilde{\chi}_1^0$ decay edge is 
significantly altered (by more than a couple GeV) --- from
$m_{\widetilde{\chi}_4^0} - m_{\widetilde{\chi}_4^0} = 185\, \hbox{GeV}$
to $151$-$156\, \hbox{GeV}$ (at this point, $18$\% of four-lepton 
events are from $\widetilde{\chi}_3^0\widetilde{\chi}_4^0$
according to Table~\ref{tab:pairperc}).
On the other hand, in plot $(b)$, where the $\widetilde{\chi}_i^0$ 
also decay to $\widetilde{\chi}_1^0$ via on-mass-shell sleptons, edges 
are shifted from 
$m_{\widetilde{\chi}_i^0} - m_{\widetilde{\chi}_1^0} = 82,124,192\, \hbox{GeV}$
to $76$-$78,101$-$107,140$-$149\, \hbox{GeV}$ for $i=2,3,4$, 
respectively\footnote{Due to the program oversight mentioned in the 
last footnote, the thicknesses of these edges shrink to 
$75.7$-$76.5,103.4$-$104.9,143.5$-$145.9\, \hbox{GeV}$, respectively.
These values are represented by the dotted lines on the plots.},
with $i=2,3,4$ decays all making noteworthy four-lepton event 
contributions.
For MSSM Point 2 = plot $(c)$, the shift in the $\widetilde{\chi}_3^0$ to 
$\widetilde{\chi}_1^0$ decay edge is only $3.5$-$5\, \hbox{GeV}$ while the 
$\widetilde{\chi}_4^0$ to $\widetilde{\chi}_1^0$ edge is virtually 
unchanged.  This accounts for $87.3$\% of the four-lepton events by 
Table~\ref{tab:pairperc}.
The situation with $\widetilde{\chi}_2^0$ is slightly complicated:
$\widetilde{\chi}_2^0$ can only decay into $\widetilde{\chi}_1^0$ via an 
on-mass-shell\footnote{Again, this feature is lost in 
HERWIG 6.5/ISAJET 7.58 .} $\widetilde{\mu}_1$, and this would lead to a 
tremendous shift in the edge position 
(from $61\, \hbox{GeV}$ to $15\, \hbox{GeV}$); however,
this is so close to the kinematical limit that decays through 
off-mass-shell $Z^{0*}$ should be competitive (again placing the edge at 
${\sim}61\, \hbox{GeV}$).  {\em But}, since $\widetilde{\chi}_2^0$ decays
lead to only a tiny fraction of the four-lepton events, note how there
is no visible edge or population discontinuity at this location (the 
innermost dashed box) on the wedgebox plot.  
Lastly, with plot $(d)$, again on-mass-shell slepton decays totally 
dominate for $i=2,3,4$, but only the $\widetilde{\chi}_2^0$ to
$\widetilde{\chi}_1^0$ decay edge is significantly shifted
(from $75.2\, \hbox{GeV}$  to 
$51.5$-$60.4\, \hbox{GeV}$\footnote{In HERWIG 6.5/ISAJET 7.58
this width shrinks to $55.6$-$57.0\, \hbox{GeV}$.}.  
But, by Table~\ref{tab:pairperc}, this decay is the most important
contributor to the signal events. 

For plot $(a)$ of Fig.\ \ref{Point2wedgeWBplot}, the expected 
$2$-$4$ wedge stands out clearly among the signal events, with 
outer edges at the expected location.  The background is mostly
from direct $\widetilde{\chi}_2^0 \widetilde{\chi}_3^0$
direct production, giving the $2$-$3$ wedge shown in green
(direct neutralino-neutralino production is predominantly
$\widetilde{\chi}_2^0 \widetilde{\chi}_3^0$ at all interesting
points in the MSSM parameter space, with direct
$\widetilde{\chi}_2^0 \widetilde{\chi}_2^0$ production always
highly suppressed \cite{EWpaper}).
The proximity of this wedge's outer edges to the red $M_Z$ lines
may complicate the experimental analysis; however, if the SM 
$Z^{0(*)}Z^{0(*)}$ background is well-modeled, a subtraction technique
to clear up this zone may be feasible.
Note that selecting only events with 
$100\, \hbox{GeV} < M(e^+e^-) \: < \: 150\, \hbox{GeV}$,
$0 < M({\mu}^+ {\mu}^-) \: < \: 50\, \hbox{GeV}$
or
$0 < M(e^+ e^-) \: < \: 50\, \hbox{GeV}$
$100\, \hbox{GeV} \: <  \: M({\mu}^+ {\mu}^-) < 150\, \hbox{GeV}$,
corresponding to the legs of the $2$-$4$ wedge lying beyond the 
$2$-$3$ wedge and the $Z^0$-line, changes the signal:background
ratio from $728$:$683$ seen on the plot to $128$:$15$.  
This is an example of a cut that can be applied {\it a posteriori}
based on the examination of the wedgebox plot --- as opposed to 
assuming {\it a priori} extra knowledge about where in the MSSM
parameter space Nature has chosen to sit.

Plot $(b)$ of Fig.\ \ref{Point2wedgeWBplot}
mainly shows a densely-populated $2$-$2$ box whose edges are well inside
the $M_Z$ lines.  A faint $2$-$3$ or $2$-$4$ wedge is also discernible
(in fact Table~\ref{tab:pairperc} shows this to be a $2$-$4$ wedge), 
while the empty upper-right corner which does not join with the 
$2$-$2$ box suggests that 
$\widetilde{\chi}_2^0 \widetilde{\chi}_4^0$ and
$\widetilde{\chi}_3^0 \widetilde{\chi}_4^0$ decays are present
while $\widetilde{\chi}_4^0 \widetilde{\chi}_4^0$ are
absent (further suggesting that said decay mode is kinematically
inaccessible, which helps pin down the relative masses of the 
heavy Higgs bosons and the heavier neutralinos). 

Plot $(c)$'s most obvious feature is an outer box, which in fact is 
a $4$-$4$ box.  Topology alone does not distinguish this from a 
plot dominated by a $3$-$3$ box or a $2$-$2$ box, though the location
of the outer edges well beyond $M_Z$ might give pause for entertaining 
the latter possibility.  A $3$-$4$ wedge may also be discerned from the 
somewhat diminished event population in the upper right-hand box in the 
plot.  Comparison of this plot with the other three quickly points out
the absence of a dense event-population in this plot.  Seeing such a 
wedgebox plot experimentally strongly hints that 
leptonic $\widetilde{\chi}_2^0$ decays are being suppressed, perhaps 
with a mass spectrum favoring sneutrino spoiler modes as noted above.

Like plot $(b)$, plot $(d)$ shows a $2$-$2$ box, but with outer
edges at a very different location.  Plot $(d)$ also has more signal
events outside of the $2$-$2$ box than does plot $(b)$, and said
events are more scattered in $(d)$.  A lot of these events are from 
$H^0,A^0$ decays into 
$\widetilde{\chi}_1^{\pm} \widetilde{\chi}_2^{\mp}$ pairs.  
Thus, the alignment of the wedgebox features to the dashed lines 
derived from neutralino features shown is less compelling.  

In both  Fig.\ \ref{Point1squareWBplot} and
 Fig.\ \ref{Point2wedgeWBplot}, note how closely the wedgebox plot 
features, obtained
by the full event generator \& detector simulation analysis, conform
to the dashed-line borders expected from the simple formula \ref{two-body}.  
This strongly supports the assertion that a wedgebox-style analysis is 
realistic in the actual experimental situation.

\section{Summary and conclusions}

Recapping the findings presented herein:

\subsection{New signals}

For many interesting choices of the basic input parameters of the MSSM,
heavier Higgs boson decay modes of the type
$H^0, A^0 \rightarrow \widetilde{\chi}_i^0 \widetilde{\chi}_j^0$,
with $i,j \ne 1$ are potentially important LHC signal modes.  
The neutralinos' subsequent leptonic decays, typified by
$\widetilde{\chi}_i^0 \rightarrow \ell^+ \ell^- \widetilde{\chi}_1^0$,
can yield a four-isolated-lepton 
(where here $\ell$ refers to electrons and/or muons) plus 
missing-transverse-energy signature.  Such leptonic neutralino decays may 
proceed via either an intermediate charged slepton or via an intermediate
$Z^{0(*)}$, where in either case this intermediate state may be on- or 
off-mass-shell.  The present study presents for the first time a 
systematic investigation of the potential for discovering such a signature 
at the LHC, including all possible such neutralino pairs:
$\widetilde{\chi}_2^0\widetilde{\chi}_2^0,
\widetilde{\chi}_2^0\widetilde{\chi}_3^0,
\widetilde{\chi}_2^0\widetilde{\chi}_4^0,
\widetilde{\chi}_3^0\widetilde{\chi}_3^0,
\widetilde{\chi}_3^0\widetilde{\chi}_4^0,
\, \hbox{and}\, \widetilde{\chi}_4^0\widetilde{\chi}_4^0$.
Other Higgs boson decays that may lead to the same signature
are also incorporated,
including: decays to chargino pairs 
$H^0,A^0 \rightarrow \widetilde{\chi}_1^{\pm} \widetilde{\chi}_2^{\mp}, 
\widetilde{\chi}_2^+ \widetilde{\chi}_2^-$, in which case 
$\widetilde{\chi}_2^{\mp}$ yields three leptons while the other 
chargino gives the fourth;
$H^0,A^0 \rightarrow \widetilde{\chi}_1^0 \widetilde{\chi}_3^0,
\widetilde{\chi}_1^0 \widetilde{\chi}_4^0$, where the 
$\widetilde{\chi}_3^0$ or
$\widetilde{\chi}_4^0$
must provide all four leptons; and
$H^0 \rightarrow h^0 h^0, Z^{0(*)} Z^{0(*)}$, 
$A^0 \rightarrow h^0 Z^{0(*)}$,
\& $H^0,A^0 \rightarrow \widetilde{\ell}^+ \widetilde{\ell}^-$, all three 
of which yield negligible contributions in all cases studied.
This surpasses previous studies which restricted virtually all of their 
attention to $H^0,A^0 \rightarrow \widetilde{\chi}_2^0 \widetilde{\chi}_2^0$, 
and also did not consider the possibility of neutralino decays to on-mass-shell 
sleptons (with the incorporation of the heaviest neutralinos as is
done herein this assumption becomes particularly restrictive).  

Naturally, at least some of the --inos must be reasonably light for this
$H^0,A^0 \rightarrow 4\ell + E_T^{\rm{miss}}$ signature to be seen.
Parameter-space scans studying the potential scope of such a signal
indicate that the --ino parameter $M_2$ needs to be relatively low
while the Higgsino mixing parameter $\mu$ need not be so constrained
(however, if $| \mu |$ is not also relatively low, then the 
signal is dominated by the $\widetilde{\chi}_2^0 \widetilde{\chi}_2^0$
mode).  Relatively light slepton masses are also quite helpful,
and the slepton mass spectrum plays a crucial r\^ole in determining
for what values of the other MSSM input parameters large rates may occur.   
Said large rates are possible throughout most of the 
phenomenologically-interesting value ranges of the Higgs-sector parameters
$M_A$ and $\tan\beta$, depending of course on the accompanying choice of 
other MSSM inputs, as the discovery regions delineated herein illustrate.

\subsection{Comparison with previous results}

\begin{figure}[!t]
\centerline{}
\begin{center}
\vskip -1.95cm
\epsfig{file=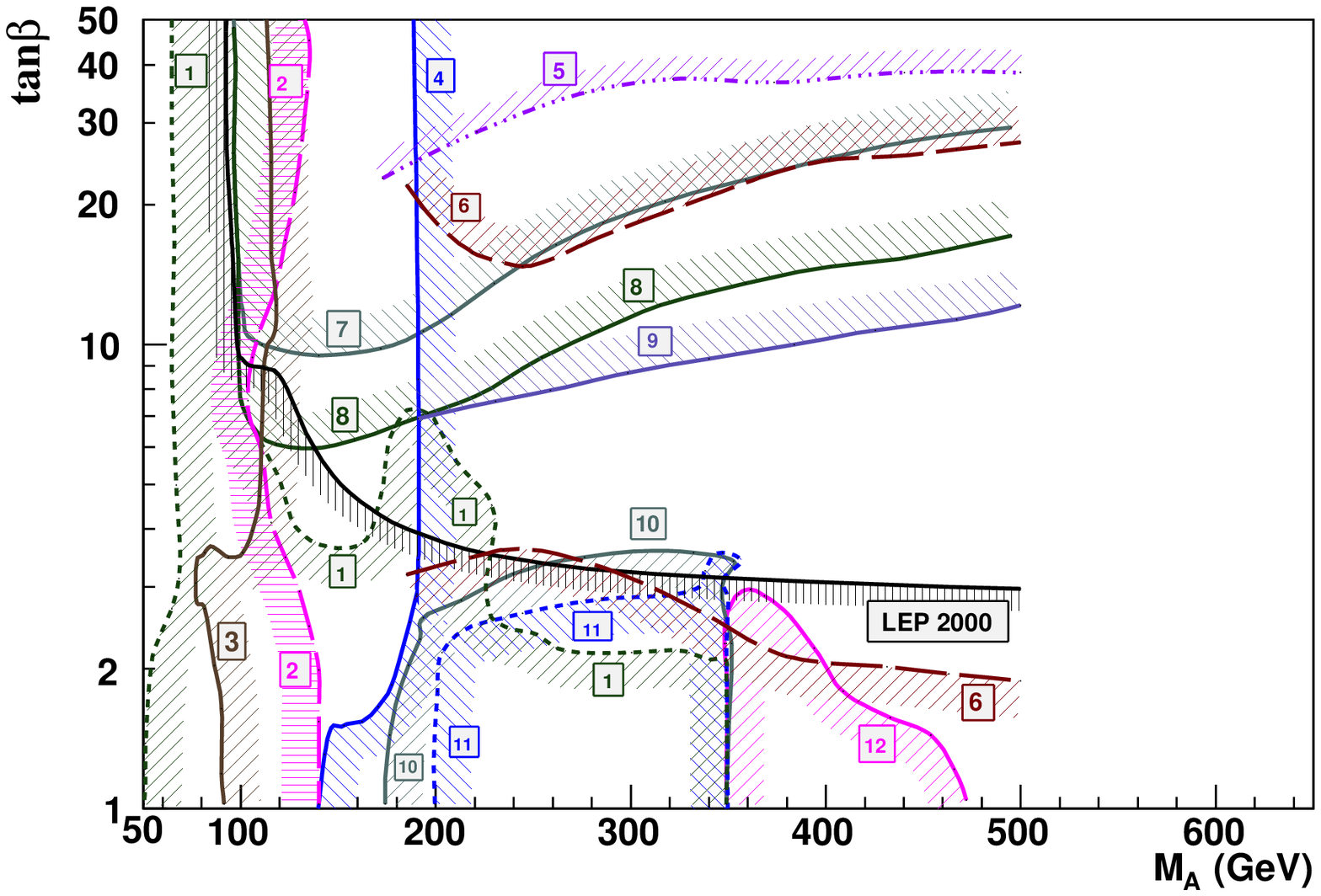,height=90mm,width=140mm}
\vskip -1.0cm
\epsfig{file=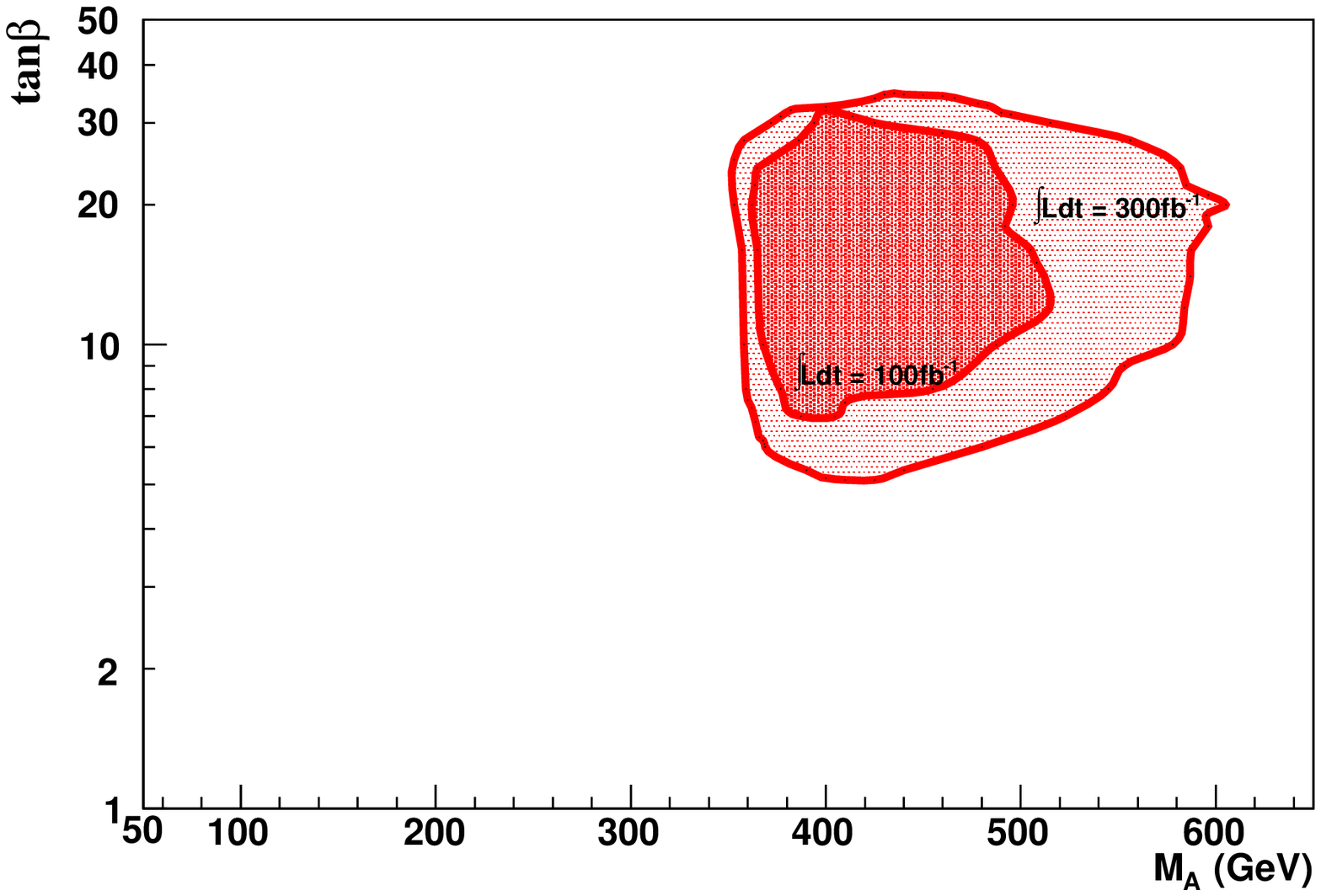,height=90mm,width=140mm}
\end{center}
\vskip -0.7cm
\caption{
Discovery regions in the ($M_A, \tan\beta $) plane,
here with a logarithmic $\tan\beta$ scale,
assuming MSSM Parameter Set 1 --ino inputs and for
${\cal L}_{int} = 100\, \hbox{fb}^{-1}$ and $300\, \hbox{fb}^{-1}$,
for the (lower plot) MSSM Higgs bosons' $4\ell$ signals from their
decays into neutralino or chargino pairs
(here $H^0,A^0$ decays to $\widetilde{\chi}_2^0 \widetilde{\chi}_2^0$ 
totally dominate).  This is
shown juxtaposed (upper plot) with $300\, \hbox{fb}^{-1}$ regions for 
MSSM Higgs boson signatures from decays to SM particles based 
upon LEP results and ATLAS simulations \protect\cite{ATLASsource}, 
where labels represent:
1. $H^0 \rightarrow Z^0Z^{0*} \rightarrow 4\, \hbox{leptons}$;
2. $t \rightarrow b H^+, \; H^+ \rightarrow \tau^+ \nu$ + c.c.;
3. $t \bar{t}h^0,\; h^0 \rightarrow b \bar{b}$;
4. $h^0 \rightarrow \gamma \gamma$ and
  $W^{\pm}h^0/tth^0, \; h^0 \rightarrow \gamma\gamma$;
5. $b \bar{b}H^0, b\bar{b}A^0$ with $H^0/A^0 \rightarrow b \bar{b}$;
6. $H^+ \rightarrow t \bar{b}$ + c.c.;
7. $H^0/A^0 \rightarrow {\mu}^+ {\mu}^-$;
8. $H^0/A^0 \rightarrow {\tau}^+ {\tau}^-$;
9. $g\bar{b} \rightarrow \bar{t}H+, \; H^+ \rightarrow \tau^+ \nu$ +
  c.c.;
10. $H^0 \rightarrow h^0h^0 \rightarrow b \bar{b}\gamma\gamma$;
11. $A^0 \rightarrow Z^0h^0 \rightarrow \ell^+ \ell^- b \bar{b}$;
12. $H^0/A^0 \rightarrow t \bar{t}$.
Note that SM discovery regions are not for
  the same input parameters: they presume a very heavy 
  sparticle spectrum; identical MSSM inputs to those used for the lower plot
  may well yield smaller SM discovery regions in a revised upper plot.
  For the $4\ell$ signals from
  $\widetilde{\chi}^0_i \widetilde{\chi}^0_j,
   \widetilde{\chi}^+_m \widetilde{\chi}^-_n$ decays,
   the MSSM Parameter Set 1 --ino/slepton parameters are
   $\mu = -500\, \hbox{GeV}$,
   $M_2 = 180\, \hbox{GeV}$, $M_1 = 90\, \hbox{GeV}$ and
   $m_{\widetilde{\ell}_{soft}} =
   m_{\widetilde{\tau}_{soft}} = 250\, \hbox{GeV}$.
   }
   \vskip -1.2cm
   \label{Point1ATLAS:discovery}
   \end{figure}

\begin{figure}[!t]
\centerline{}
\begin{center}
\vskip -1.50cm
\epsfig{file=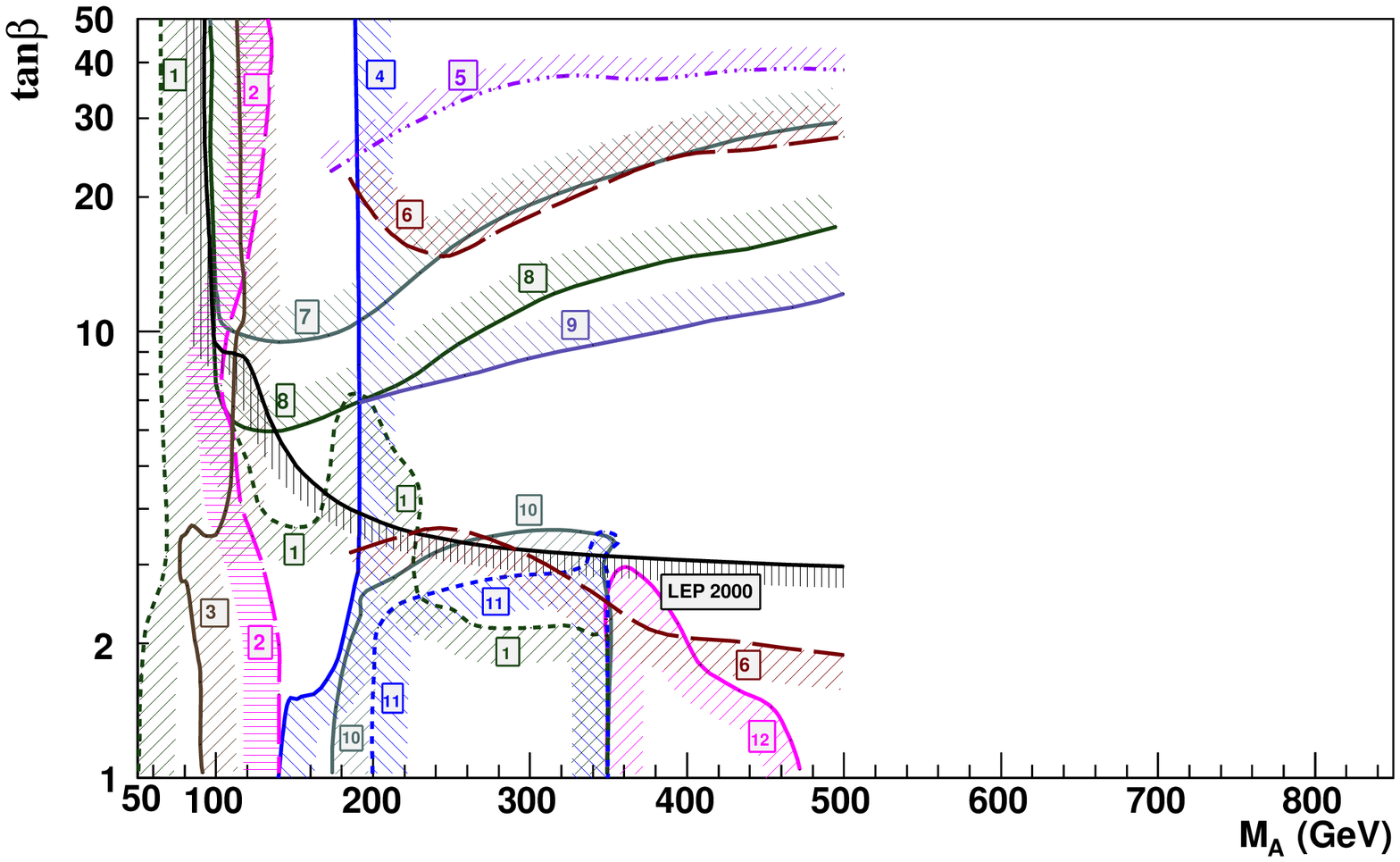,height=90mm,width=160mm}
\vskip -1.0cm
\epsfig{file=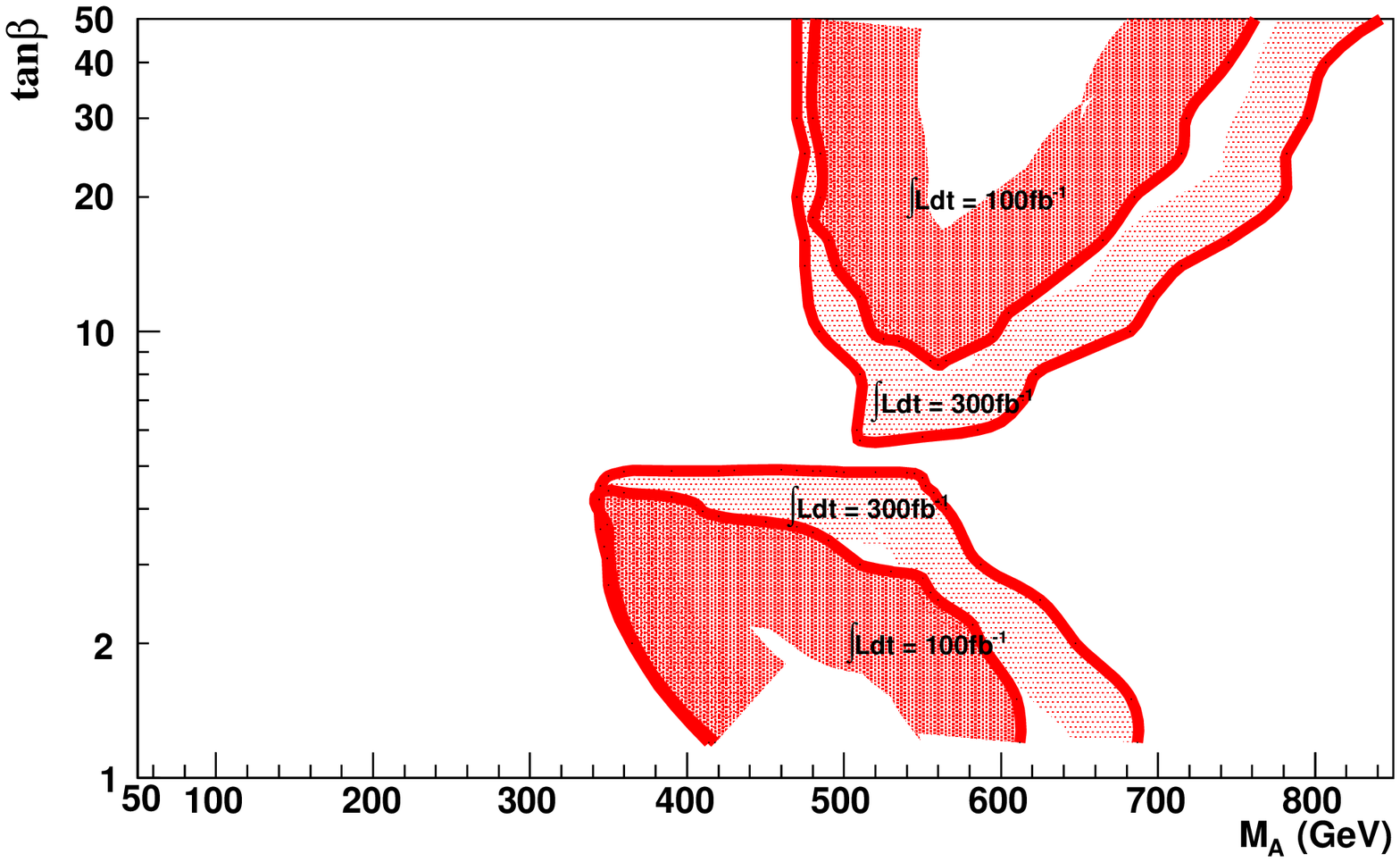,height=90mm,width=160mm}
\end{center}
\vskip -0.75cm
\caption{
Discovery regions in the ($M_A, \tan\beta $) plane,
here with a logarithmic $\tan\beta$ scale,
assuming MSSM Parameter Set 2 --ino inputs and for
${\cal L}_{int} = 100\, \hbox{fb}^{-1}$ and $300\, \hbox{fb}^{-1}$,
for the (lower plot) MSSM Higgs bosons' $4\ell$ signals from their
decays into neutralino or chargino pairs
(here Higgs boson decays to higher-mass neutralinos
typically dominate).  This is
shown juxtaposed (upper plot) with $300\, \hbox{fb}^{-1}$ regions for
MSSM Higgs boson signatures from decays to SM particles as in Fig.\ 11.
For the $4\ell$ signals from
$\widetilde{\chi}^0_i \widetilde{\chi}^0_j,
 \widetilde{\chi}^+_m \widetilde{\chi}^-_n$ decays,
 the MSSM Parameter Set 2 --ino/slepton parameters are
 $\mu = -200\, \hbox{GeV}$,
 $M_2 = 200\, \hbox{GeV}$, $M_1 = 100\, \hbox{GeV}$,
 $m_{\widetilde{\ell}_{soft}} = 150\, \hbox{GeV}$ and
 $m_{\widetilde{\tau}_{soft}} = 250\, \hbox{GeV}$.
 Here Higgs boson decays to a variety of higher mass --inos
 (see text) constitute the majority of the signal events.
 Note that, as in Fig.\ 11, since ATLAS discovery regions presume
 a very heavy sparticle spectrum, SM discovery regions made for the 
 same MSSM input parameters as used in the lower plot 
 may well yield smaller SM discovery regions in a revised upper plot.
}
\vskip -0.80cm
\label{Point2ATLAS:discovery}
\end{figure}

To clearly demonstrate the potential importance of the
$H^0,A^0 \rightarrow 4\ell + E_T^{\rm{miss}}$ signature in the hunt for
the heavier Higgs bosons, Figs.\ \ref{Point1ATLAS:discovery} and
\ref{Point2ATLAS:discovery} again show the discovery regions associated
with MSSM Point 1 and MSSM Point 2 neutralino input parameter sets
(as depicted before in Figs.\ \ref{fig7:discovery} and
\ref{fig8:discovery}, respectively), but this time with a logarithmic
scale for $\tan\beta$ {\em and} also showing the
expected reaches, assuming $300\, \hbox{fb}^{-1}$ of integrated luminosity
at the LHC, of Higgs boson decay modes into SM daughter particles
as developed by the ATLAS collaboration \cite{ATLASsource}\footnote{ATLAS
collaboration discovery region contour lines in
Figs.\ \ref{Point1ATLAS:discovery} and \ref{Point2ATLAS:discovery}
have been remade to match as closely as possible those in the original plot.}.
Clearly, the new neutralino decay mode signature can extend the discovery
reach for the heavier MSSM Higgs bosons to much higher values of $M_A$, and
also offer at least partial coverage of the so-called `decoupling region'
where only the lightest Higgs state $h^0$ could be established in the past
(through its decays into SM objects) and where said $h^0$ may be difficult
to distinguish from the sole Higgs boson of the minimal SM.  Thus, a more
complete analysis of the $H^0, A^0 \rightarrow \widetilde{\chi}_i^0
\widetilde{\chi}_j^0$ modes as is presented here may be crucial to the
establishment of an extended Higgs sector.  The inclusion of the
heavier neutralinos, $\widetilde{\chi}_3^0$ and $\widetilde{\chi}_4^0$,
absent in previous studies, is essential in extending the reach of the
$H^0,A^0 \rightarrow 4\ell + E_T^{\rm{miss}}$ signature up to the
higher Higgs boson masses unattainable by the SM decay modes.

It should be noted that the ATLAS discovery contours presented in
Figs.\ \ref{Point1ATLAS:discovery} and \ref{Point2ATLAS:discovery}
are {\em not} obtained using the same choice of MSSM input parameters as
are the $H^0, A^0 \rightarrow \widetilde{\chi}_i^0
\widetilde{\chi}_j^0$ discovery regions developed in the present work.
In fact, the ATLAS discovery regions used input choices designed to
eliminate, or at least minimize, the Higgs boson decays into sparticles.
Thus, the reach of the ATLAS discovery contours essentially represents the
maximum expanse in the MSSM parameter space achievable through these
Higgs boson decays to SM particles under the (unsubstantiated) assumption
of a very heavy sparticle sector.  Stated another way:  were the ATLAS
discovery regions to be generated for the same set of neutralino input
parameters as the
$H^0, A^0 \rightarrow \widetilde{\chi}_i^0 \widetilde{\chi}_j^0$ discovery
regions presented herein, the former may well {\em shrink} in size
(and certainly {\em not increase}), further emphasizing the
importance of thoroughly studying the
$H^0,A^0 \rightarrow 4\ell + E_T^{\rm{miss}}$ signature.
It would certainly be desirable to re-do the SM-like signature reaches of
MSSM Higgs bosons in the presence of light sparticle spectra identical to
those 
   
   \clearpage

\clearpage

\parindent=0pt

studied herein for the Higgs-to-sparticle decay channels; however,
this is clearly beyond the scope and capabilities of this study.  It also
must be emphasized that the diminution of the expected signatures from SM
decay modes of the MSSM Higgs bosons was investigated in \cite{PRD1} and
thus is fairly well-established as well as inherently sensible.

\parindent=15pt

Previous studies exploring Higgs-to-sparticle decay channels,
whether for neutral Higgs bosons ({\it e.g.}, CMS \cite{CMS1}) or for
charged Higgs bosons ({\it e.g.}, ATLAS \cite{Tord}, CMS \cite{EPJC2}),
--- and comparing, to some extent, SM and SUSY decay modes --- have not
re-scaled the reaches of previously-studied SM decay channels
(done by the same collaboration) to allow a reasonable comparison to the
new-found sparticle decay modes; nor have the SM decay modes been re-analyzed
for the same set of MSSM input parameters.  Yet clearly such comparisons are
absolutely essential to gauge the scope and impact of the new sparticle-decay
channels.  Certainly, the comparisons presented in
Figs.\ \ref{Point1ATLAS:discovery} and \ref{Point2ATLAS:discovery}
are less than optimal; however, they are far from un-informative.

It is also important to keep in mind that the assumptions inherent in the
ATLAS (and CMS) discovery regions for the SM decay modes of the MSSM
Higgs bosons are no less restrictive than the choices of MSSM input
parameters made to generate the two $4\ell + E_T^{\rm{miss}}$ discovery
regions in this study.  The parameter space scans of Sect.\ 2 further
enable the reader to put the two discovery regions shown here into a wider
perspective.

\subsection{Production and decay phenomenology of the signal}

The new $H^0,A^0 \rightarrow 4\ell + E_T^{\rm{miss}}$ discovery regions have
been mapped out using a full event generator-level analysis utilizing
HERWIG coupled with a detector simulation on a par with experimental 
analyses.  All significant backgrounds have been included in the analysis,
some for the first time in the study of such a signature.  The importance 
of the restriction on jet activity employed herein is particularly
noteworthy.  Without such a cut the Higgs signal could be swamped by 
the cascade decays of colored sparticles (gluinos and squarks), unless
said sparticles are {\it a priori} assumed to be quite heavy (at or above
the TeV scale).  The ultimate limit of this type of jet cut,
to demand that events be `hadronically quiet' quickly springs to mind
as an attractive search category.   Yet care must be taken here since, 
in Higgs boson production via $gg \rightarrow H^0, A^0$ and $b \bar{b} 
\rightarrow H^0 ,A^0$, jets emerge in the final state alongside the Higgs 
bosons due to PS effects, though such additional jets tend to be rather soft 
and collinear to the beam directions.  
In addition, rather than emulating Higgs boson production 
via $gg\rightarrow H^0, A^0$ and $b\bar{b} \rightarrow H^0 ,A^0$, one could 
instead consider $gg \rightarrow gg H^0 ,gg A^0$ and 
$gg \rightarrow b \bar{b} H^0 ,b \bar{b} A^0$ processes, in which case
one might worry about stronger jet activity emerging.
The true signal rate is the sum of these and the previous process types,
after making a correction for the overlap (as discussed previously).  
HERWIG simulations of $gg \rightarrow b \bar{b} H^0 ,b \bar{b} A^0$ at 
selected points in the parameter space indicate that the these processes 
are in fact removed by the jet cut imposed herein.  To better optimize the 
level of hadronic activity that should be allowed, full implementation of 
$2 \rightarrow 3$ loop processes ($gg \rightarrow gg H^0 ,gg A^0$ and other
channels yielding two light jets and a $H^0, A^0$ in the final state)
into HERWIG must be completed (work in progress \cite{SM}). 

The BRs of $H^0$ and $A^0$ to the assorted --ino pairs can certainly
differ markedly in regions where the signal is large, as seen for
instance in Table 3; thus one must not assume that the two contribute 
a roughly equal number of events to the $4\ell + E_T^{\rm{miss}}$ signal 
rate.  
On the other hand, results also show that only in quite narrow low-$M_A$ 
threshold regions within the discovery areas (wherein the small 
$M_H$-$M_A$ mass difference is crucial) do events due to one or the 
other Higgs boson (in this case the lighter $A^0$) totally dominate, 
producing in excess of $90$\% of the signal events.
General statements beyond this concerning the $H^0$ and $A^0$ admixture 
present in the signal seem elusive.  Throughout the 
$\widetilde{\chi}_2^0 \widetilde{\chi}_2^0$-dominated
discovery region of Fig.\ \ref{fig7:discovery}, $A^0$ produced the
majority of the events (though in some cases only slightly more than
$H^0$); whereas in Fig.\ \ref{fig8:discovery} there were substantial 
zones in which $H^0$ events dominated (as well as large segments wherein
the two Higgs boson contributions were within ${\sim}20$\% of each other).
Finally, though the cuts did typically eliminate slightly more $H^0$ 
events than $A^0$ events, this effect was of little significance.

\subsection{The topology of the signals}

Note that in comparing the signal with the MSSM backgrounds, the present 
study follows the standard procedure of comparing signal and background
rates at the same point in the MSSM parameter space.  One could well
ask whether or not larger backgrounds at a different point in parameter 
space could lead to the number of excess events attributed to the 
signal at the designated point in the MSSM parameter space.  
One way of addressing this issue is to look at the distribution of the 
signal+background events on a $M(e^+ e^-)$ {\it vs.} $M(\mu^+ \mu^-)$
wedgebox plot in addition to merely asking what is the raw rate.   
To wit, analyses of selected points in parameter space, again at the 
full event generator + detector simulation level, are presented
illustrating that: (1) small changes in the MSSM input parameters
can lead to significant topological changes in the pattern observed on the 
wedgebox plot; (2) the signal and background events often have 
markedly different distribution patterns on the wedgebox plot, pointing
toward the possibility of further purifying cuts (perhaps in 
conjunction with extra information garnered from other studies or
additional assumptions to clarify of what one is obtaining a purer
sample) such as the example presented for plot $(a)$ of 
Fig.\ \ref{Point2wedgeWBplot}; and (3) the composition of the 
$H^0,A^0 \rightarrow 4\ell + E_T^{\rm{miss}}$ signal, that is, what
percentages are due to  
$H^0, A^0 \rightarrow \widetilde{\chi}_i^0 \widetilde{\chi}_j^0$
for different $i$ and $j$, may be ascertained to some level.  The
basic topological features of the wedgebox plot provide strong,
often easily interpreted, leads as to which modes are the dominant 
contributors.  The locations of the edges of such features on the 
wedgebox plot also provide information about the sparticle spectrum.
The densities of event points in each component of wedgebox checkerboard
can also be used to distinguish wedgebox plots with the same 
topological features/edges, such as, for instance, telling a wedgebox plot 
with a $2$-$3$ wedge and a $2$-$2$ box from one with only a $2$-$3$ wedge.
Further, these point density distributions may be used to reconstruct 
information about the relative production rates of the different 
$H^0, A^0 \rightarrow \widetilde{\chi}_i^0 \widetilde{\chi}_j^0$
processes, though extracting such `dynamical' information may well be far 
more complicated than is the task of extracting `kinematical'
information about the sparticle spectrum from the locations of the edges.
All of this is further complicated by the remaining background 
events, and a more holistic study looking at both the Higgs boson
produced signal and the MSSM backgrounds together may be most
appropriate \cite{EWpaper}.

\section*{Note}

Motivated in part by the earlier archival submission of this work, a 
similar analysis was eventually
carried out by a member of ATLAS \cite{simonetta}, also aiming at
mapping out MSSM Higgs boson discovery regions via 
$H^0, A^0 \rightarrow \widetilde{\chi}_i^0 \widetilde{\chi}_j^0$ decays.
Results of this ATLAS analysis are essentially consistent with those 
presented herein, though the actual shapes of the discovery regions obtained
differ somewhat.  These differences are in part attributable to
adopting different selection criteria and employing different simulation 
tools.  Of particular note are the $t\bar t$ and $b\bar bZ^{0(*)}$ 
backgrounds which are quite significant in the case of the ATLAS analysis
but yield no background events in this study\footnote{Simulations 
of 40 million $b\bar bZ^{0(*)}$ ($t\bar t$) events yielded 1(0) event(s) 
passing the set of selection cuts.}.
This is mainly due to the more stringent lepton isolation criteria adopted 
for this study which are very effective at removing leptons produced in these 
two would-be background processes from $B$-mesons decays.
The restrictions on $E_T^\ell$, which are absent from \cite{simonetta}, 
also aid in removing residual background events.

\section*{Acknowledgments}

The authors thank the organizers of the 2003 Les Houches workshop 
in association with which earlier stages of this work were performed.
We also thank Guang Bian for assistance in preparing a couple 
of the figures.  Communications with Simonetta Gentile are gratefully 
acknowledged.
This work was supported in part by National Natural Science Foundation 
of China Grant No.\ 10875063 to MB and a Royal Society Conference Grant to SM, 
who is also supported
in part by the program `Visiting Professor - Azione D - Atto Integrativo 
tra la Regione Piemonte e gli Atenei Piemontesi'.

\def\pr#1 #2 #3 { {\rm Phys. Rev.}            {\bf #1}, #3 (#2)}
\def\prd#1 #2 #3{ {\rm Phys. Rev. D}          {\bf #1}, #3 (#2)}
\def\prl#1 #2 #3{ {\rm Phys. Rev. Lett.}      {\bf #1}, #3 (#2)}
\def\plb#1 #2 #3{ {\rm Phys. Lett. B}         {\bf #1}, #3 (#2)}
\def\npb#1 #2 #3{ {\rm Nucl. Phys. B}         {\bf #1}, #3 (#2)}
\def\prp#1 #2 #3{ {\rm Phys. Rep.}            {\bf #1}, #3 (#2)}
\def\zpc#1 #2 #3{ {\rm Z. Phys. C}            {\bf #1}, #3 (#2)}
\def\epjc#1 #2 #3{ {\rm Eur. Phys. J. C}      {\bf #1}, #3 (#2)}
\def\mpl#1 #2 #3{ {\rm Mod. Phys. Lett. A}    {\bf #1}, #3 (#2)}
\def\ijmp#1 #2 #3{{\rm Int. J. Mod. Phys. A}  {\bf #1}, #3 (#2)}
\def\ptp#1 #2 #3{ {\rm Prog. Theor. Phys.}    {\bf #1}, #3 (#2)}
\def\jhep#1 #2 #3{ {\rm J. High Energy Phys.} {\bf #1}, #3 (#2)}
\def\jphg#1 #2 #3{ {\rm J. Phys. G}           {\bf #1}, #3 (#2)}

\end{document}